\documentclass[a4paper,
		     11pt,
		     DIV21,
		     twocolumn,
		     oneside,
		     bibliography=totoc,
		     headsepline=false,
]{scrartcl}

\usepackage[T1]{fontenc}

\usepackage{amsmath}
\usepackage{amsfonts}
\usepackage{amsthm}
\usepackage{amscd}
\usepackage{grffile}
\usepackage{tikz} 
\usepackage{eurosym} 
\usepackage{graphicx}
\usepackage{psfrag}
\usepackage{listings}
\usepackage{curves}
\usepackage{calc}
\usepackage{picinpar}
\usepackage{enumerate}
\usepackage{algorithmic}
\usepackage{algorithm}
\usepackage{bm}
\usepackage{multibib}
\usepackage{hyperref}
\usepackage{textcase}
\usepackage{nicefrac}
\usepackage{stmaryrd}

\setkomafont{title}{\normalfont\Large}
\setkomafont{sectioning}{\normalfont}
\setkomafont{section}{\scshape}
\setkomafont{subsection}{\itshape}
\setkomafont{paragraph}{\itshape}

\title{A fully-coupled discontinuous Galerkin method for two-phase flow in porous media with discontinuous capillary pressure}

\author{\large\textit{Peter Bastian} \\
\large Heidelberg University, Interdisciplinary Center for Scientific Computing\\
\large Im Neuenheimer Feld 368, D-69120 Heidelberg, Germany\\
\large {\ttfamily peter.bastian@iwr.uni-heidelberg.de}
}

\date{\large\today}

\begin{document}

\maketitle

\begin{abstract}
In this paper we formulate and test numerically a fully-coupled discontinuous Galerkin (DG) me\-thod
for incompressible two-phase flow with discontinuous capillary pressure. The spatial discretization uses the
symmetric interior penalty DG formulation with weighted averages and is based on a wetting-phase
potential / capillary potential formulation of the two-phase flow system.  After discretizing in time with
diagonally implicit Runge-Kutta schemes the resulting systems of nonlinear algebraic equations are solved
with Newton's method and the arising systems of linear equations are solved efficiently and in parallel
with an algebraic 
multigrid method. The new scheme is investigated for various test problems from the literature
and is also compared to a cell-centered finite volume scheme in terms of accuracy and
time to solution.
We find that the method is accurate, robust and efficient.
In particular no post-processing
of the DG velocity field is necessary in contrast to results reported by several authors 
for decoupled schemes. Moreover, the
solver scales well in parallel and three-dimensional problems with up to nearly 100 million 
degrees of freedom per time step have been computed on 1000 processors.
\end{abstract}

\section{Introduction}

Modelling and simulation of multi-phase flow in porous media has important applications
ranging from secondary oil recovery \cite{Peaceman,AzizSettari} to 
nuclear waste repositories \cite{BourgeatJurakSmai2009} and CO$_2$ sequestration \cite{CO2Benchmark2009}.
In this paper we concentrate on the flow of two incompressible and immiscible phases
as a basic model which reduces, upon certain manipulation of the equations, to two coupled
nonlinear, time-dependent partial differential equations of elliptic-parabolic type providing several formidable difficulties for 
the analysis as well as for the numerical solution.

One complication of the model that we do allow is heterogeneity in the capillary pressure
function which can have a decisive effect on the flow of the fluids and requires careful numerical
treatment \cite{Kueper1,Kueper2}.
Modeling of two-phase flow with different capillary pressure curves in different
parts of the domain (termed ``discontinuous capillary pressure'' in the following)
has been studied extensively in the literature \cite{Bear72, VanDuijn1995, Molenaar97, VanDuijn1998}.
The mathematically correct form of the interface conditions has been derived through a regularisation
technique in \cite{VanDuijn1995}.
Significant advances in the model analysis have been achieved in \cite{Bertsch2003,Buzzi2009,Cances2009,doi:10.1137/11082943X}.

The treatment of discontinuous capillary pressures in numerical simulators has
been studied for around two decades in the context of a wide variety of different numerical schemes and
is still an active field of research.
Kueper and Frind \cite{Kueper1,Kueper2} used finite difference schemes and compared their
numerical simulations against experimental results.
Standard Galerkin and Petrov-Galerkin method are compared against upwind finite volume methods
by Helmig and Huber in \cite{Helmig97,Helmig1998697}. The authors find that upwinding is crucial and that standard Galerkin
and Petrov-Galerkin methods may produce unphysical results. Upwinding, however, implements the
required interface conditions only approximately.
Discontinuous capillary pressures in the context of various cell-centered finite volume schemes
with two-point flux approximation are studied in \cite{VanDuijn1995,MZA:8194530,Brenner2013} while
an extension to multi-point flux on hexahedral, non-conforming meshes has been given recently in \cite{Wolff2013}.
The vertex-centered finite volume method with an exact treatment of the interface conditions
has been introduced in \cite{Bastian99,BH99} and has been extended to fractured porous media in \cite{Kluft2004}.
The mixed finite element method is a very accurate, efficient and popular method for numerically solving
heterogeneous elliptic problems. The extension of this method to two-phase flow with capillary pressure
discontinuities has been presented for the multi-dimensional case in \cite{Hoteit200856}.

Discontinuous Galerkin (DG) finite element methods \cite{DGProceedings00,Riviere2008,DiPietroErn2012}
are a class of non-conforming
finite element schemes having several advantages: They are able to
achieve higher order of convergence while being locally conservative, are able to handle
full permeability tensors, unstructured, non-conforming meshes and can be used to
increase the ratio of floating point operations to memory accesses which is becoming increasingly 
important for modern computer architectures.
Moreover, DG schemes have been designed for elliptic \cite{RiviereWheelerGirault99,DGUnified02,Ern01042009}, parabolic \cite{LDG98,AizingerDawsonCockburnCastillo01,BastianLangCouplex} 
and hyperbolic problems \cite{DGV}. For a recent comparison of a wide range of discretization schemes
for 3d elliptic benchmark problems see \cite{FVCA6}.

Application of DG methods to incompressible, two-phase flow started about 10 years
ago with the work in \cite{Cisc2002,BastianRiviereTwoPhase,RiviereEccomas2004,KlieberRiviere2006}
which all use decoupled formulations where, per time step,
first a pressure equation is solved and then the saturation is advanced by an explicit time-stepping scheme
(also called IMPES: implicit pressure, explicit saturation).
For the pressure equation primal DG methods based on symmetric (SIPG)
or non-symmetric (NIPG) interior penalties (see \cite{DGUnified02}) are used.
If the saturation equation is hyperbolic upwinding and limiters are used to remove
unphysical oscillations and to ensure convergence to the correct solution.
In \cite{BastianRiviereTwoPhase} this is combined with an $H(\text{div})$ velocity reconstruction technique that
has been introduced and analyzed in \cite{BastianRiviere}.
Eslinger \cite{Eslinger2005} presented a decoupled scheme based on the local discontinuous Galerkin
scheme (LDG, \cite{LDG98})
and a Kirchhoff transformation of the nonlinear diffusion terms which can handle compressible fluids as well as
discontinuous capillary pressure. 
Decoupled formulations combining a mixed finite element method for the pressure equation
with an explicit DG method for the saturation equation have been presented in \cite{Nayagum2004,Hoteit200856}.
The paper of Hoteit and Firoozabadi \cite{Hoteit200856} introduced a formulation of
two-phase flow based on phase potentials and a new velocity variable that enables the
treatment of discontinuous capillary pressures within a mixed method.
The first fully coupled DG schemes for 2d horizontal two-phase flow have been put forward
in \cite{Epshteyn2007383,Riviere2008}. The authors compare two different formulations based on either
total fluid conservation or both phase conservation equations and do \textit{not} use 
upwinding nor slope limiting. The implicit Euler method is used for time discretization.
More recently, a decoupled DG scheme in one space dimension based on a global pressure / saturation 
formulation has been introduced in \cite{Ern20101491,Ern2012348}.
Discontinuous capillary pressure functions and consequently discontinuities in global pressure as
well as saturation are taken into account by a careful design of the penalty terms within an SIPG approach. Heterogeneity
and anisotropy in absolute permeability is taken care of by weighted averages introduced in \cite{Ern01042009}.
The authors emphasize that an $H(\text{div})$ reconstruction \cite{ErnHdiv2007,ErnMozolevskiSchuh2009}
of total velocity is strictly required
by the scheme and provide numerical evidence that unphysical oscillations occur if this is not done.
The saturation equation is discretized in time by the \textit{implicit} Euler method (albeit being decoupled)
and upwinding but no limiters are employed. The first DG method for compositional two-phase
flow has been presented by the same authors in \cite{ErnMozolevski2012}.

Due to the incompressibility constraint any simulator for the two-phase flow system requires
the solution of large, sparse linear systems. In the decoupled approach this is one system per time
step while in the fully-coupled approach non-linear algebraic systems need to be solved iteratively
resulting in the solution of several linear systems per time step. The multigrid method \cite{Bibel} 
is among the fastest iterative methods for solving linear systems of equations arising from the discretization of elliptic PDEs.
Multigrid methods are most developed for the systems arising from low-order finite element and finite
volume discretizations but in the last decade they have been extend to systems arising from
DG discretizations. Geometric multigrid applied to DG for the Poisson equation is analyzed in \cite{kanschat03:dg_multilevel,brenner05:_conver_multig_algor_inter_penal_method} while
heterogeneous elliptic problems have been treated recently in \cite{ayuso11:multilevel} and
smoothed aggregation multigrid solvers have been presented in \cite{PLH09a,olson_schroder_2011_jcp}

In this paper we present a fully-coupled symmetric interior penalty DG method for incompressible two-phase
flow based on a formulation using wetting-phase potential and capillary potential as primary
variables. As equations, total fluid conservation and conservation of the non-wetting phase with
a reconstruction of the same velocity variable as in Hoteit and Firoozabadi \cite{Hoteit200856} are used.
Discontinuity in capillary pressure functions is taken into account by incorporating the interface conditions
into the penalty terms for capillary potential. Heterogeneity in absolute permeability
ist treated by the weighted averages from \cite{Ern01042009,Ern20101491}.
The higher-order DG discretization in space is complemented by higher-order diagonally implicit Runge-Kutta methods in time.
The large-scale linear systems arising within the Newton scheme per Runge-Kutta stage are solved
with an algebraic multigrid method \cite{amg4dg} based on subspace correction \cite{xu92:iterative_methods_by_space_decomp}.

The rest of this paper is structured as follows: In section 2 we describe the two-phase flow
model and the specific formulation used. The DG discretization and fully-coupled solution
approach is introduced in section 3 while section 4 gives details about the algebraic
multigrid solver for the linear systems. Section 5 reports numerical results on four test cases
in several variants. A conclusion on the findings of this paper is provided in section 6.


\section{Two-phase Flow Model}

\subsection{Model Formulation}

The system of immiscible and incompressible flow of two phases $\alpha\in\{w,n\}$ in a domain $\Omega\subset
\mathbb{R}^d$ and time interval $\Sigma=(0,T)$ reads:
\begin{subequations} \label{eq:two_phase}
\begin{align}
\Phi \partial_t s_\alpha + \nabla \cdot v_\alpha &= q_\alpha \quad \text{in $\Omega\times\Sigma$} ,
\label{eq:mass_conservation}\\
v_\alpha &= -\lambda_\alpha K (\nabla p_\alpha - \rho_\alpha g \nabla d ) , \label{eq:darcy}\\
p_n - p_w &= \pi(s_w,x) ,\label{eq:capi_pressure}\\
s_w + s_n &= 1. \label{eq:sat_alg}
\end{align}
\end{subequations}
Here $\Phi$ is porosity,
$s_\alpha(x,t)$, $p_\alpha(x,t)$ are the unknown saturation and pressure depending on position $x$ and time $t$, 
$v_\alpha$ is the velocity of phase $\alpha$,
$q_\alpha$ are external sources and sinks, phase mobility $\lambda_\alpha=
k_{r\alpha}(s_\alpha)/\mu_\alpha$ is relative permeability divided by dynamic viscosity, $K(x)$ is the absolute permeability tensor, 
$\rho_\alpha$ is (constant)
mass density, $g$ is the gravitational acceleration, $d(x)$ is depth and $\pi(s_w,x)$ is the capillary pressure - 
saturation relationship. We are particularly interested in the case of discontinuous capillary pressure where 
the dependence of $\pi(s_w,x)$ on the position $x$ is discontinuous.
This results in discontinuous saturation
requiring special treatment \cite{VanDuijn1995,Molenaar97,BH99}.

A large variety of formulations of the two-phase flow system \eqref{eq:two_phase} have
been proposed in the literature based on the following options:
\begin{enumerate}
\item Provided the algebraic system \eqref{eq:capi_pressure}, \eqref{eq:sat_alg} is
invertible, two out of the four unknown functions $p_w, p_n, s_w$ and $s_n$ can be eliminated.
This results in various pressure-saturation and pressure-pressure formulations.
\item The system of variables can be transformed into new variables resulting in
global pressure based formulations \cite{chavent,Ern20101491} and potential based formulations \cite{Hoteit200856}.
\item The PDE system \eqref{eq:mass_conservation}, \eqref{eq:darcy}
can be taken as is or rearranged into a total fluid conservation equation coupled to one
phase conservation equation. Either the phase velocity or the total velocity $v_t = v_w+v_n$
can be used in the remaining phase conservation equation.
\end{enumerate}

Our formulation of the system \eqref{eq:two_phase} is based in part on the formulation
given by Hoteit and Firoozabadi in \cite{Hoteit200856}. Introducing the wetting-phase potential and capillary potential
as primary variables
\begin{equation}
\phi_{w} = p_w - \rho_w g d, \quad \phi_c = p_n-p_w  - (\rho_n-\rho_w)gd \label{eq:potentials}
\end{equation}
we can write the total velocity and the non-wetting phase velocity as
\begin{align}
v_t &= v_a - \lambda_n K \nabla \phi_c,  \label{eq:total_velo_potential} \\
v_n &= f_n v_a- \lambda_n K \nabla \phi_c, \label{eq:phase_velo_potential}
\end{align}
with the newly introduced velocity
\begin{equation} \label{eq:va}
v_a = - \lambda_t K \nabla \phi_w ,
\end{equation}
the total mobility $\lambda_t=\lambda_w+\lambda_n$ and the fractional flow function
$f_n = \lambda_n/\lambda_t$. 
Note that saturation can be
computed from the capillary potential provided the capillary pressure-saturation relationship
is invertible at a given position $x$:
\begin{equation}
s_w(x,t) = \psi(\phi_c(x,t) + (\rho_n-\rho_w)gd(x),x)
\end{equation}
where $\psi(\pi(s_w,x),x)=s_w$. Summing \eqref{eq:mass_conservation} for $\alpha=w,n$ 
and using \eqref{eq:sat_alg} we obtain the total fluid conservation equation
\begin{equation}
\nabla\cdot v_t = q_t \label{eq:total_fluid_conservation}
\end{equation}
where $q_t = q_w+q_n$. As second equation we use conservation of the nonwetting-phase:
\begin{equation}
\Phi \partial_t(1-\psi(\phi_c)) + \nabla\cdot v_n = q_n . \label{eq:nw_fluid_conservation}
\end{equation}
Inserting \eqref{eq:total_velo_potential} into \eqref{eq:total_fluid_conservation}
and \eqref{eq:phase_velo_potential} into 
\eqref{eq:nw_fluid_conservation} we obtain the final form of our formulation:
\begin{align}
- \nabla \cdot \left ( \lambda_t K \nabla \phi_w  + \lambda_n K \nabla \phi_c \right) &= q_t, \tag{A}\\
-\Phi\partial_t \psi(\phi_c) + \nabla\cdot ( f_n v_a - \lambda_n K \nabla \phi_c) &= q_n .\tag{B}
\end{align}
The first equation is elliptic with respect to $\phi_w$ with non-degenerate coefficient $\lambda_t$ depending
on $\phi_c$. The second equation is non-linear degenerate parabolic in $\phi_c$ and is coupled to the first equation
through the velocity $v_a$. In regions where $f_n=1$ (i.e. $s_n=1$) the two equations (A), (B)
coincide and the system becomes singular. Therefore we require that the wetting phase does not vanish.

Note that the differences of this formulation to the one given in \cite{Hoteit200856}
are two-fold: (i) we use capillary potential instead of saturation as a primary variable and (ii) we
use conservation of non-wetting phase instead of conservation of wetting phase
\begin{equation}
\Phi\partial_t s_w + \nabla\cdot(f_w v_a) = q_w \label{eq:FiroozabadiPhaseEquation}
\end{equation} 
which is always hyperbolic. In case of dominating capillary diffusion these diffusive
effects need to be incorporated via the velocity field $v_a$ whereas in our formulation
a diffusion term is present. Therefore we consider our formulation more suited to the case of
dominating capillary diffusion. 
On the other hand, if capillary diffusion effects are small, we need to rely on on
the non-wetting phase flow being in the same direction as the wetting-phase.
An advantage of pressure-pressure formulations is that 
part of the nonlinearity is moved from the diffusion term to the time derivative and that they
provide a set of persistent variables for two-phase compositional flow in case of phase appearance
and disappearance \cite{co2_2011}.

The equations (A), (B) are supplemented by boundary conditions 
\begin{align*}
\phi_w &= \Phi_w \quad\text{on $\Gamma_w^D$}, & v_t\cdot\nu&=J_t  &&\text{on $\Gamma_w^N$},\\
\phi_c &= \Phi_c \quad\text{on $\Gamma_n^D$}, & v_n\cdot\nu&=J _n &&\text{on $\Gamma_n^N$}
\end{align*}
with $\Gamma_w^D$ having non-zero measure and initial condition $\phi_c(x,0) = \phi_c^0(x)$.

\subsection{Interface Conditions}

We assume the domain $\Omega$ is partitioned into subdomains $\Omega^{(i)}$
with different capillary pressure saturation relationships: $\pi(s_w,x) = \pi^{(i)}(s_w) \  \forall x\in\Omega^{(i)}$.
For illustration consider two subdomains $\Omega^{(l)}$, $\Omega^{(h)}$ with corresponding curves
$\pi^{(l)}$, $\pi^{(h)}$ of Brooks-Corey \cite{BC} type and $p_e^{(l)}=\pi^{(l)}(1)<p_e^{(h)}=\pi^{(h)}(1)$, 
i.e. $\pi^{(l)}$ having a smaller
entry pressure:
\begin{center}
\begin{tikzpicture}[scale=0.7]
\filldraw[very thick, fill=white!90!black] (0,0)  rectangle (5,3);
\filldraw[very thick,white!70!black] (1.5,1)  rectangle (3.5,1.5);
\draw[very thick] (1.5,1)  rectangle (3.5,1.5);
\node[below] at (2.5,0) {$\Omega$};
\node at (0.7,0.5) {$\Omega^{(l)}$};
\node at (3.5,2.0) {$\Omega^{(h)}$};
\draw[->] (3.0,1.8) -- (2.5,1.25);
\end{tikzpicture}
\begin{tikzpicture}[scale=0.5]
\draw[->] (-1.5,0) -- (6.5,0) node[below right] (n) {$s_w$};
\draw[->] (0,-0.1) -- (0,6) node[left] (n) {$p_c$};
\draw[very thick] (0.6,6) .. controls (2,3.5) and (4,2.5) .. (6,1.5);
\draw[very thick] (0.1,6) .. controls (2,2) and (4,1) .. (6,0.5);
\draw[very thin] (-0.1,1.5) -- (6,1.5);
\node[left] at (-0.1,1.9) {$p_e^{(h)}$};
\draw[very thin] (-0.1,0.5) -- (6,0.5);
\node[left] at (-0.1,0.7) {$p_e^{(l)}$};
\node[below] at (0,0) {$0$};
\draw[very thin] (6,-0.1) -- (6,1.5);
\node[below] at (6,0) {$1$};
\draw[very thin] (3.6,-0.1) -- (3.6,1.6);
\node[below] at (3.8,0) {$s_w^\ast$};
\node at (1.0,2.3) {$\pi^{(l)}$};
\node[right] at (3.8,3.4) {$\pi^{(h)}$};
\end{tikzpicture}
\end{center}
In \cite{VanDuijn1995} it was derived through a regularization argument that at
the interface of the two regions capillary pressure $p_c=p_n-p_w$ is continuous
if, evaluted from the region $\Omega^{(l)}$, it is larger than the entry pressure of the region
$\Omega^{(h)}$. This corresponds to a critical saturation 
$s_w^\ast$ given by $\pi^{(l)}(s_w^\ast)=p_e^{(h)}$.
Otherwise capillary pressure is
discontinuous and evaluated from the region $\Omega^{(h)}$ is equal to the entry pressure.
Saturation is always discontinuous and takes on the two values $s_w^{(l)}$, $s_w^{(h)}$ given by:
\begin{equation} \label{eq:if_condition_normal}
\left\{ \begin{array}{ll}
\pi^{(h)}(s_w^{(h)}) = \pi^{(l)}(s_w^{(l)}) & \text{if $\pi^{(l)}(s_w^{(l)})\leq p_e^{(h)}$ }\\
\pi^{(h)}(s_w^{(h)}) = p_e^{(h)} & \text{else}
\end{array} \right. 
\end{equation}
In addition, wetting-phase pressure is continuous (assuming the wetting phase is always present)
and the fluxes in normal direction of both phases (and consequently the total flux) are continuous
at the interface.

The interface condition \eqref{eq:if_condition_normal} can be reformulated in terms of the
capillary potential $\phi_c$ by defining the entry potentials $\phi_e^{(i)}(x) = p_e^{(i)}-(\rho_n-\rho_w)gd(x)$
and setting:
\begin{equation} \label{eq:if_condition_potential}
\left\{ \begin{array}{ll}
\phi_c^{(h)} = \phi_c^{(l)} & \text{if $\phi_c^{(l)}\geq \phi_e^{(h)}$} \\
\phi_c^{(h)} = \phi_e^{(h)}    & \text{if $\phi_c^{(l)}< \phi_e^{(h)}$}
\end{array} \right. 
\end{equation}

\section{Discontinuous Galerkin Discretization}

\subsection{Notation}

For the formulation of the DG discretization we employ the notation of \cite{Ern20101491}.
By $\{\mathcal{T}_h\}_{h>0}$ we denote a family of shape regular triangulations of the domain
$\Omega$ consisting of elements $T$ which are either simplices or parallelipipeds (this condition
is only introduced for ease of notation) in $d=1,2,3$ space dimensions.
The diameter of $T$ is $h_T$ and $\nu_T$ is its unit outer normal vector. $F$ is an interior face if it is the
intersection of two elements $T^-(F), T^+(F)\in\mathcal{T}_h$ and $F$ has non-zero measure in $\mathbb{R}^d$.
All interior faces are collected in the set $\mathcal{F}_h^i$. Likewise, $F$ is a boundary face if it is
the intersection of some $T^-(F)\in\mathcal{T}_h$ with $\partial\Omega$ and has non-zero measure.
All boundary faces make up the set $\mathcal{F}_h^{\partial\Omega}$ and we set 
$\mathcal{F}_h = \mathcal{F}_h^i \cup \mathcal{F}_h^{\partial\Omega}$.
The diameter of $F\in\mathcal{F}_h$ is $h_F$ and with each $F\in\mathcal{F}_h$ we associate
a unit normal vector $\nu_F$ oriented from $T^{-}(F)$ to $T^+(F)$ in case of interior faces
and coinciding with $\nu_{T^-(F)}$ in case of boundary faces.

It is assumed that the finite element mesh $\mathcal{T}_h$ resolves the boundaries of
the subdomains $\Omega^{(i)}$. By $\mathcal{F}_h^{\Gamma}\subseteq\mathcal{F}_h^i$
we denote the interior faces located at media discontinuities. For any face $F\in\mathcal{F}_h^{\Gamma}$
the normal direction $\nu_F$ is chosen such that it is oriented from the element with higher entry pressure
to the element with lower entry pressure.

The DG finite element space of degree $p$ on the mesh $\mathcal{T}_h$ is
\begin{equation}
V_h^p = \left\{ v\in L^2(\Omega) : \forall T\in\mathcal{T}_h, v|_T\in \mathcal{P}_p\right\}
\end{equation}
where $\mathcal{P}_p$ is either $\mathbb{P}_p$, the set of polynomials of total degree $p$
or $\mathbb{Q}_p$, the set of polynomials of maximum degree $p$. 
A function $v\in V_h^p$ is two-valued on an interior face $F\in\mathcal{F}_h^i$ and
by $v^-$ we denote the restriction from $T^-(F)$ and by $v^+$ the restriction from $T^+(F)$.
For any point $x\in F \in \mathcal{F}_h^i$ we define the jump
\begin{equation}
\llbracket v \rrbracket (x) = v^-(x)-v^+(x)
\end{equation}
and the weighted average
\begin{equation}
\{ v \}_\omega (x) = \omega^- v^-(x) - \omega^+ v^+(x)
\end{equation}
for some weights $\omega^- + \omega^+ = 1$, $\omega^\pm \geq 0$.
A particular choice of the weights depending on the
absolute permeability tensor $K$ has been introduced in \cite{Ern2008,Ern01042009}. Assuming that
$K^\pm$ is constant on $T^\pm(F)$ they set
\begin{align*}
\omega^- &= \frac{\delta_{K\nu}^+}{\delta_{K\nu}^- + \delta_{K\nu}^+}, &
\omega^+ &= \frac{\delta_{K\nu}^-}{\delta_{K\nu}^- + \delta_{K\nu}^+} 
\end{align*}
with $\delta_{K\nu}^{\pm} =\nu_F^t K^\pm \nu_F$. 
The definitions of jump and average are extended to $x \in F \in \mathcal{F}_h^{\partial\Omega}$:
\begin{equation}
\llbracket v \rrbracket (x) = \{ v \}_\omega (x) = v^-(x) .
\end{equation}
Finally, we denote for any domain $Q$ by
\begin{equation*}
(v,w)_Q = \int_Q v\cdot w \ dx
\end{equation*}
the $L^2$ scalar product of two (possibly vector-valued) functions, by 
$|Q| = (1,1)_Q$ the measure of the set $Q$ and by 
$\langle a, b \rangle = 2ab/(a+b)$
the harmonic mean of two numbers.

\subsection{Total Fluid Conservation Equation}

The discrete weak form defining the DG method for the total fluid
conservation equation (A) is given for a test function $w$ and fixed time $t$ by
\begin{equation}\label{eq:weakformA}
\begin{split}
a_h&(\phi_{wh},\phi_{ch},w)  = \\
&\sum_{T\in\mathcal{T}_h} 
(\lambda_t K \nabla\phi_{wh} + \lambda_n K \nabla\phi_{ch}, \nabla w)_T \\
& -\hspace{-1em} \sum_{F\in\mathcal{F}_h^{i}\cup\mathcal{F}_h^{Dw}} 
\hspace{-1em} (\nu_F\cdot\{ \lambda_t K \nabla\phi_{wh} + \lambda_n K \nabla\phi_{ch} \}_\omega, \llbracket w \rrbracket)_F \\
& -\theta \hspace{-1em}\sum_{F\in\mathcal{F}_h^{i}\cup\mathcal{F}_h^{Dw}} 
 \hspace{-1em}(\nu_F\cdot\{ \lambda_t K \nabla w\}_\omega , \llbracket \phi_{wh} \rrbracket)_F \\
& + \hspace{-1em}\sum_{F\in\mathcal{F}_h^{i}\cup\mathcal{F}_h^{Dw} } 
\hspace{-1em}\gamma_{F,w} (\llbracket w \rrbracket , \llbracket \phi_{wh} \rrbracket)_F \\
\end{split}
\end{equation}
where we have split the domain boundary into Dirchlet and Neumann parts
\begin{align*}
\mathcal{F}_h^{D\alpha} = \{ F\in \mathcal{F}_h^{\partial\Omega} \ : \ F\subseteq \Gamma_\alpha^D \}, \\
\mathcal{F}_h^{N\alpha} = \{ F\in \mathcal{F}_h^{\partial\Omega} \ : \ F\subseteq \Gamma_\alpha^N \}
\end{align*}
and $\theta\in\{-1,0,+1\}$ results in the non-symmetric, incomplete and symmetric version
of the interior penalty DG method. The penalty factor $\gamma_{F,w}$ is crucial for the 
performance of the method and is chosen for interior faces as
\begin{align*}
\gamma_{F,w} &= m \langle \lambda_t^- \delta_{K\nu}^- , \lambda_t^+ \delta_{K\nu}^+ \rangle \frac{p (p+d-1) |F|}{\min(|T^-(F)|,|T^+(F)|)}
\end{align*}
and for boundary faces as
\begin{align*}
\gamma_{F,w} &= m \lambda_t^- \delta_{K\nu}^- \frac{p (p+d-1) |F|}{|T^-(F)|)}
\end{align*}
with a user-defined parameter $m$. A typical choice is $m=20$ for the examples shown
below. The definition of the penalty parameter takes into account the coefficient
of the elliptic equation, space dimension, polynomial degree and element form. It 
combines the choices from \cite{Ern2008,Ern01042009} and \cite{HoustonHartmann2008}.

The right hand side linear form incorporating source/sink term, Dirichlet and Neumann boundary conditions
is
\begin{equation}\label{eq:rhsA}
\begin{split}
l_h(w) &= \sum_{T\in\mathcal{T}_h} (q,w)_T 
- \hspace{-0.5em}\sum_{F\in\mathcal{F}_h^{Nw}} (J_t,w)_F \\
&\qquad -\theta \sum_{F\in\mathcal{F}_h^{Dw}} 
 (\nu_F\cdot (\lambda_t K \nabla w) , \Phi_{w})_F \\
&\qquad + \sum_{F\in\mathcal{F}_h^{Dw} } 
\gamma_{F,w}  (w ,\Phi_w )_F\\
\end{split}
\end{equation}

The transport equation equation (A) is coupled to the equation for 
total fluid conservation through the velocity $v_a$ defined in \eqref{eq:va}.
Within an element $T$ the discrete approximation of $v_a$ is computed from the discrete function
$\phi_{wh}$.  At interior and boundary faces $F$ its normal component
is computed as follows:
\begin{equation*}
V_a = 
\left\{ \begin{array}{ll}
\gamma_{F,w} \llbracket \phi_{wh} \rrbracket - \nu_F\cdot\{\lambda_t K \nabla \phi_{wh}\}_\omega &
\text{$F\in\mathcal{F}_h^i$} \\
\gamma_{F,w}  (\phi_{wh}-\Phi_w) -\nu_F\cdot \lambda_t K \nabla \phi_{wh} &
\text{$F\in\mathcal{F}_h^{Dw}$} \\
-\nu_F\cdot \lambda_t K \nabla \phi_{wh} & \text{$F\in\mathcal{F}_h^{Nw}$}
\end{array} \right . .
\end{equation*}
We emphasize that the velocity field used in this way 
in the transport equation is not in $H(\text{div},\Omega)$. 

\subsection{Non-wetting phase conservation equation}

In order to solve the time-dependent problem we follow the ``method of lines'' approach
discretizing first in space and then in time. The discrete weak formulation for the spatial
derivatives of the right hand side of equation (B) then reads:
\begin{equation}\label{eq:weakformB}
\begin{split}
b_h(&\phi_{wh}, \phi_{ch}, z) =  \\
&- \sum_{T\in\mathcal{T}_h} (f_n v_a - \lambda_n K \nabla\phi_{ch} , \nabla z)_T \\
&+\sum_{F\in\mathcal{F}_h^{i}\cup\mathcal{F}_h^{Dn}}
( \langle f_n^{\uparrow,-}, f_n^{\uparrow,+}\rangle \, V_a , \llbracket z \rrbracket )_F \\
& -\sum_{F\in\mathcal{F}_h^{i}\cup\mathcal{F}_h^{Dn}} 
 (\nu_F\cdot \{\lambda_n K\nabla\phi_{ch}\}_\omega ,\llbracket z \rrbracket )_F\\
& - \theta \hspace{-1em}\sum_{F\in \mathcal{F}_h^{i}\cup\mathcal{F}_h^{Dn}} 
 \hspace{-1em} (\nu_F\cdot\{\lambda_n K \nabla z\}_\omega ,  J(\phi_{ch}) )_F \\
& + \hspace{-1em}\sum_{F\in\mathcal{F}_h^{i}\cup\mathcal{F}_h^{Dn}} 
\hspace{-1em}\gamma_{F,n} (\llbracket z \rrbracket ,  J(\phi_{ch}))_F
\end{split}
\end{equation}
In line 3, $f_n^{\uparrow,\pm}$ denotes the upwind evaluation of the
fractional flow function which is obtained as follows. First, capillary
potential is evaluated through upwinding
\begin{equation}
\phi_{ch}^\uparrow = \left\{\begin{array}{ll}
\phi_{ch}^- & V_a \geq 0\\
\phi_{ch}^+ & \text{else}
\end{array}\right. .
\end{equation}
Then the saturations on either side are evaluated by inverting the 
correponding capillary pressure-saturation function
\begin{equation}
s_w^{\uparrow,\pm}(x,t) = \psi^\pm(\phi_{ch}^\uparrow + (\rho_n-\rho_w)gd(x),x)
\end{equation}
and with these the fractional flow function is computed on either side
\begin{equation}
f_n^{\uparrow,\pm} = \frac{ \frac{k_{rn}^\pm(1-s_w^{\uparrow,\pm})}{\mu_n}}
{ \frac{k_{rw}^\pm(s_w^{\uparrow,\pm})}{\mu_w} + \frac{k_{rn}^\pm(1-s_w^{\uparrow,\pm})}{\mu_n} } .
\end{equation}
Then in line 3 of equation \eqref{eq:weakformB} the flux is computed by taking the harmonic average of 
the two values of the fractional flow function on either side of the face.

In order to incorporate the interface condition \eqref{eq:if_condition_potential}
for capillary potential the penalty terms make use of the extended jump function
\begin{equation*}
J(\phi_{ch}) = \left\{ \begin{array}{ll}
\phi_{ch}^{(h)} - \phi_{ch}^{(l)}  & \phi_{ch}^{(l)}\geq \phi_e^{(h)}, F\in \mathcal{F}_h^\Gamma  \\
\phi_{ch}^{(h)} - \phi_e^{(h)} & \phi_{ch}^{(l)}<\phi_e^{(h)}, F\in \mathcal{F}_h^\Gamma \\
\phi_{ch}^- - \phi_{ch}^+ & F\in \mathcal{F}_h^i \setminus \mathcal{F}_h^\Gamma \\
\phi_{ch}^- & F\in \mathcal{F}_h^{Dn} 
\end{array} \right. .
\end{equation*}
It enforces \eqref{eq:if_condition_potential} (weakly) at media discontinuities
and defaults to the standard jump term at all other faces.

The factor used in the interior penalty term now employes the arithmetic
average of mobilites
\begin{align*}
\gamma_{F,n} &= m \frac{\lambda_n^-+\lambda_n^+}{2} \langle  \delta_{K\nu}^- , \delta_{K\nu}^+\rangle
\frac{p (p+d-1) |F|}{\min(|T^-(F)|,|T^+(F)|)}
\end{align*}
which is important to get the correct front propagation in case
of discontinuous initial conditions, cf. the discussion in \cite{Ern2012348}.
The user-defined parameter $m$ is typically chosen to be the same as for 
the total fluid conservation equation. 

Finally, the right hand side linear form for the transport equation reads:
\begin{equation}\label{eq:rhsB}
\begin{split}
r_h(z) &=  \sum_{T\in\mathcal{T}_h} ( q_n , z )_T
- \sum_{F\in\mathcal{F}_h^{Nn}} (J_n , z )_F \\
&\qquad -\theta \sum_{F\in\mathcal{F}_h^{Dn}} 
( \nu_F\cdot (\lambda_n K \nabla z), \Phi_{c} )_F\\
&\qquad + \sum_{F\in\mathcal{F}_h^{Dn} } 
\gamma_{F,w}  (z ,\Phi_c )_F\\
\end{split}
\end{equation}

\subsection{Fully-coupled Solution approach}

Following the method of lines approach the semi-discrete weak formulation
now consists of the following problem:
Find $\phi_{wh}(t), \phi_{ch}(t) : \Sigma \to V_h^p$ such that
\begin{equation}
\begin{split}
a_h(\phi_{wh}(t),\phi_{ch}&(t),w) + \partial_t \left( \Phi(1-\psi(\phi_{ch})), z \right)_\Omega  \\
&+ b_h(\phi_{wh}, \phi_{ch}, z) = l_h(w) + r_h(z)
\end{split}
\end{equation}
for all $t\in\Sigma$ and $w,z \in V_h^p$.
This equation comprises a large system of ordinary differential equations that
is now discretized using diagonally implicit Runge-Kutta schemes. In particular, we employ the
one step $\theta$ scheme which includes the implicit Euler and the Crank-Nicolson method
and the Alexander schemes of order two and three described in \cite{alexander:77}.

Within each Runge-Kutta stage a large, nonlinear system of algebraic equations
needs to be solved. This is done iteratively using Newton's method with line search
globalization strategy and inexact (iterative) solution of the Jacobian system, see
\cite{BH99} for details. The Jacobians
are generated numerically using first-order finite differences. 
In the following section we describe how the linear systems are solved.

\section{Solution of Linear Systems}\label{Sec:Solver}

Let the linear system that is to be solved in each step of Newton's method
be denoted by
\begin{equation}\label{eq:linear_system}
A x = b . 
\end{equation}
The matrix $A$ is large, sparse and contains a block that stems from the discretization
of an elliptic PDE with varying coefficients \cite{BH99}. Therefore, the condition
number is expected to be of order $\kappa(A)=O(h^{-2})$ which requires robust and
efficient preconditioners to be able to solve large-scale problems
(see Section \ref{sec:solver_results} for a quantitative assessment of this claim).

Multigrid methods are among the most efficient preconditioners for solving
linear systems arising from the discretization of elliptic and parabolic PDEs \cite{TrottenbergBook2001}.
In particular, algebraic multigrid is well suited to handle problems with
varying coefficients \cite{ruge87:multig_method_amg}. Our preconditioner 
is based on the aggregation-based algebraic multigrid variant introduced
independently by several authors in the 1990s \cite{braess95:amg,PVanek_JMandel_MBrezina_1996a,Raw}.
Features of this method are its robustness for elliptic problems with varying
coefficients, its applicability to systems of partial differential equations and its
parallel scalability to a large number of processors \cite{blattamg}. A comparison
of geometric multigrid and two algebraic multigrid variants (including our implementation)
for large-scale anisotropic
elliptic problems has been given recently in \cite{Eike2013}.

Although algebraic multigrid methods may be constructed that can directly be applied
to linear systems arising from discontinuous Galerkin discretizations \cite{johannsen05:nipg_multigrid}
our method exploits the fact that the standard conforming finite element space
\begin{equation}
W_h = \left\{ v\in C^0(\Omega) : \forall T\in\mathcal{T}_h, \, v|_T\in \mathcal{P}_1\right\}
\end{equation}
is a subspace of the DG finite element space $V_h^p$ (provided the mesh is conforming).
Since low-frequency errors can be represented well in $W_h$ it is sufficient to apply a 
standard single-grid preconditioner to the fine grid DG system and combine it multiplicatively with
an algebraic multigrid preconditioner seeking a correction in the subspace $W_h$
in the sense of \cite{xu92:iterative_methods_by_space_decomp}.
This method has been discussed in \cite{amg4dg} for heterogeneous elliptic
problems in a sequential implementation. Here it is applied to the full two-phase flow
problem in a parallel implementation.

The error propagation matrix $E$ of a generic linear iterative method
$x^{(k+1)} = x^{(k)} + B (b-A x^{(k)})$ for solving \eqref{eq:linear_system}
employing  the preconditioner $B$ is given by $E = I-BA$.
The error propagation matrix $E_C$ of the combined AMG/DG preconditioner  
can be written as 
\begin{equation}
\begin{split}
E_{C} &= (I-B_{DG}A)^{\nu_2} \\
&(I- R^T B_{AMG} R A)
(I-B_{DG})^{\nu_1} 
\end{split}
\end{equation}
where $B_{DG}$ is a single grid preconditioner for the DG system $A$, e.g. block Gau\ss-Seidel,
block SSOR or block ILU with one block corresponding to all degrees of freedom
associated with a mesh element. The entries of the restriction matrix $R$ are given by
the representation of the basis functions $\varphi_i$ of $W_h$ w.r.t. the basis functions
$\psi_j$ of $V_h^p$:
\begin{equation} \label{eq:restriction_matrix}
\varphi_i^{CG} = \sum_{j=1}^{n_{DG}} r_{ij} \psi_j^{DG}.
\end{equation}
Finally, $B_{AMG}$ is the AMG preconditioner for the matrix 
$$A_{CG} = R A R^T.$$
The combined preconditioner can be derived in a fully algebraic way
except the step \eqref{eq:restriction_matrix} where information about the
finite element basis functions is needed.

\section{Numerical Results}

\subsection{Remarks on the Implementation}

All methods discussed in this paper have been realized within the 
\textit{Distributed and Unified Numerics Environment (DUNE)} \cite{Dune2008a,Dune2008b}
and DUNE-PDELab \cite{pdelabalgoritmy}.
DUNE is a flexible software framework that provides standardized interfaces to various
parallel, hierarchical mesh representations, sparse linear algebra operations and finite
element basis functions. The DUNE-PDELab module, which is based on the DUNE
framework, allows the implementation of discretization schemes with relatively small
coding effort (e.g. the complete DG scheme for two-phase flow takes less than 1000 lines
of C++ code) and provides time discretizations and solvers in a reusable form.

\subsection{Test Case 1: Van Duijn--De Neef Problem}

In \cite{VanDuijn1998} the authors derived an analytical solution for a one-dimensional
two-phase flow problem with heterogeneous capillary pressure. This problem has been
used as a test problem in \cite{Hoteit200856,Ern20101491}. Here we use the same parameters as
Ern et al. in \cite{Ern20101491}.

The domain $\Omega=(0,1.2)$ is divided into two subdomains
$\Omega^{(l)}=(0,0.6)$ and $\Omega^{(r)}=(0.6,1.2)$. The parameters
for the two-phase problem are $\Phi=1$, $\rho_w, \rho_n=1$, $\mu_w,\mu_n=1$.
Relative permeabilites are of Brooks-Corey type  \cite{BC} with $\lambda=2$
and are idential in both subdomains:
\begin{equation*}
k_{rw}(s_w) = s_w^\frac{2+3\lambda}{\lambda}\hspace{-0.5em}, 
\quad k_{rn}(s_n) = s_n^2\left( 1-(1-s_n)^\frac{2+\lambda}{\lambda} \right) .
\end{equation*}
In the implementation we set $k_{r\alpha}=0$ if $s_\alpha<0$ and
$k_{r\alpha}=1$ if $s_\alpha>1$. 
Absolute permeability is heterogeneous. For test case 1a we set
$$ K^{(l)} = 1, \qquad K^{(r)} = 0.25$$
while for test case 1b we set
$$ K^{(l)} = 1, \qquad K^{(r)} = 0.64.$$
The capillary pressure function is also of Brooks-Corey type with $\lambda=2$
in both subdomains 
\begin{equation}\label{eq:BrooksCoreyPc}
 \pi(s_w,x) = p_e(x) s_w^{-1/\lambda}
\end{equation}
and entry pressures given by $p_e(x) = \sqrt{1/K(x)}$.

Our formulation is based on the inverse of $\pi$ which is regularized by replacing it with straight lines
if $p_c<p_e$ or $p_c>R p_e$ for some parameter $R>1$:
\begin{equation}\label{eq:PcRegularization}
\psi(p_c) = \left\{\begin{array}{ll}
1-\frac{\lambda}{p_e}(p_c-p_e) & p_c<p_e \\
\frac{1}{R^{\lambda}}-\frac{\lambda(p_c-R p_e)}{R^{1+\lambda}p_e}  & p_c > R p_e\\
\left(p_e/p_c\right)^{\lambda} & \text{else}
\end{array}\right. .
\end{equation}
We used $R=4$ for test case 1a and $R=6$ for test case 1b.

As boundary conditions we set $\phi_{wh}(0)=0$, $\phi_{ch}(0)=1+10^{-4}$
and $\nu\cdot v_t(1.2)=0$, $\nu\cdot v_n(1.2)=0$. The initial condition was
$\phi_{wh}(x)=0$, $\phi_{ch}(x)=1+10^{-4}$ for $x\in\Omega^{(l)}$ and
$\phi_{ch}(x)=1.5R$ for $x\in\Omega^{(r)}$ (which corresponds to $s_w=0$) .

\begin{figure*}
\includegraphics[width=0.499\textwidth]{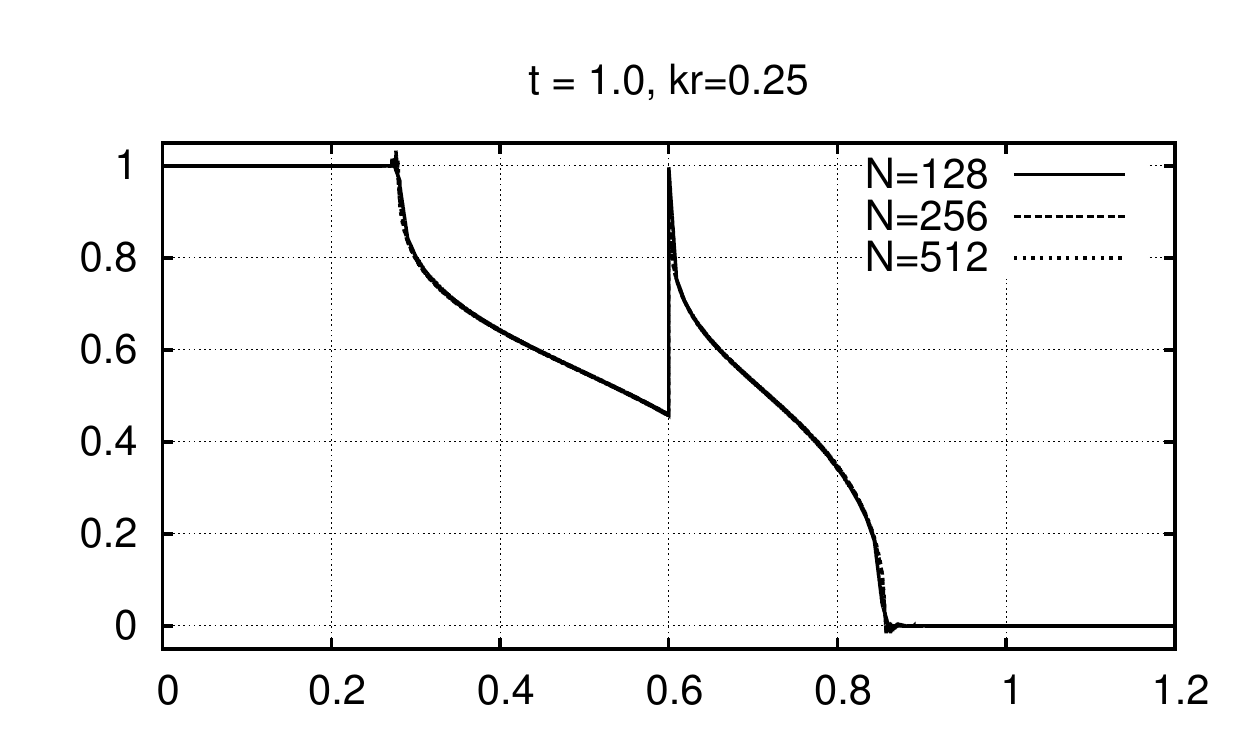}\hfill
\includegraphics[width=0.499\textwidth]{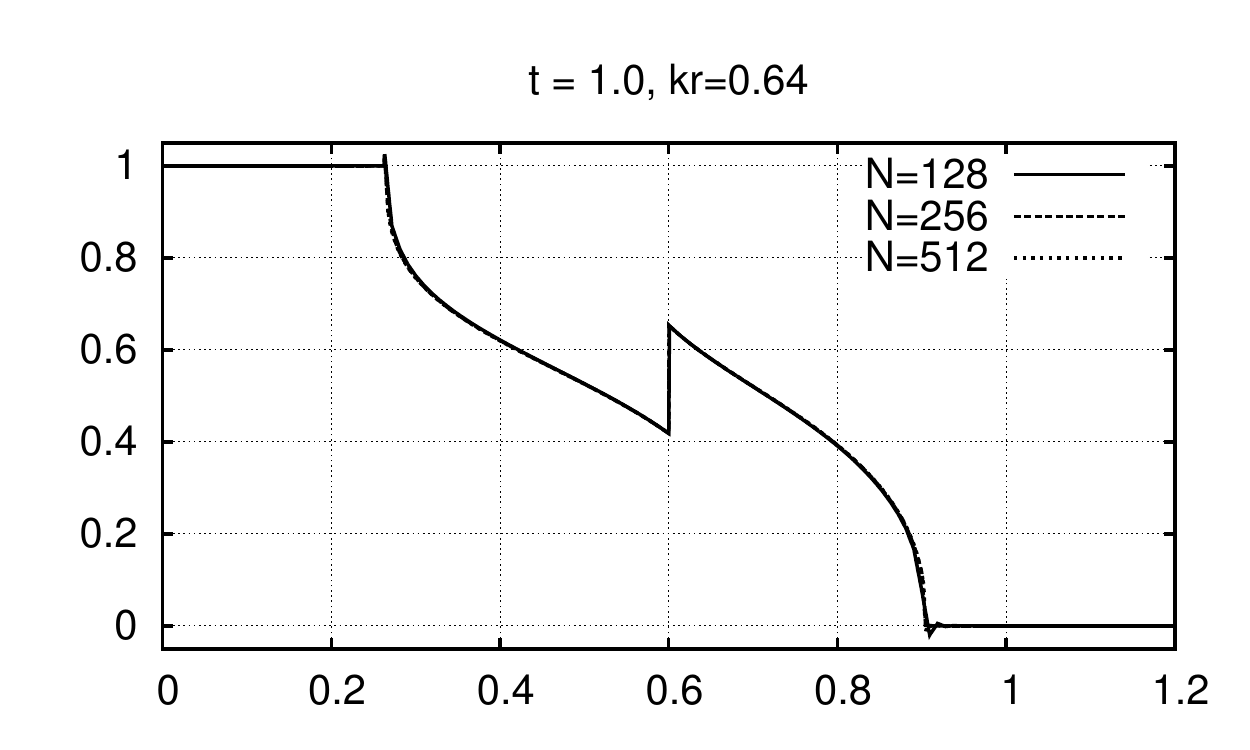}\\[-5mm]
\includegraphics[width=0.24\textwidth]{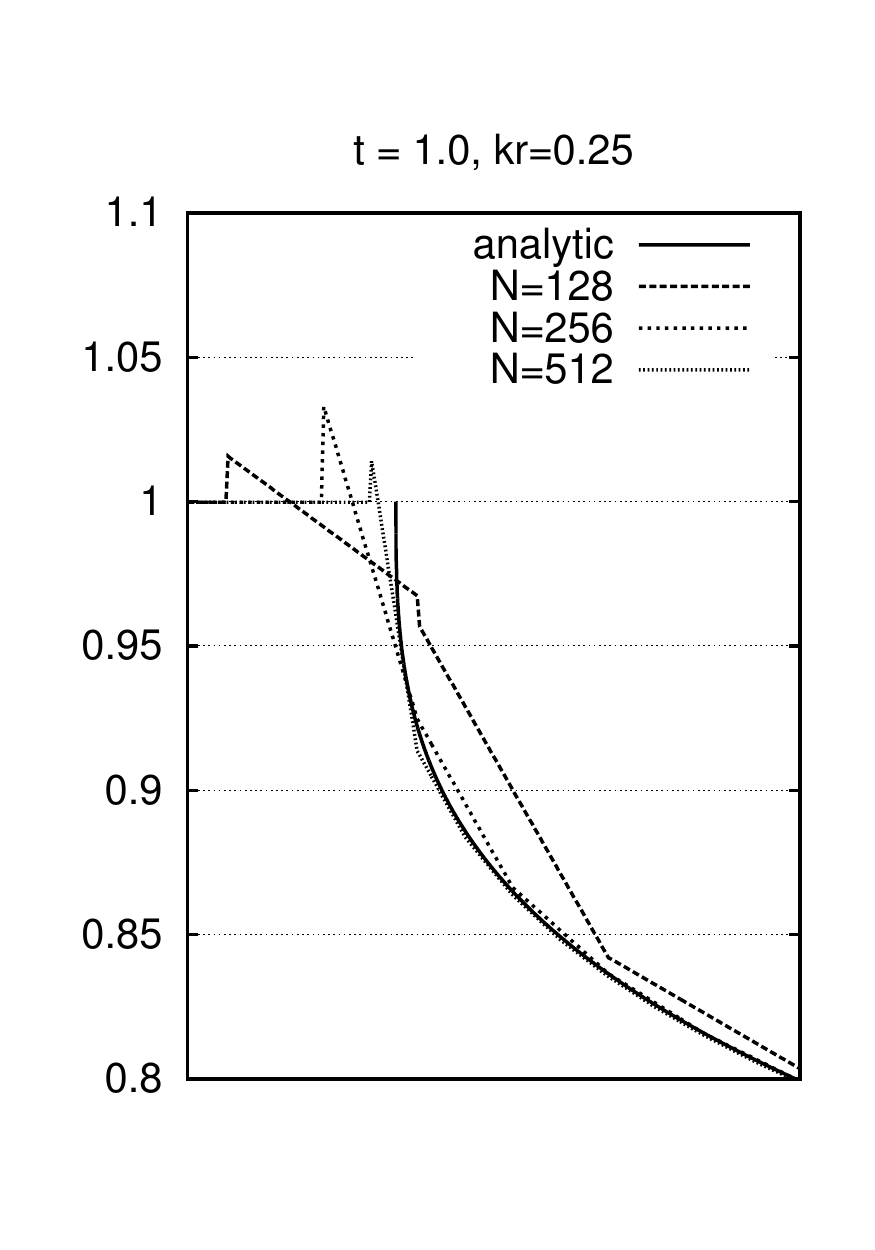}\hfill
\includegraphics[width=0.24\textwidth]{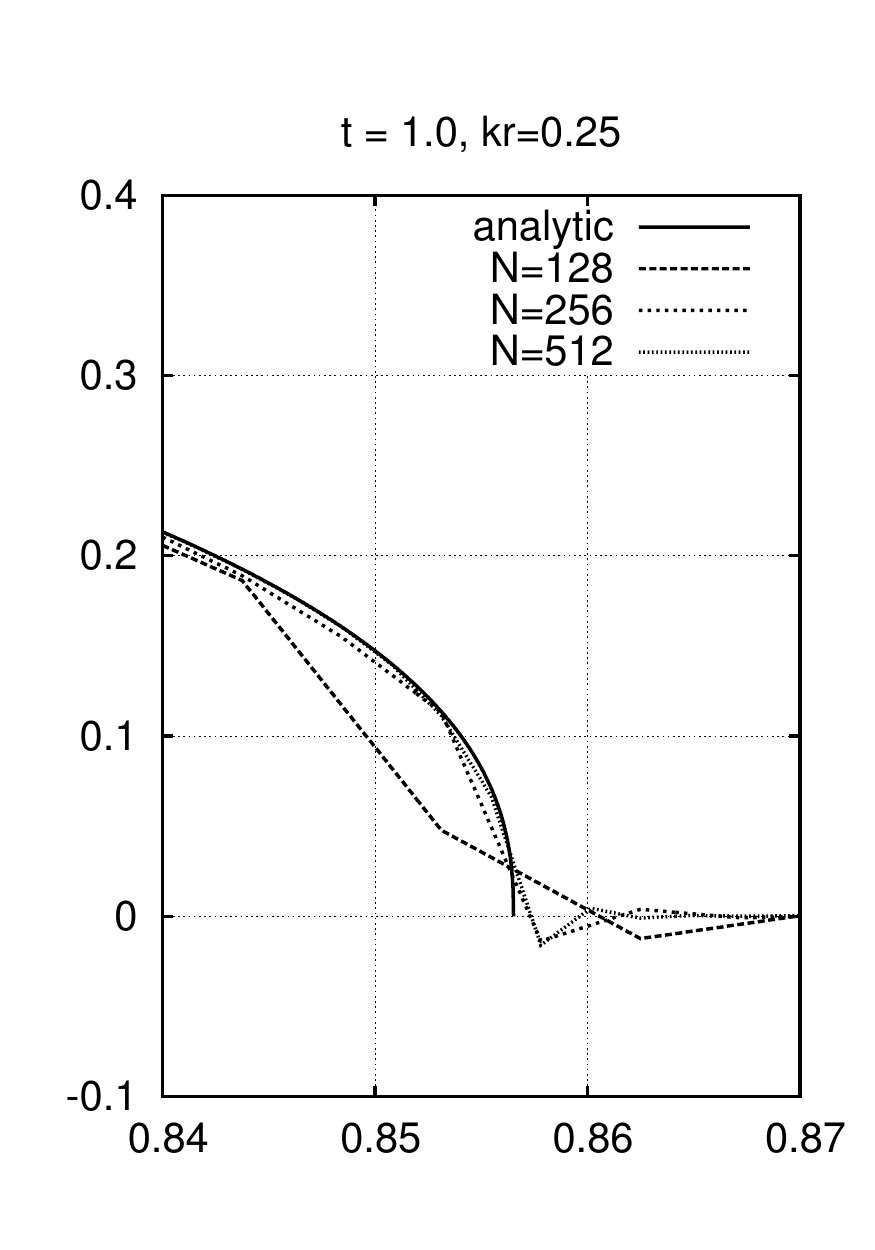}\hfill
\includegraphics[width=0.24\textwidth]{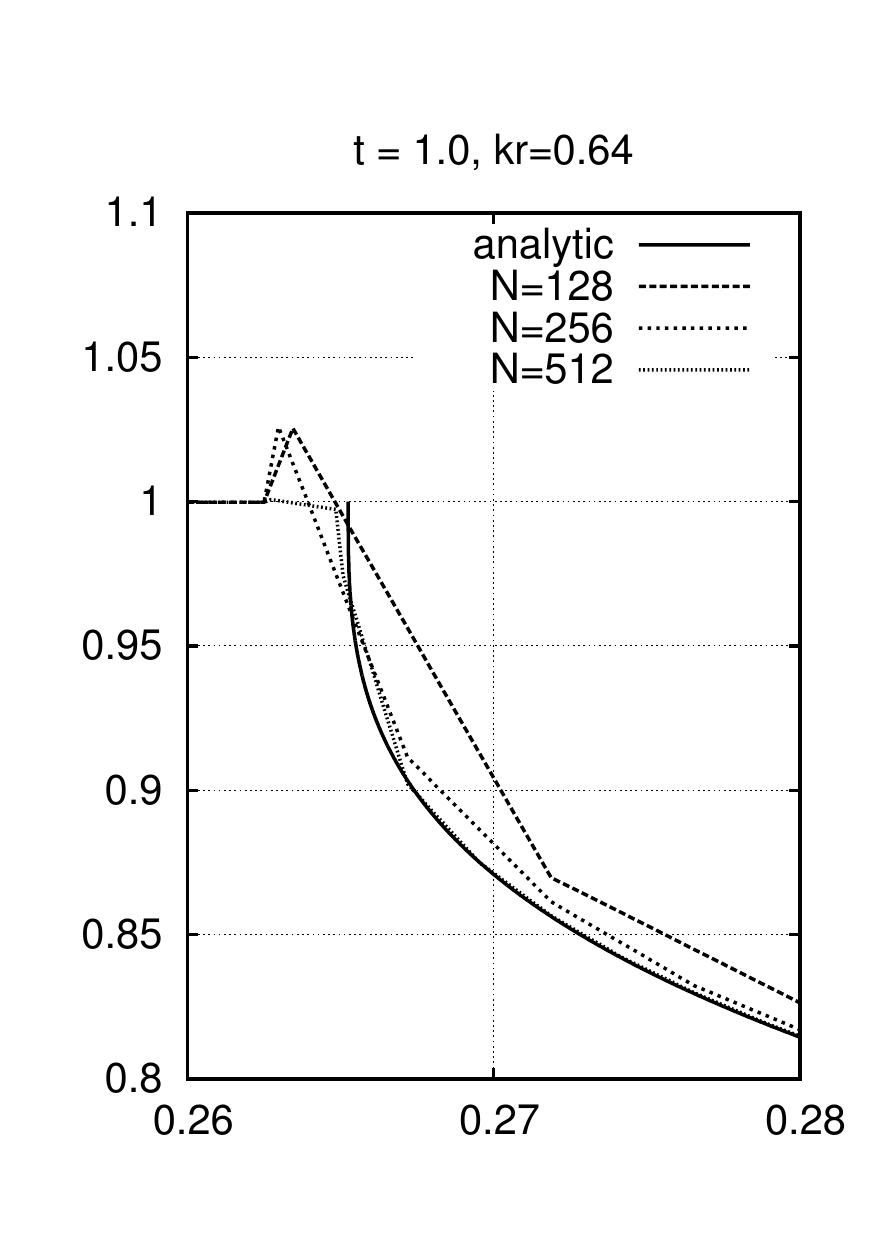}\hfill
\includegraphics[width=0.24\textwidth]{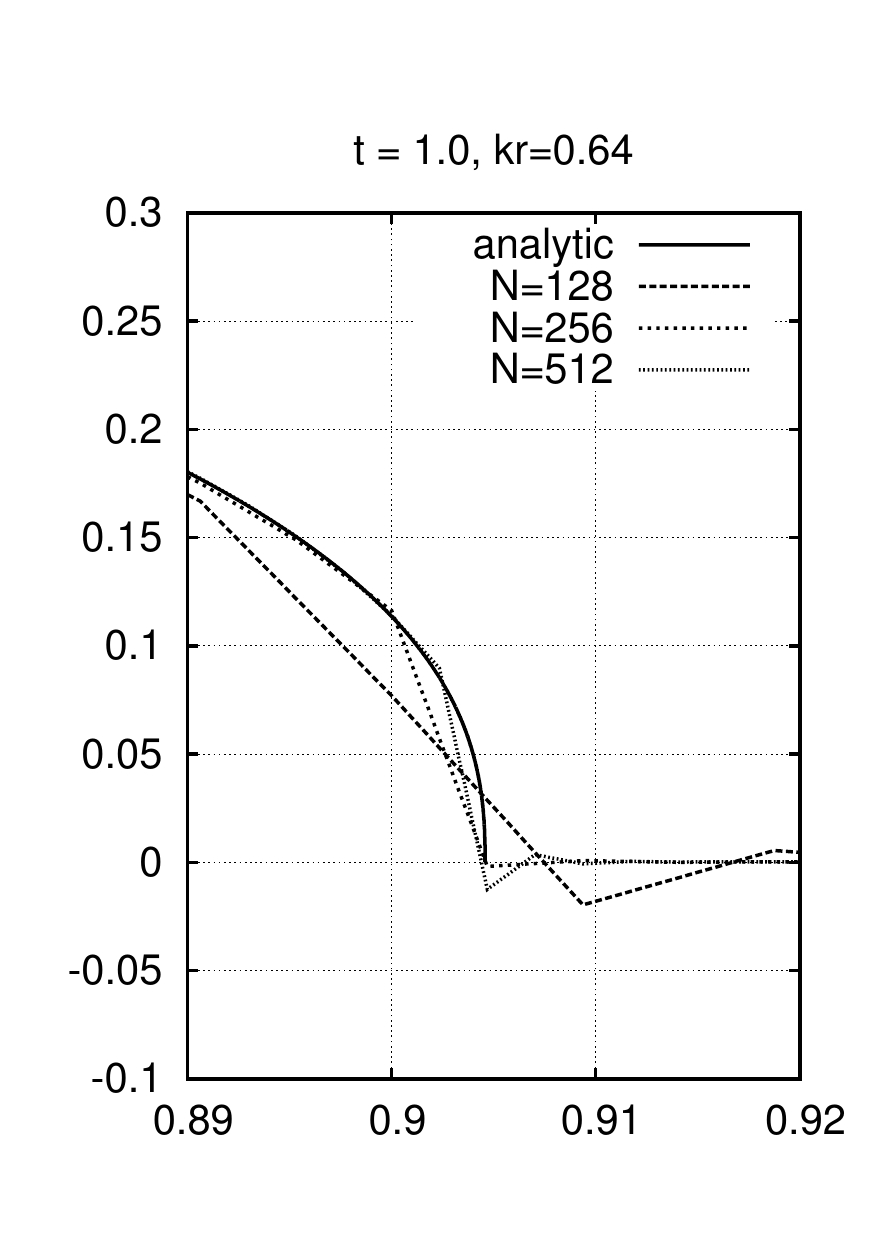}\\[-10mm]
\caption{Wetting phase saturation in test case 1 at time $t=1$.
First row shows case 1a (left) and case 1b (right). Second row shows details of the
solution close to the free boundary for case 1a and 1b.}
\label{fig:test_case_1}       
\end{figure*}

Numerical results for test case 1a/b are shown in Figure \ref{fig:test_case_1} at time $t=1$ 
for three different spatial meshes employing 128, 256 and 512 equidistant elements.
In all computations polynomial degree $p=1$ for both variables
and the second order Alexander scheme in time has been used.
The figure shows excellent agreement of the numerical solution with the analytical
solution even on the coarsest mesh. Comparing with the results given in \cite{Ern20101491}
we find that our solution is already more accurate on the coarse meshes. Note that in
\cite{Ern20101491} only the saturation equation is solved based on a fractional
flow formulation and equidistant time steps were taken. In our simulations we solve
the fully-coupled two phase problem where the time step size was chosen adaptively by
the algorithm depending on the convergence of the Newton method. The actual
number of time steps taken is reported in Table \ref{tab:test_case_1}. The average time step
size corresponds roughly to the size of the time steps taken in \cite{Ern20101491}.

Figure \ref{fig:test_case_1} shows that the media discontinuity is captured well by
the scheme without any oscillations. The second row of Figure \ref{fig:test_case_1} shows details
of the solutions in the vicinity of the free boundary. The results indicate that the
scheme converges towards the analytic solution and the steep front is well captured.
There are small oscillations at the free boundary that are reduced as the mesh is refined.

\begin{table}
\caption{Number of time steps for test case 1 to compute the time interval $(0,1)$.}
\label{tab:test_case_1}       
\begin{center}
\begin{tabular}{rrrr}
\noalign{\smallskip}\hline\noalign{\smallskip}
$N$ & $\Delta t_{\max}$ & $k_r=0.64$ & $k_r=0.25$ \\
\noalign{\smallskip}\hline\noalign{\smallskip}
128 & $1\cdot 10^{-2}$  & 187 & 184\\
256 & $5\cdot 10^{-3}$ & 296 & 282 \\
512 & $2.5\cdot 10^{-3}$  & 484 & 497 \\
\hline\noalign{\smallskip}
\end{tabular}
\end{center}
\end{table}

\subsection{Test Case 2: Ern et al. Problem}

This problem is again taken from \cite{Ern20101491} where it was used to illustrate
the importance of the $H(\text{div})$ reconstruction of total velocity in the
decoupled scheme based on a fractional flow formulation. Here we use it to
illustrate that such a reconstruction is \textit{not} necessary in our fully-coupled approach.
Note, that in \cite{Ern2012348} a corrigendum concerning this test case was published.

The domain $\Omega=(0,2)$ is one-dimensional and is partitioned
into two subdomains $\Omega^{(l)}=(0,1)$ and $\Omega^{(r)}=(1,2)$.
The parameters are $\Phi = 0.2$, $\rho_w, \rho_n=1$, $\mu_w,\mu_n=1$ and
$K=1$. Relative permeabilites are of Brooks-Corey type with $\lambda=2$ in both subdomains.
The capillary pressure-saturation functions are
\begin{equation}\label{eq:pc_ern}
\pi^{(l)} = 5(1-s_w)^2, \quad \pi^{(r)} = 4(1-s_w)^2+1
\end{equation}
resulting in a critical saturation $s_w^\ast=1/\sqrt{5}$.
These functions are somewhat unusual and 
the fact $\frac{d}{ds_w}\pi^{(i)}(1)=0$ results in an infinite slope
of the inverse capillary pressure-saturation
relationship $\psi$ at entry pressure.
Therefore, $\psi^{(i)}$ is regularized for $p_c<\tilde{p}_e^{(i)} = \pi^{(i)}(1) + 10^{-2}$
by a straight line in a $C^1$ fashion.

The boundary conditions are $\phi_{wh}(0)=1.8$, $\phi_{wh}(2)=0$,
$\phi_{ch}(0)=0$, $\nu\cdot v_n(2)=0$ and the initial conditions are
\begin{equation*}
\phi_{ch}(x) = \left\{\begin{array}{rl}
5\cdot (0.9)^2 & 0.1 < x < 0.9 \\
1 & x>1 \\
0 & \text{else}
\end{array}\right . .
\end{equation*}

\begin{figure*}
\includegraphics[width=0.499\textwidth]{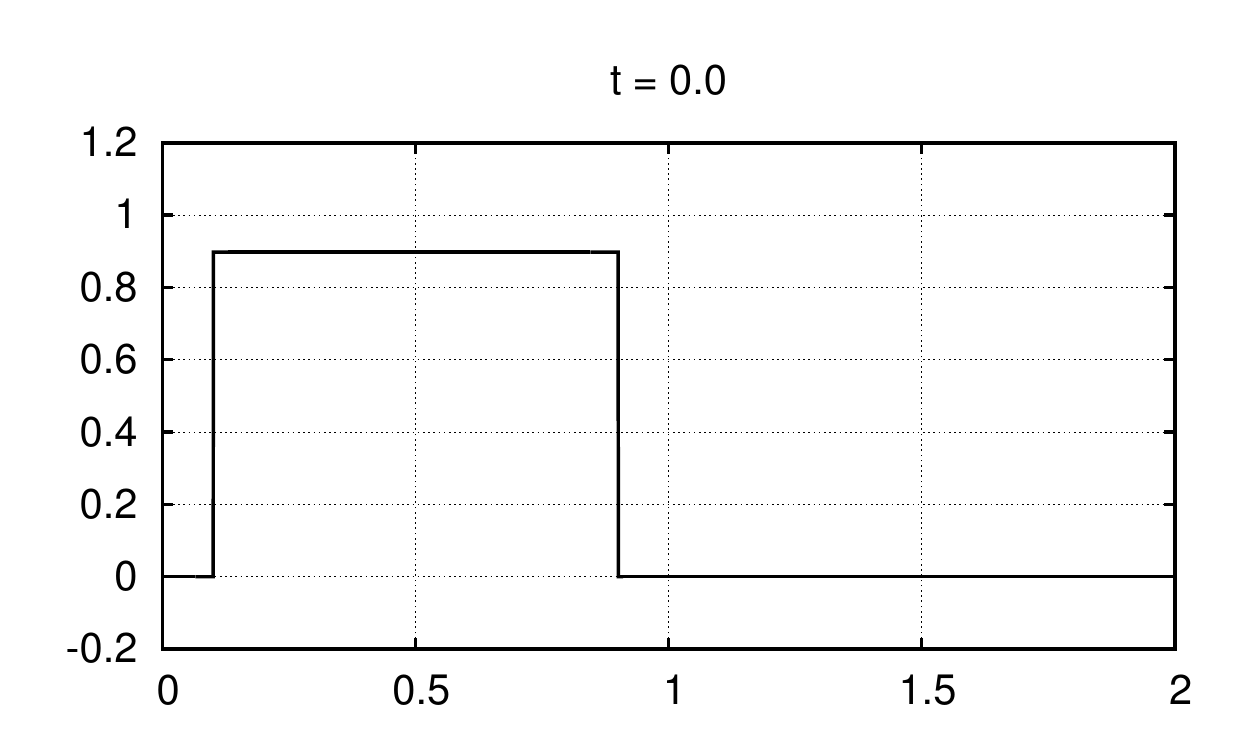}\hfill
\includegraphics[width=0.499\textwidth]{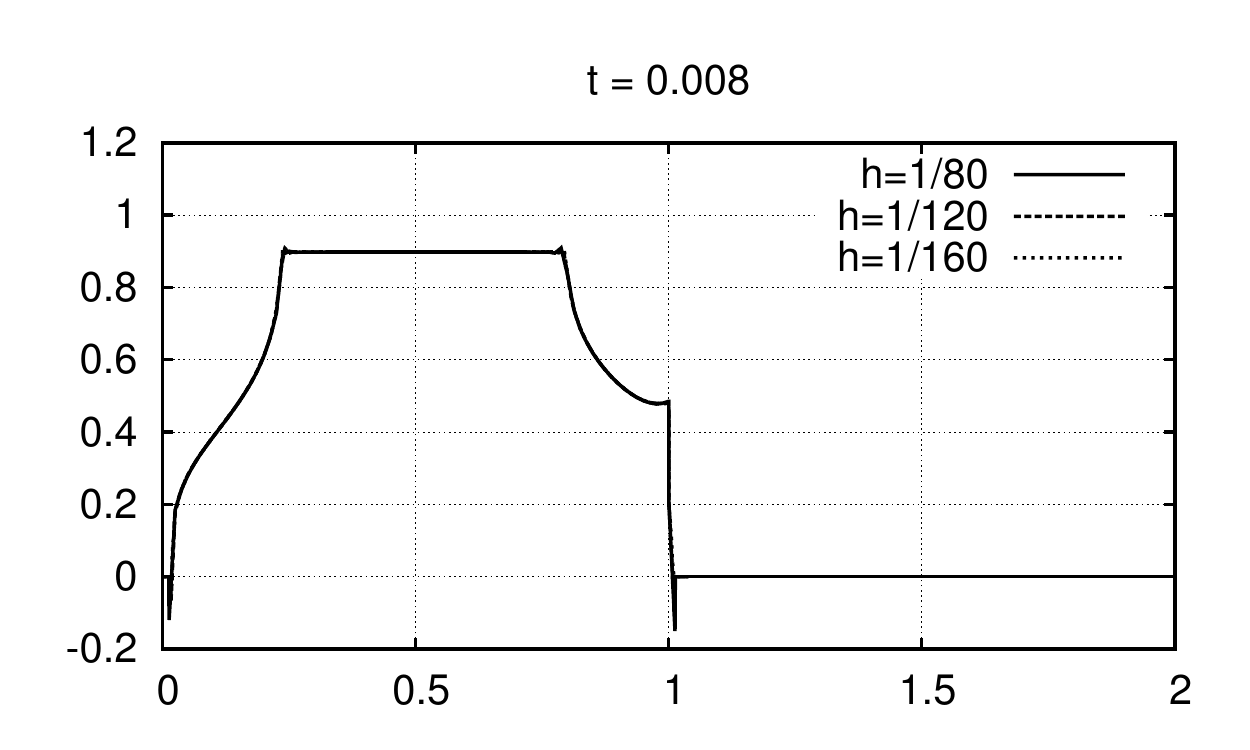}\\[-6mm]
\includegraphics[width=0.499\textwidth]{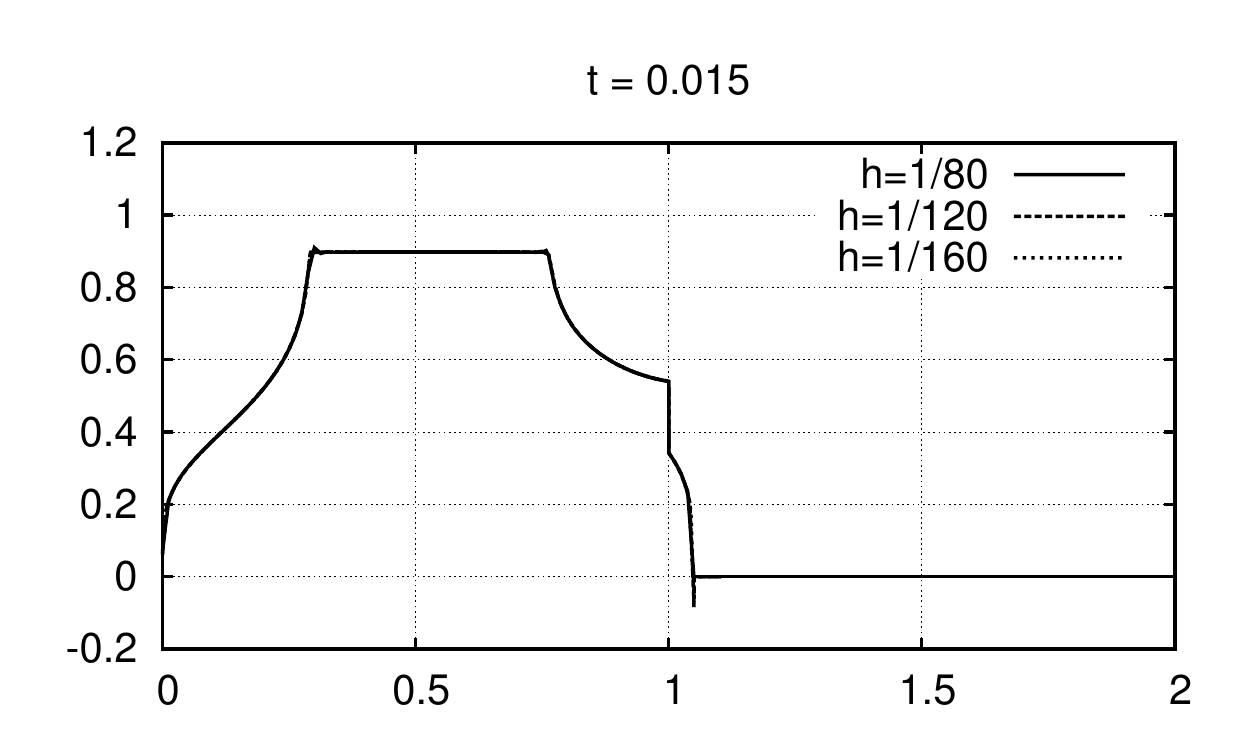}\hfill
\includegraphics[width=0.499\textwidth]{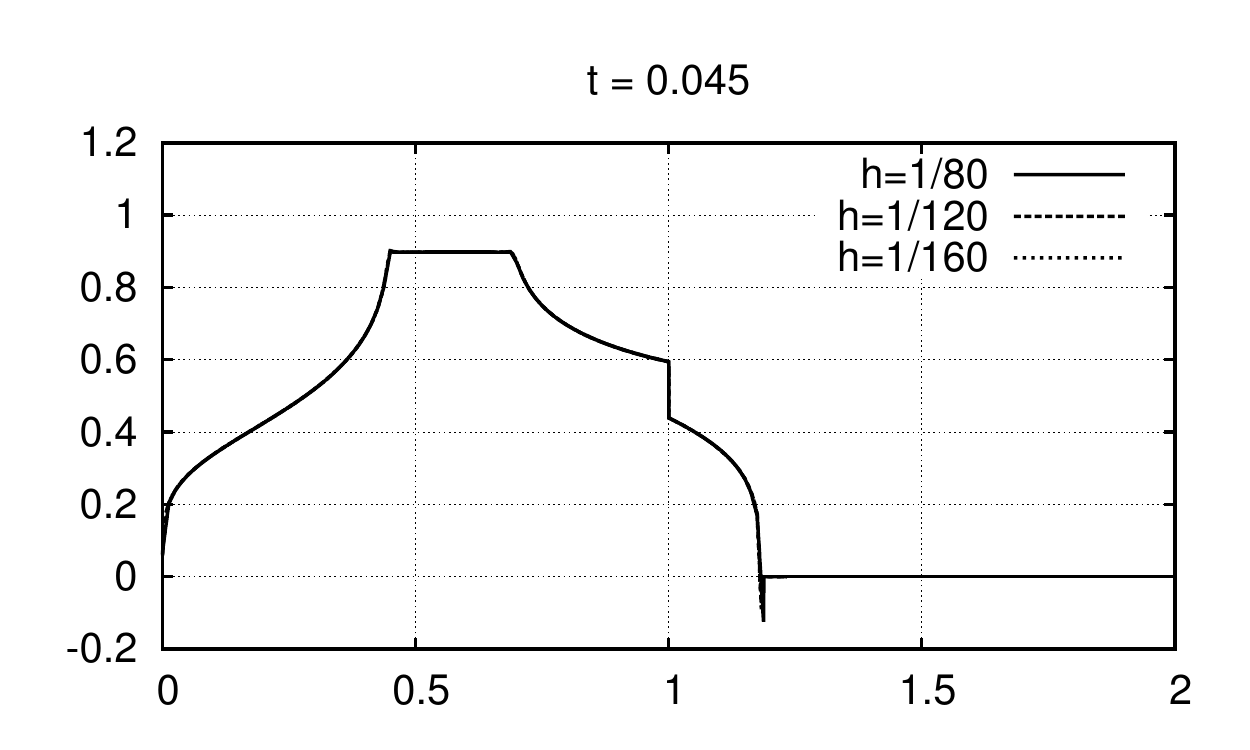}\\[-6mm]
\includegraphics[width=0.499\textwidth]{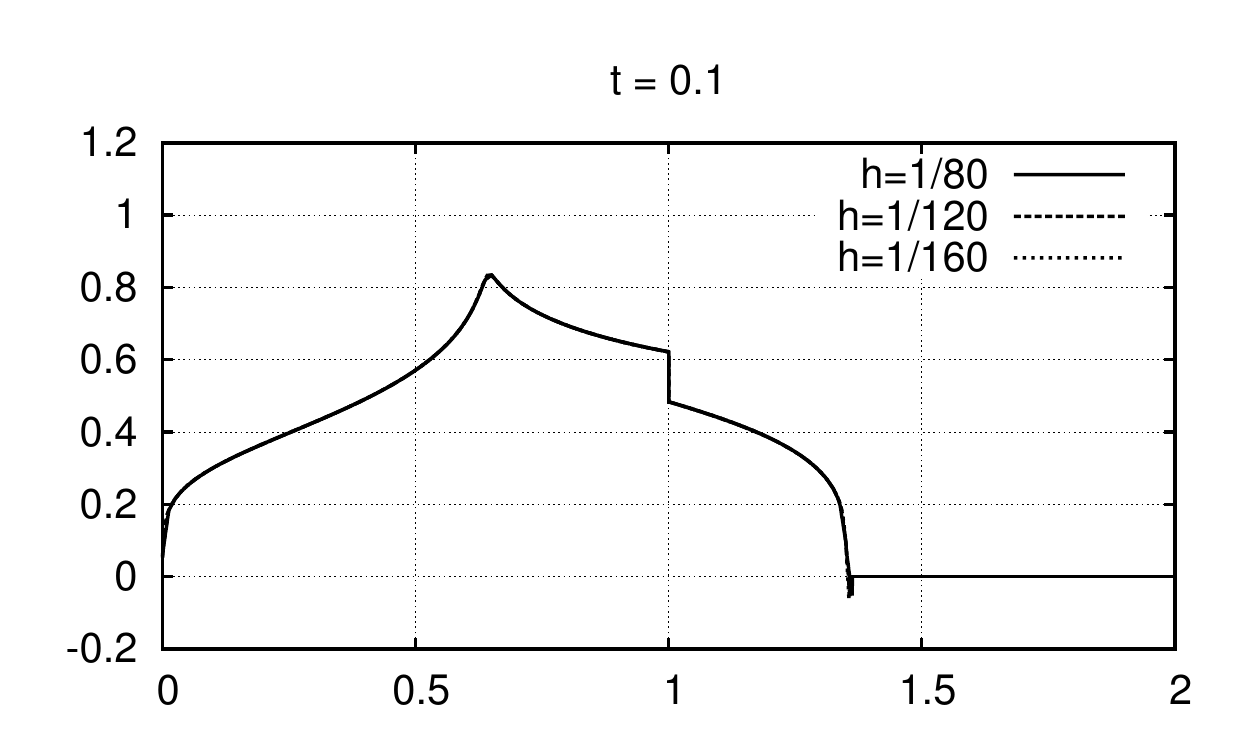}\hfill
\includegraphics[width=0.499\textwidth]{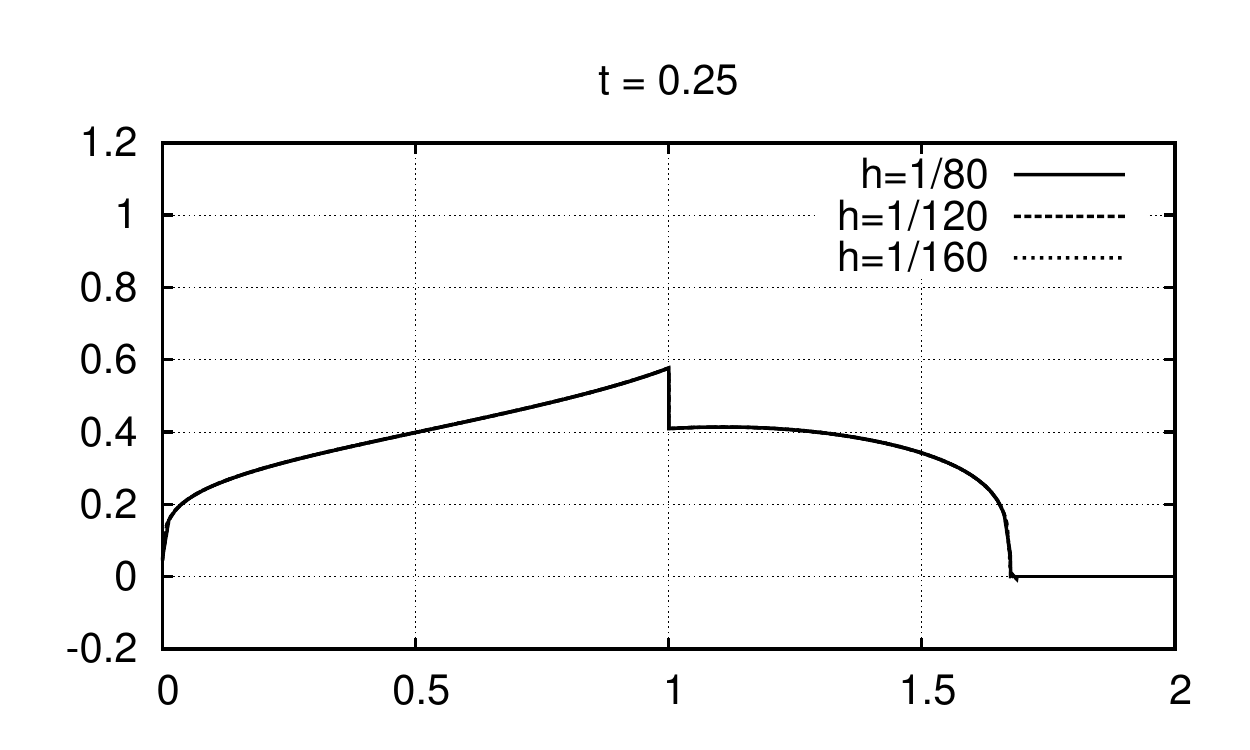}\\[-10mm]
\caption{Comparison of saturation of non-wetting phase in test case 2 for different mesh sizes. 
Time step size was 
$\Delta t=1\cdot 10^{-3}$ for $h=1/80$, 
$\Delta t=5\cdot 10^{-4}$ for $h=1/120$ and 
$\Delta t=2.5\cdot 10^{-4}$ for $h=1/160$.
Solution is shown at times $t=0, 0.008, 0.015, 0.045, 0.1, 0.25$ (from top to bottom, left to right).}
\label{fig:test_case_2_all}       
\end{figure*}

\begin{figure*}
\includegraphics[width=0.24\textwidth]{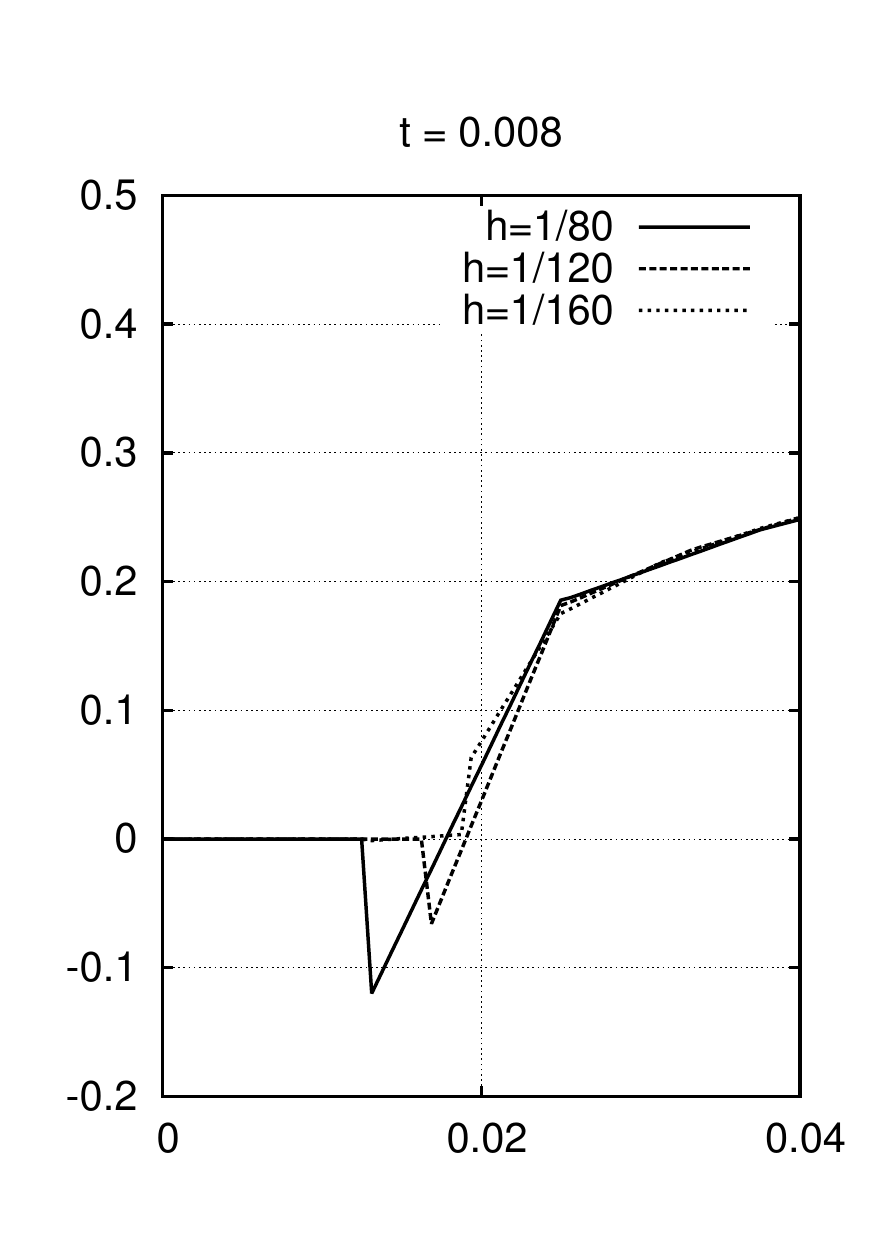}\hfill
\includegraphics[width=0.24\textwidth]{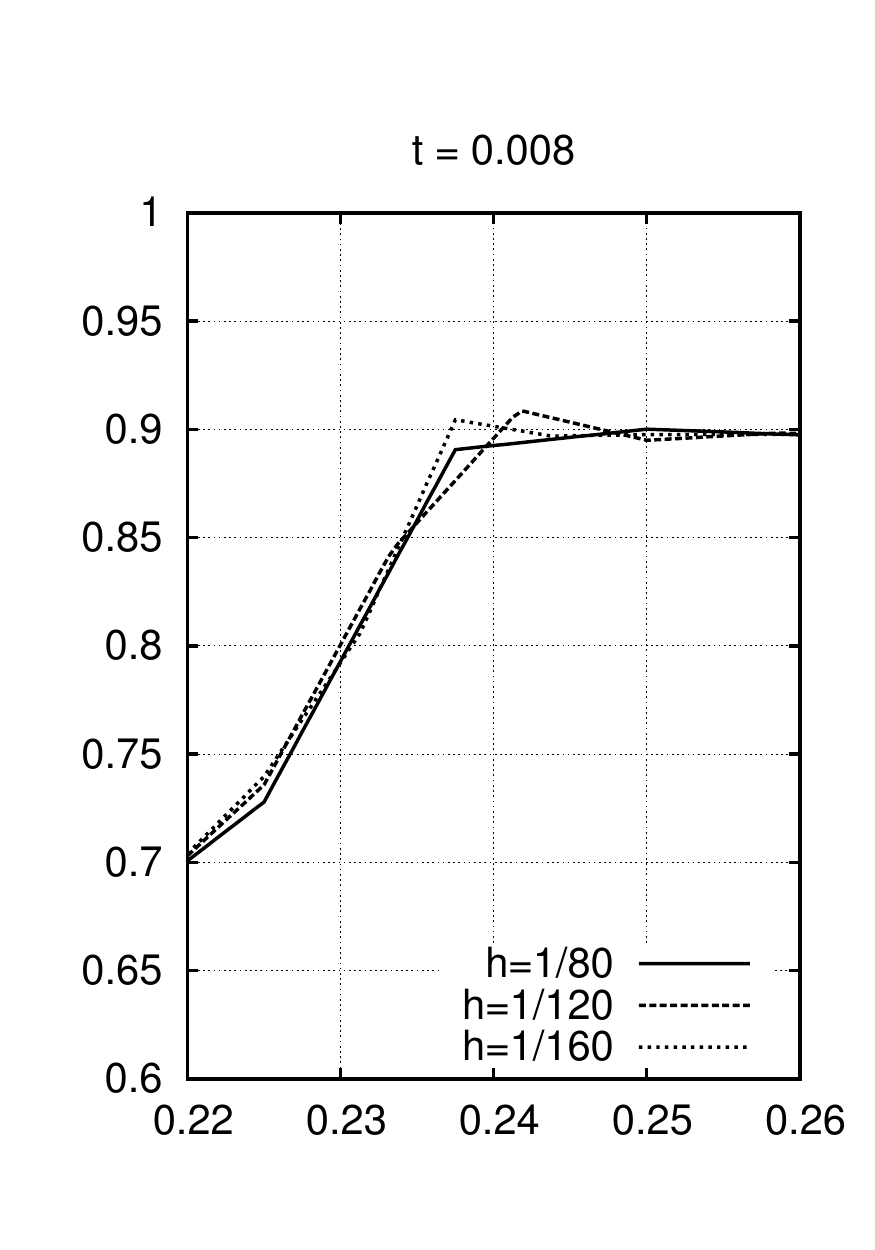}\hfill
\includegraphics[width=0.24\textwidth]{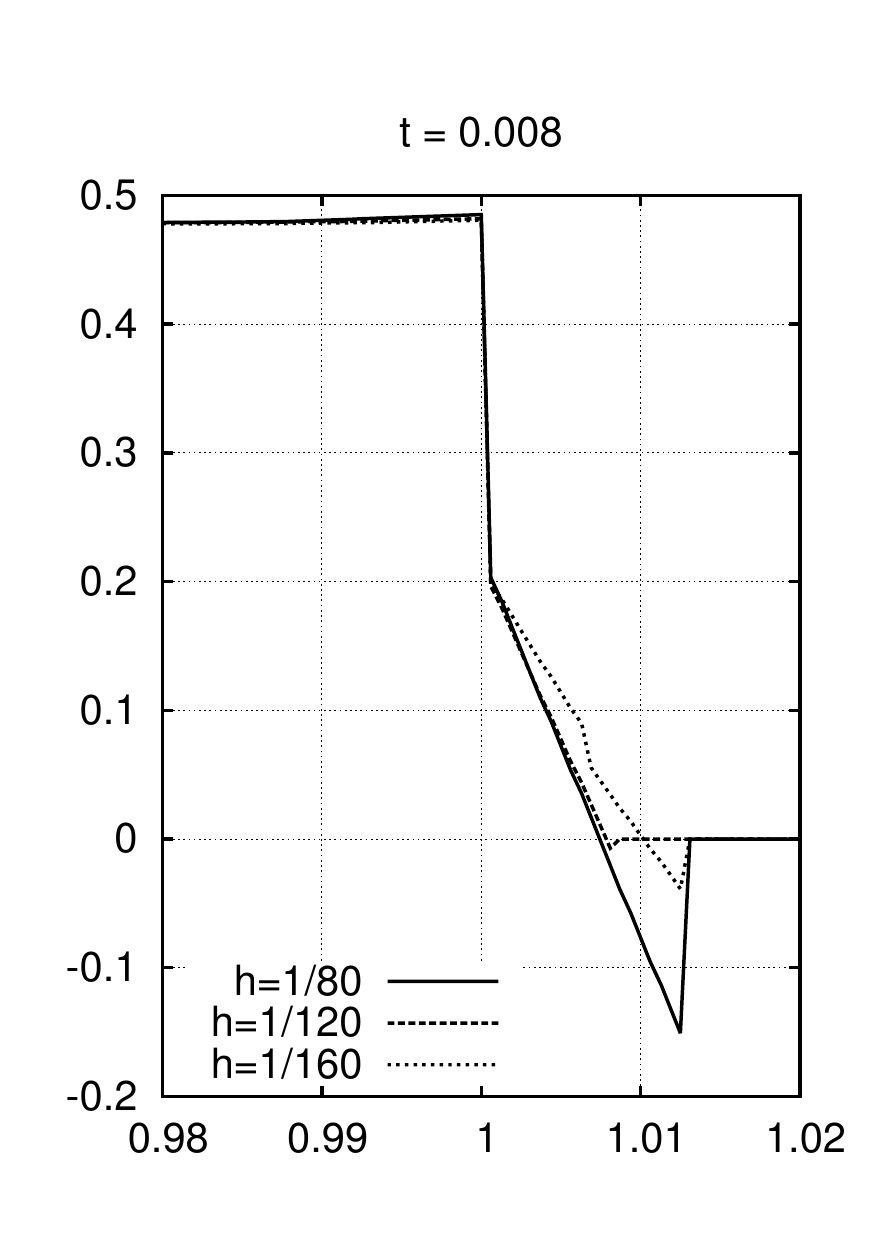}\hfill
\includegraphics[width=0.24\textwidth]{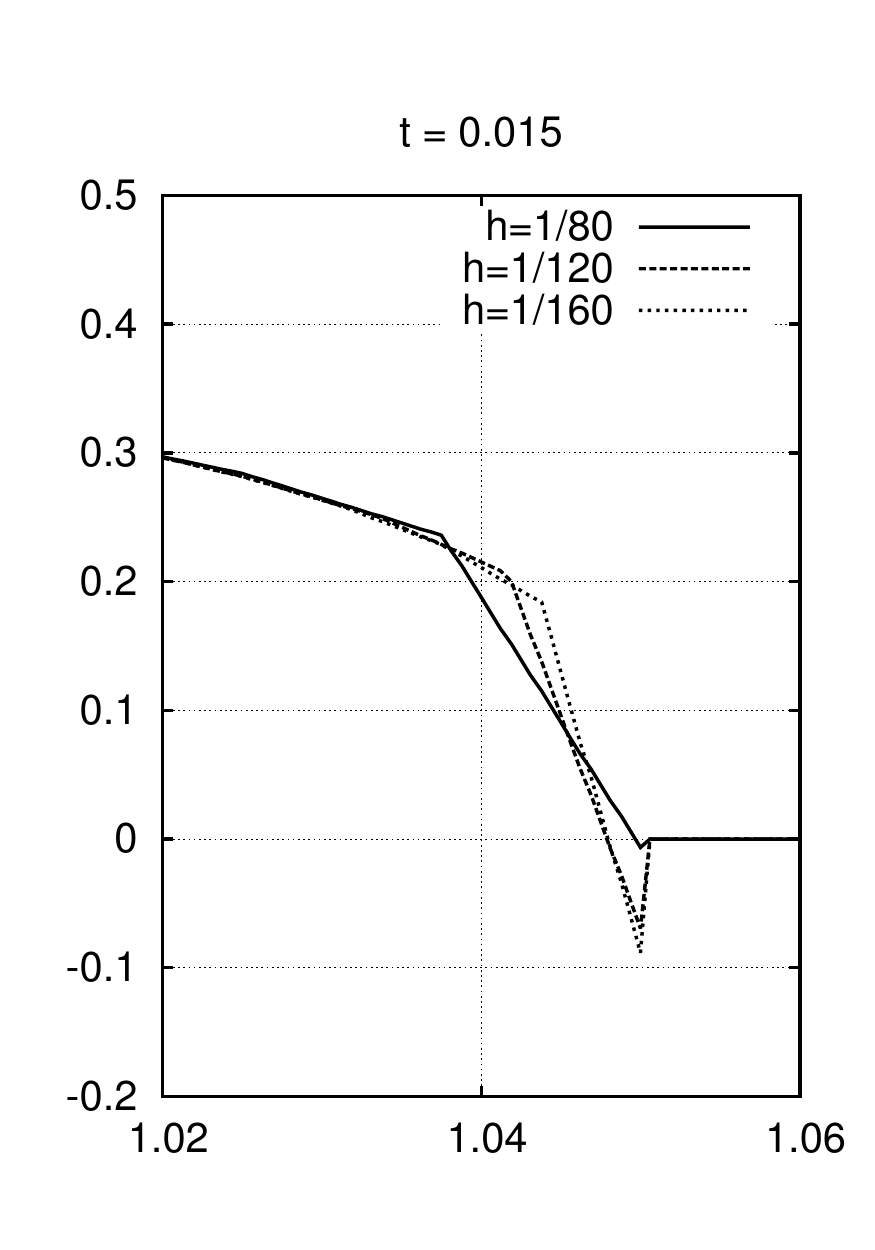}\\[-10mm]
\caption{Details of the solutions in test case 2. From left to right: Bottom part of left front at $t=0.008$, 
upper part of left front at $t=0.008$, bottom part of right front at $t=0.008$,
bottom part of right front at $t=0.015$.}
\label{fig:test_case_2_detail}       
\end{figure*}

Figure \ref{fig:test_case_2_all} shows the non-wetting phase saturation at various times
obtained with polynomial degree $p=1$ and the second-order Alexander scheme. 
Time steps were equidistant and are given in the figure legend. Due
to the pressure gradient the non-wetting phase is pushed to the right. When the critical saturation is
reached the right subdomain is infiltrated and a saturation discontinuity persists
at the media interface. The left front moves to the left since capillary diffusion
dominates the convective flux, at least initially. As in the first test case we observe convergence
of the solution under mesh refinement and a corresponding reduction of the 
oscillations in the vicinity of the free boundary. 

In comparison with the results
given in \cite{Ern20101491,Ern2012348} we see less variations between the
solutions on different refinement levels indicating that the solution is more accurate 
already on coarser grids. Computations with first-order fully-implicit Euler (not provided)
did not show significant differences which suggests that this error might be a splitting error
of the decoupled scheme. Moreover, we emphasize again that the results in Figure \ref{fig:test_case_2_all}
were obtained without $H(\text{div})$ reconstruction.

\subsection{Test Case 3: 2d DNAPL Infiltration}

In this section we consider a two-dimensional (vertical) DNAPL infiltration problem with
two different rock types similar to the one described in \cite{BH99}. This problem tests the
ability of the method to realize the interface conditions, handling of gravitational effects
and various boundary conditions. The quality of the solution on coarse meshes, its scalability
to very fine meshes and the performance relative to a cell-centered finite volume scheme are 
also evaluated.

\subsubsection{Problem Setup}

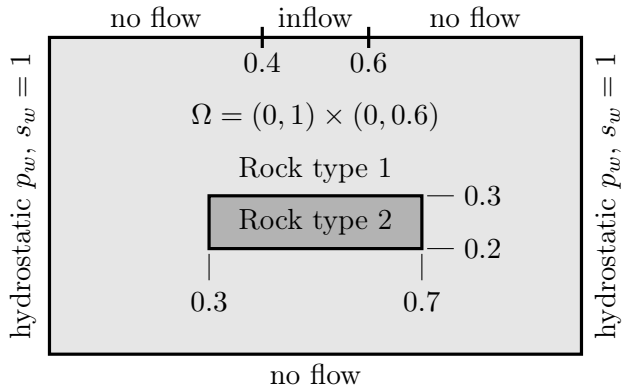
\begin{figure}
\begin{center}
\begin{tikzpicture}[scale=0.7]
\filldraw[very thick,fill=white!90!black] (0,0)  rectangle (10,6);
\filldraw[very thick,fill=white!70!black] (3,2)  rectangle (7,3);
\node at (5,4.5) {$\Omega=(0,1)\times(0,0.6)$};
\node at (5,3.5) {Rock type 1};
\node at (5,2.5) {Rock type 2};
\draw (3.0,1.4) -- (3,1.9); \node at (3,1) {$0.3$};
\draw (7.0,1.4) -- (7,1.9); \node at (7,1) {$0.7$};
\draw (7.1,2) -- (7.6,2); \node[right] at (7.6,2) {$0.2$};
\draw (7.1,3) -- (7.6,3); \node[right] at (7.6,3) {$0.3$};
\draw[very thick] (4,5.8) -- (4,6.2); \node[below] at (4,5.9) {$0.4$};
\draw[very thick] (6,5.8) -- (6,6.2); \node[below] at (6,5.9) {$0.6$};
\node[above] at (5,6) {inflow};
\node[above] at (2,6) {no flow};
\node[above] at (8,6) {no flow};
\node[below] at (5,0) {no flow};
\node[rotate=90] at (10.5,3) {hydrostatic $p_w$, $s_w=1$};
\node[rotate=90] at (-0.5,3) {hydrostatic $p_w$, $s_w=1$};
\end{tikzpicture}
\vspace{-4mm}
\caption{Geometry and boundary conditions for the DNAPL infiltration problem.}
\label{fig:dnapl2dsetup}
\end{center}
\end{figure}

The geometry and boundary conditions are given in Figure \ref{fig:dnapl2dsetup}. 
At the inflow boundary on the top a flux of $0.075$ $[kg\  s^{-1} m^{-2}]$ of the 
non-wetting phase into the domain is prescribed. At all other parts of the top and bottom
boundary no flow conditions are prescribed for both phases (at the inflow boundary no flow
is also prescribed for the wetting phase). At the left and right boundary full saturation of the wetting phase, $s_w=1$,
and hydrostatic conditions for the pressure $p_w$ are prescribed.

The mass densities of the fluids are $\rho_w=1000$ $[kg\ m^{-3}]$ and $\rho_n=1460$ $[kg\ m^{-3}]$
while the dynamic viscosities are $\mu_w=10^{-3}$ $[Pa\ s]$ and $\mu_n=0.9\cdot 10^{-3}$ $[Pa\ s]$.
The relative permeabilities are chosen as quadratic functions and are the same for both rock types:
\begin{equation*}
k_{rw}(s_w) = s_w^2, \qquad k_{rn}(s_n) = s_n^2 .
\end{equation*}
If $s_\alpha<0$ we set $k_{r\alpha}(s_\alpha)=0$ and if $s_\alpha>1$ we set $k_{r\alpha}(s_\alpha)=1$.
The capillary pressure saturation function is of Brooks-Corey type 
with parameters given in Table \ref{tab:rock_type_parameters} and
the regularization given in \eqref{eq:PcRegularization} is applied with $R=4$.
Finally, porosity and absolute permeability are given in Table \ref{tab:rock_type_parameters} as well.

\begin{table}
\caption{Parameters for the two rock types in the DNAPL infiltration problem.}
\label{tab:rock_type_parameters}       
\vspace{-4mm}
\begin{center}
\begin{tabular}{lll}
\hline\noalign{\smallskip}
Parameter & Rock type 1 & Rock type 2\\
\noalign{\smallskip}\hline\noalign{\smallskip}
Porosity $\Phi$ & 0.4 & 0.4\\
Abs. perm. $K$ $[m^2]$ & $6.64\cdot 10^{-11}$ & $3.32\cdot 10^{-11}$\\
rel. perm. & \multicolumn{2}{c}{quadratic, see text} \\ 
$p_e$ $[Pa]$ & 755 & 1163 \\
$\lambda$ & 2.5 & 2\\
\hline\noalign{\smallskip}
\end{tabular}
\end{center}
\end{table}

\subsubsection{Accuracy on Coarse Meshes}

\begin{figure*}
\includegraphics[width=0.33\textwidth]{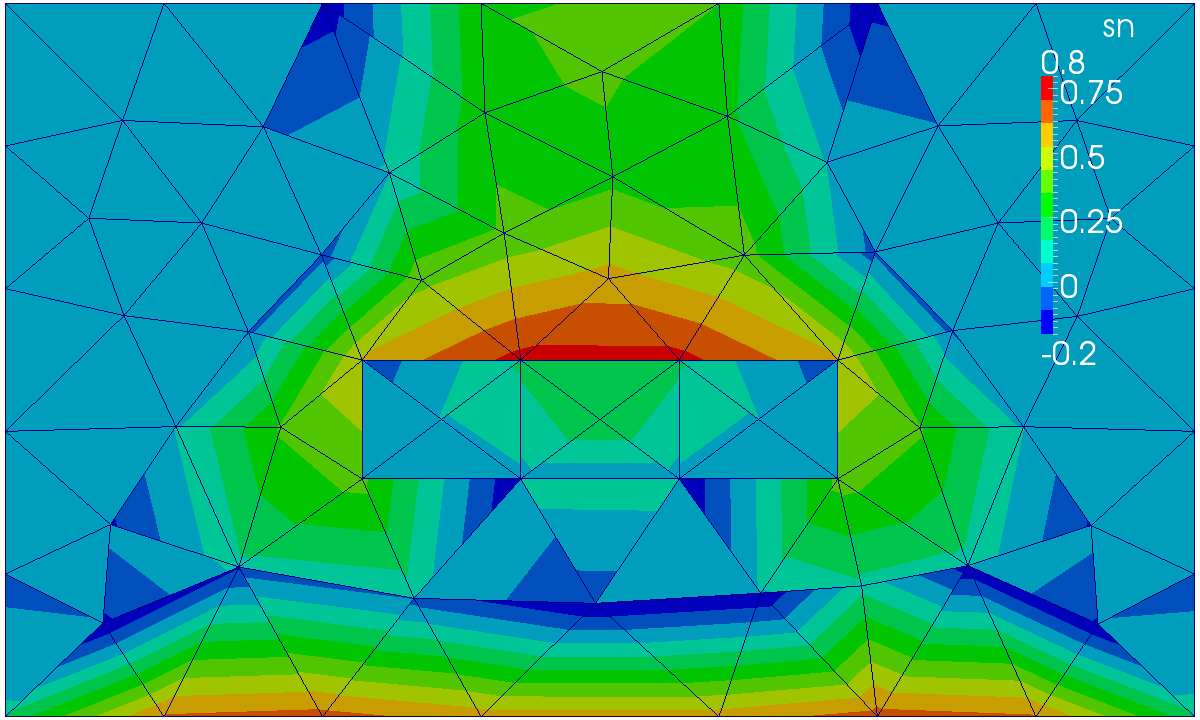}\hfill
\includegraphics[width=0.33\textwidth]{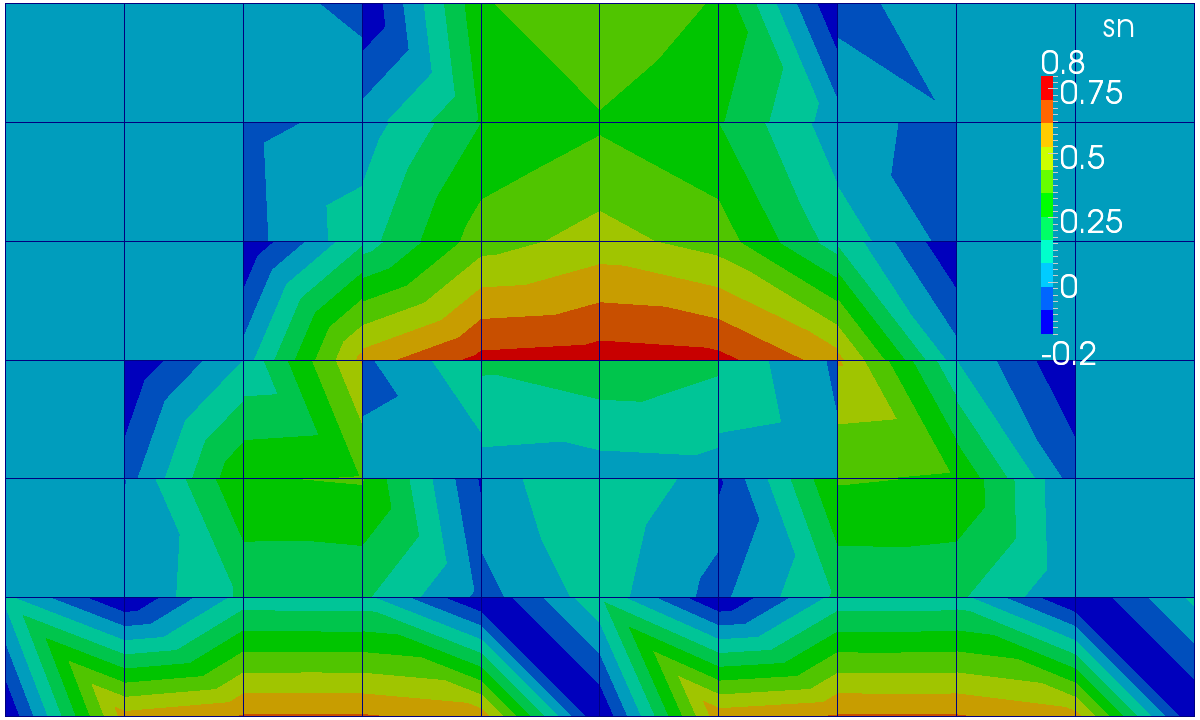}\hfill
\includegraphics[width=0.33\textwidth]{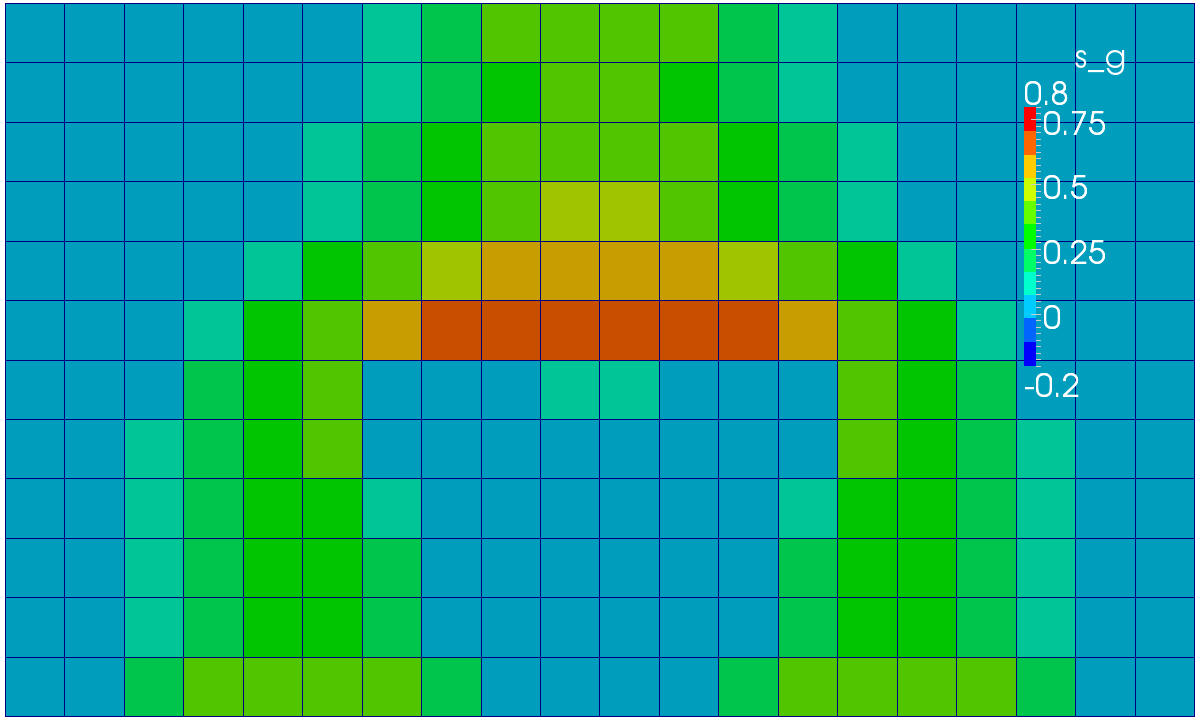}\\
\includegraphics[width=0.33\textwidth]{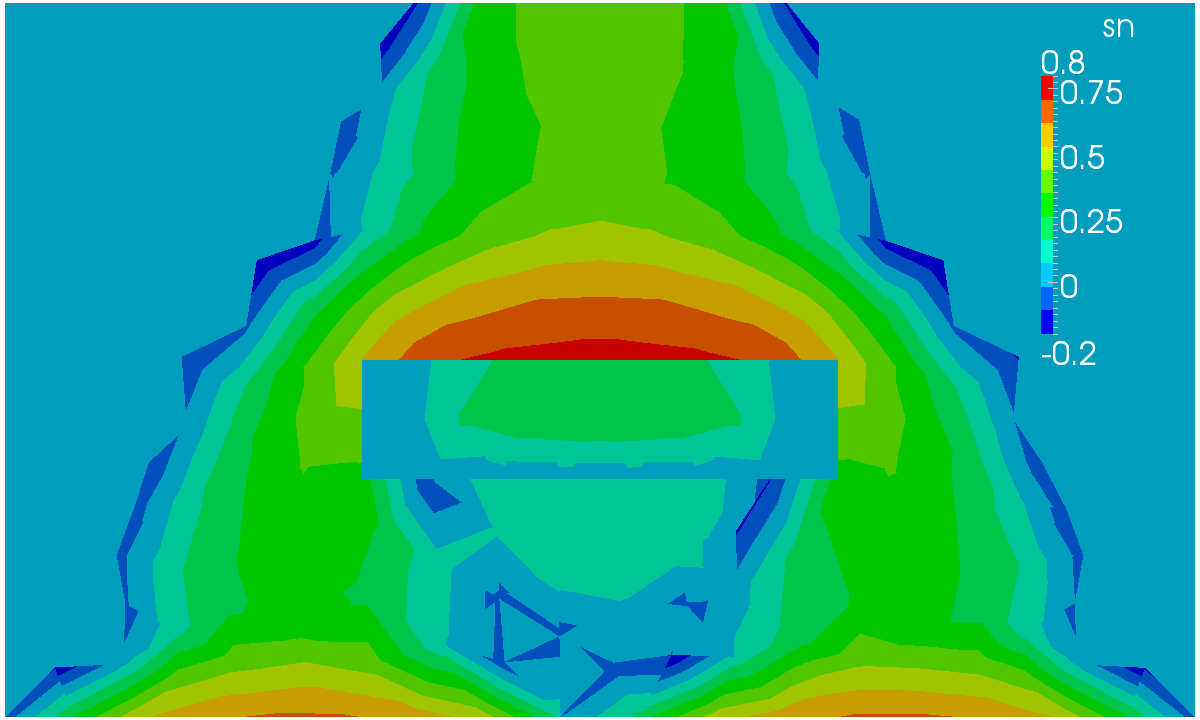}\hfill
\includegraphics[width=0.33\textwidth]{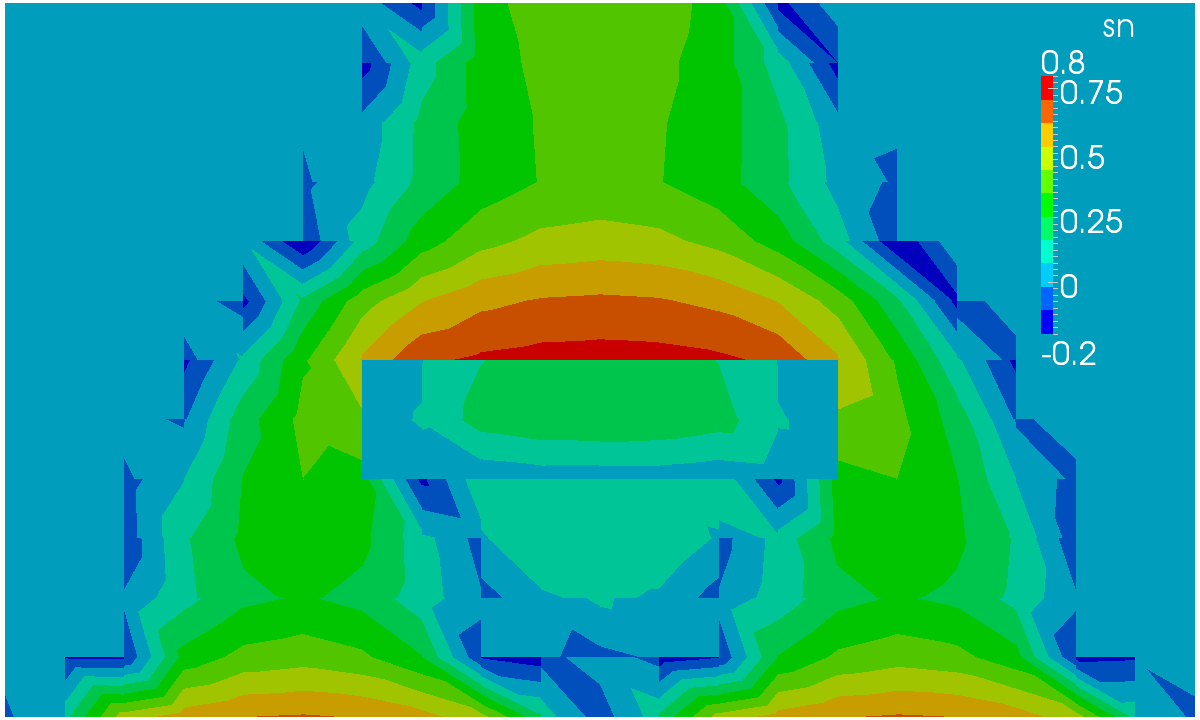}\hfill
\includegraphics[width=0.33\textwidth]{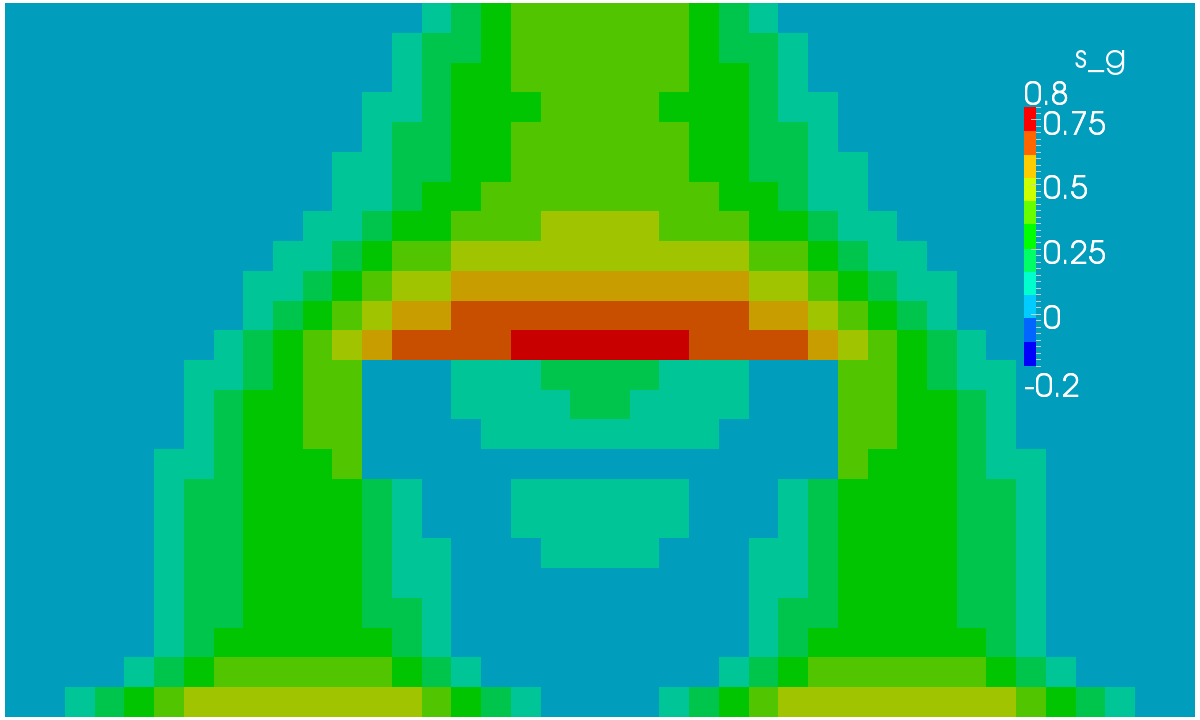}\\
\includegraphics[width=0.33\textwidth]{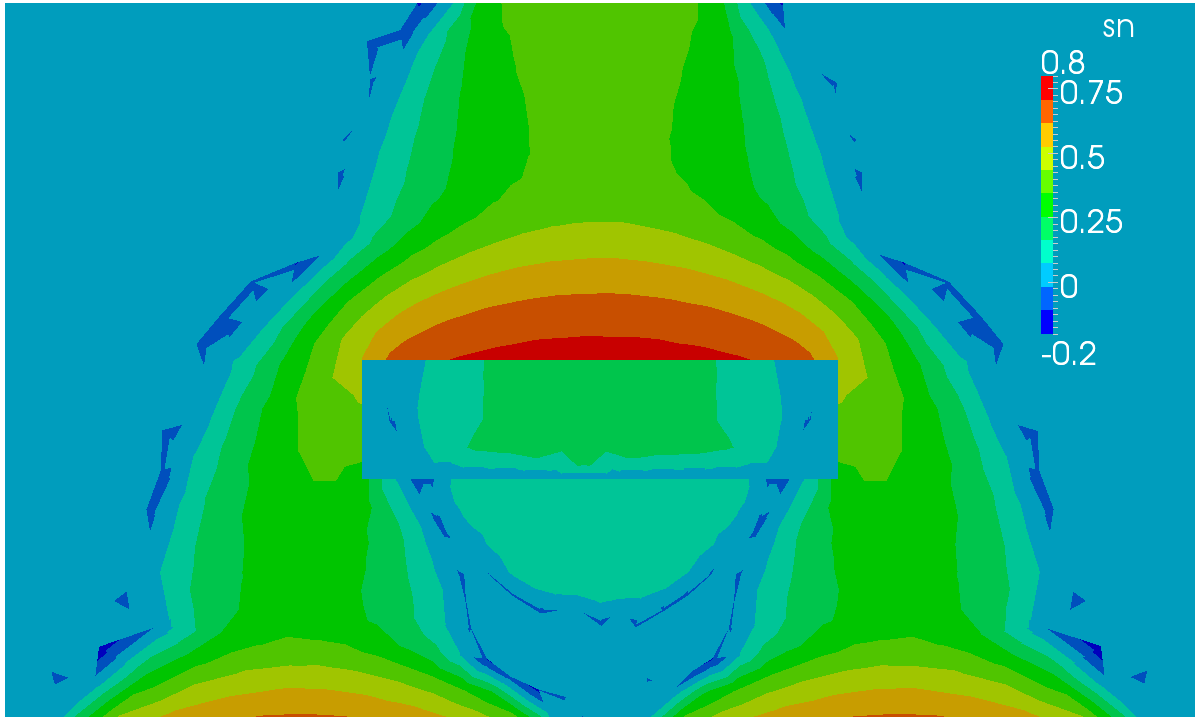}\hfill
\includegraphics[width=0.33\textwidth]{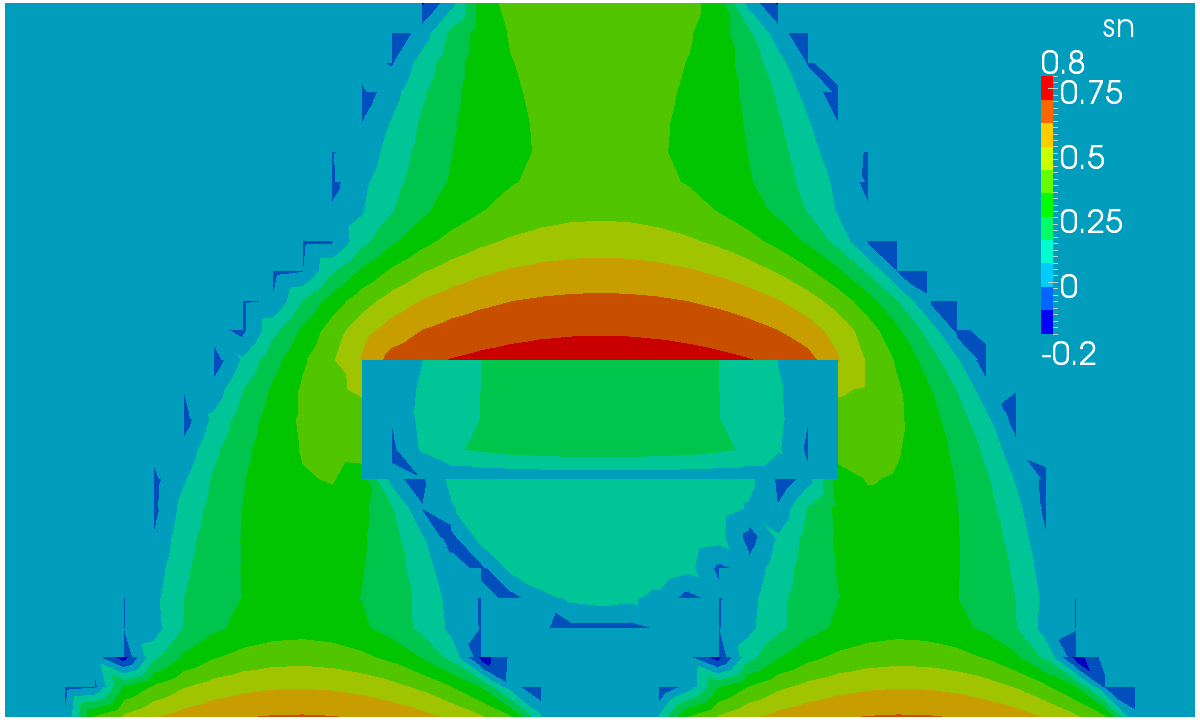}\hfill
\includegraphics[width=0.33\textwidth]{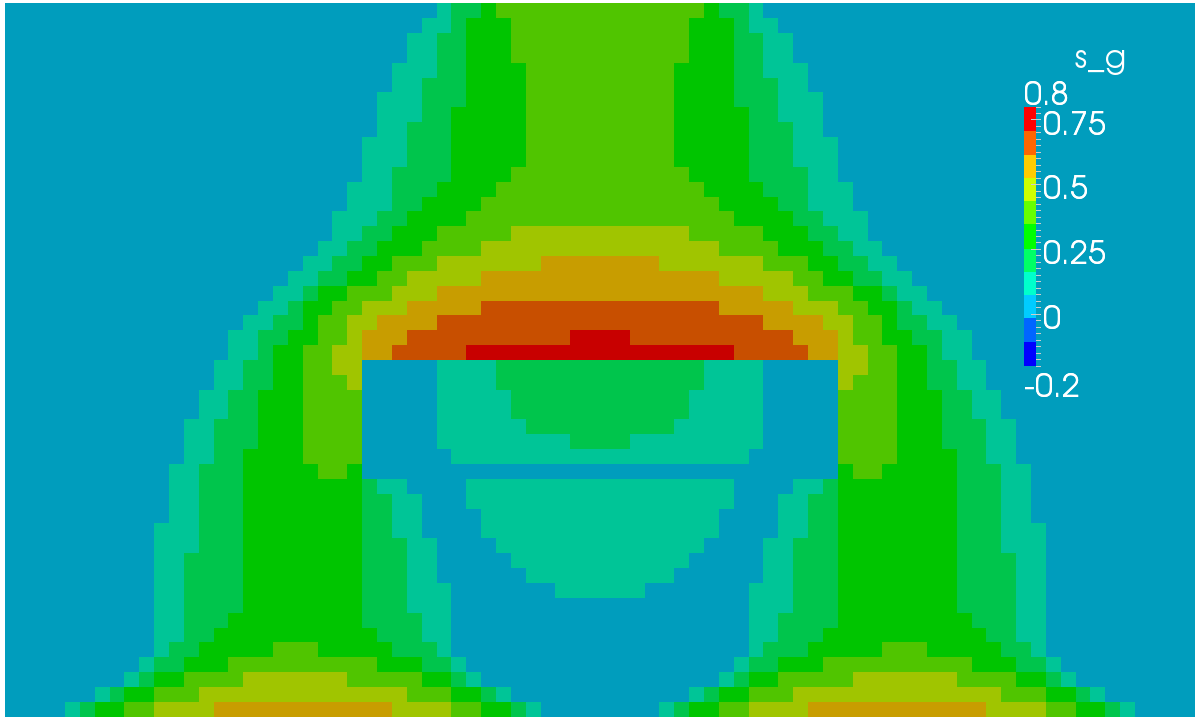}

\smallskip
\hrule
\smallskip

\includegraphics[width=0.33\textwidth]{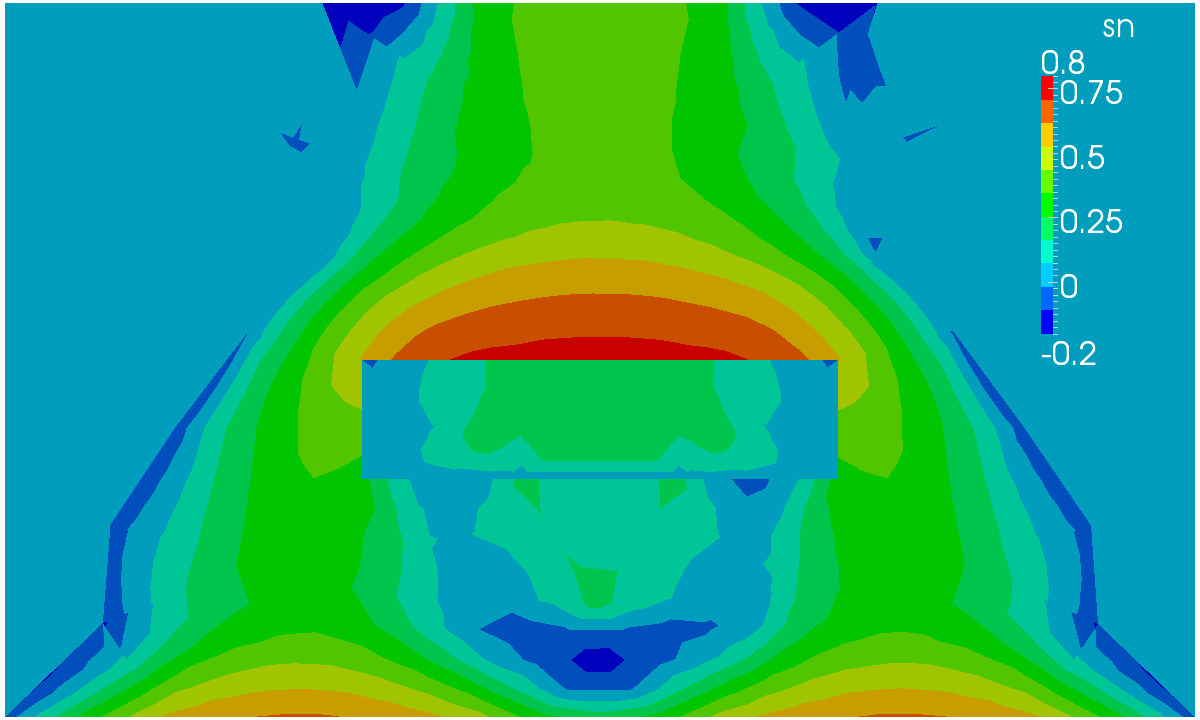}\hfill
\includegraphics[width=0.33\textwidth]{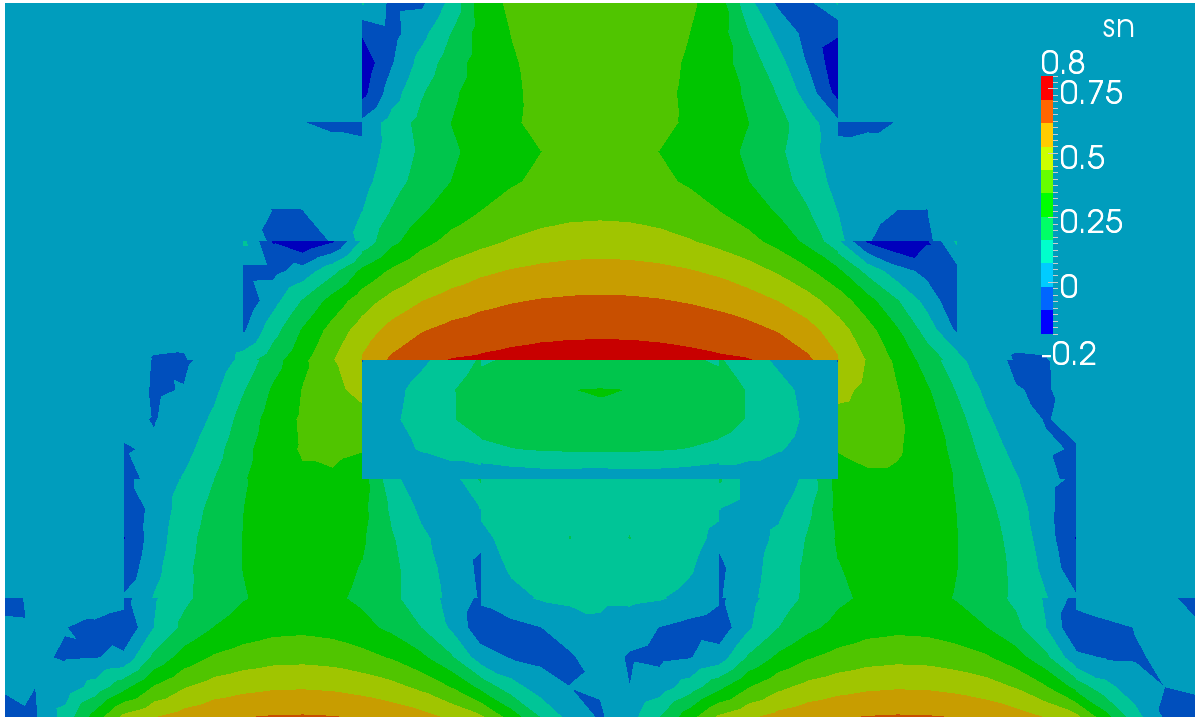}\hfill
\includegraphics[width=0.33\textwidth]{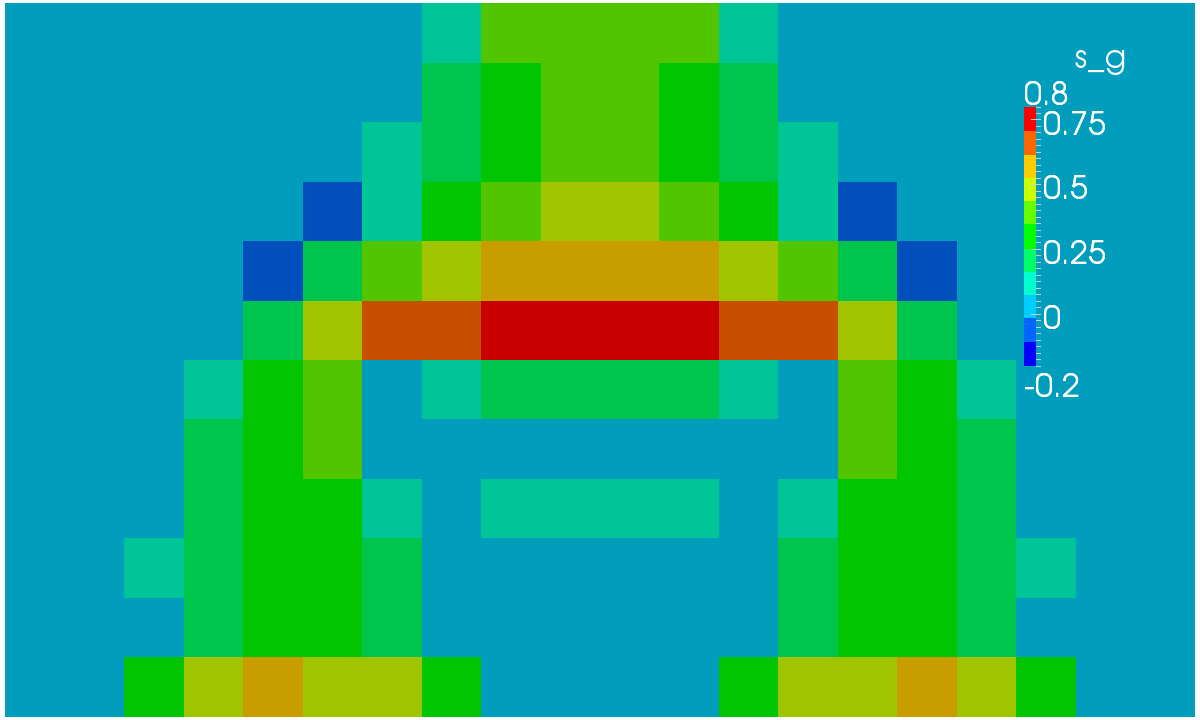}\\
\includegraphics[width=0.33\textwidth]{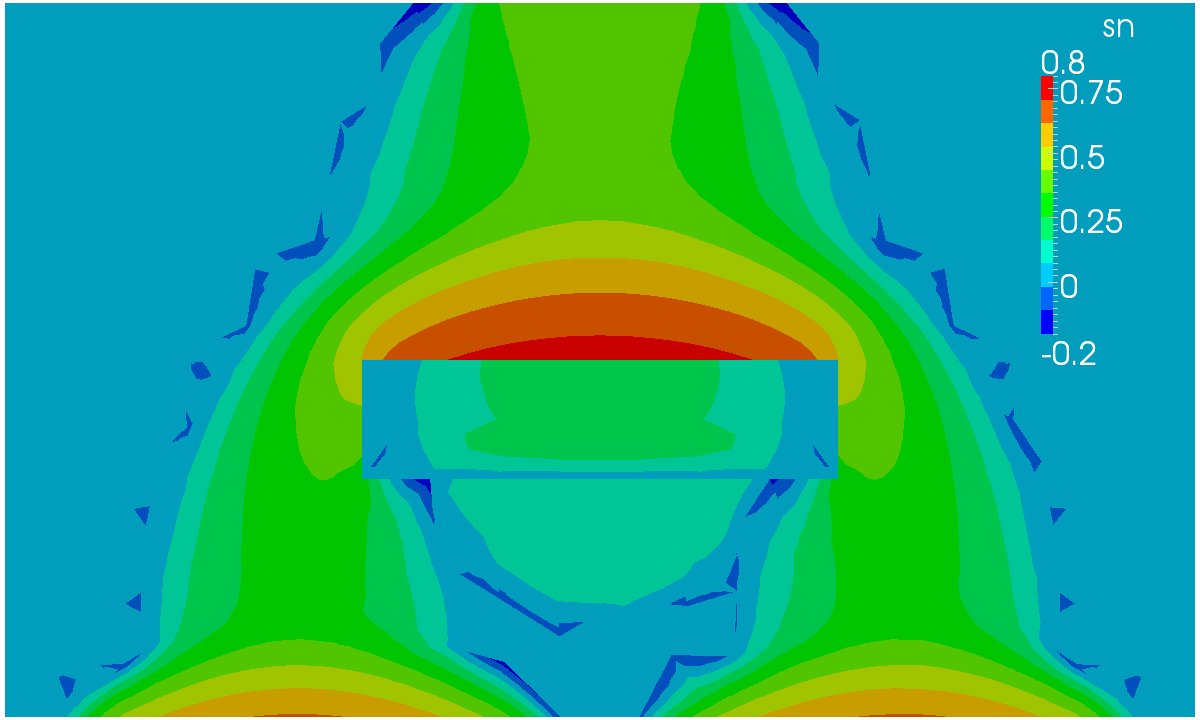}\hfill
\includegraphics[width=0.33\textwidth]{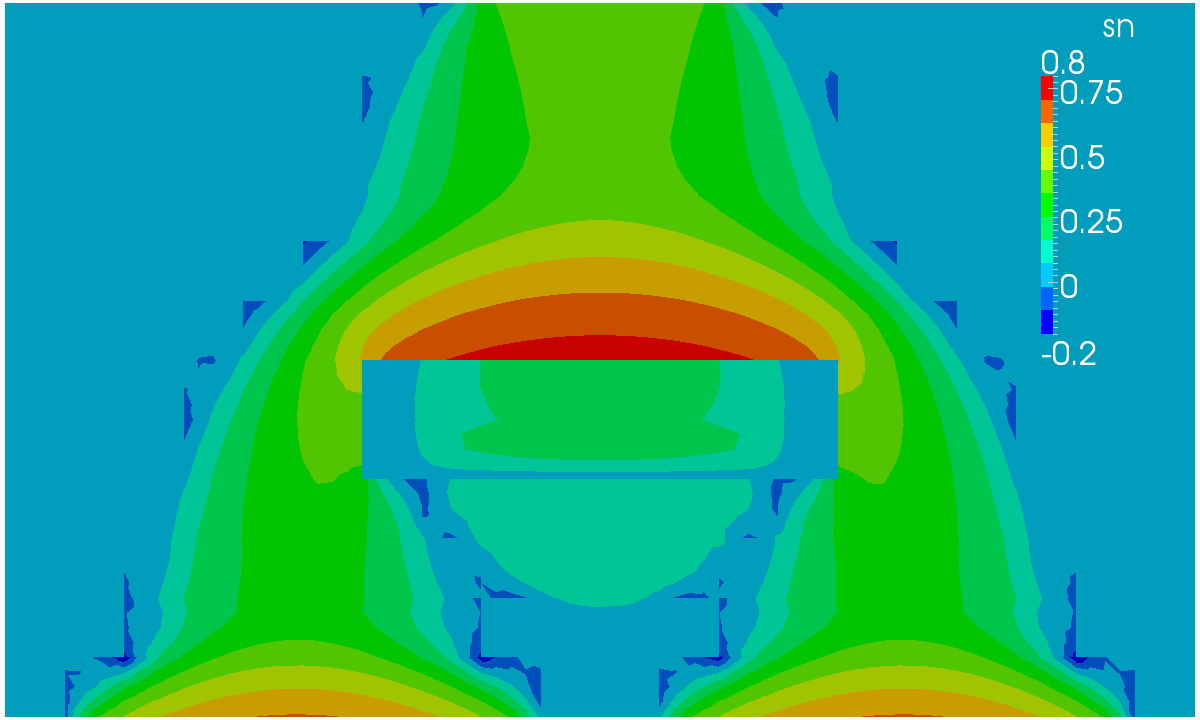}\hfill
\includegraphics[width=0.33\textwidth]{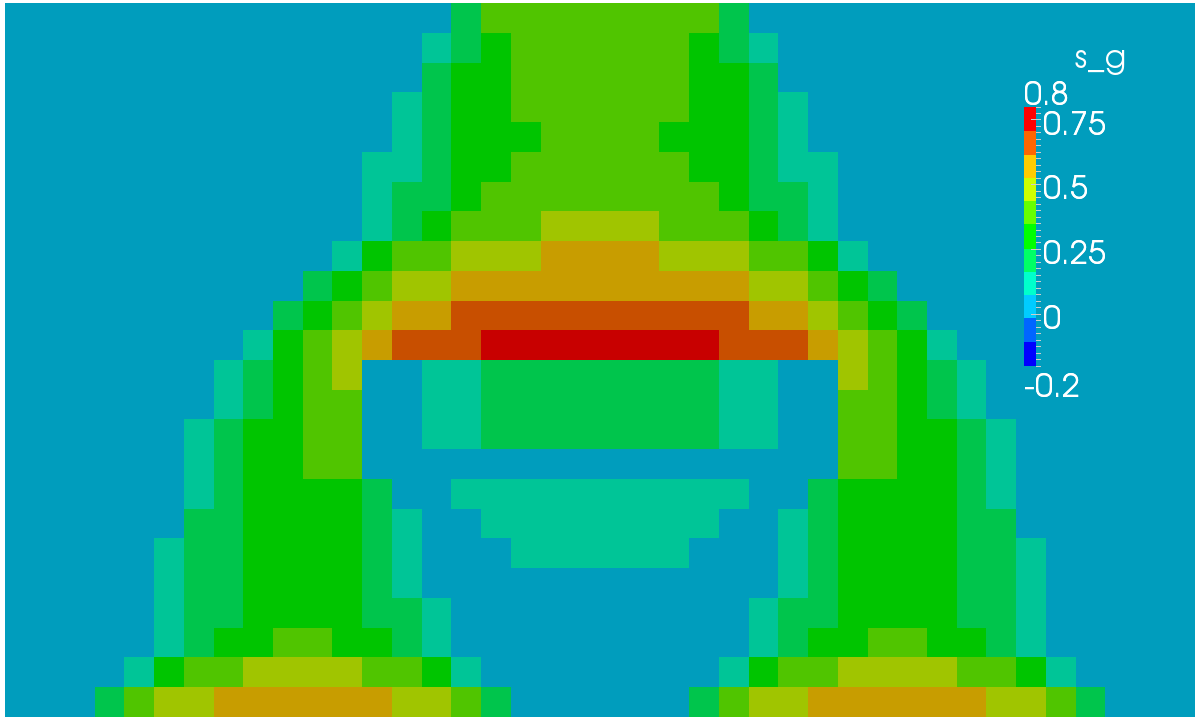}\\
\includegraphics[width=0.33\textwidth]{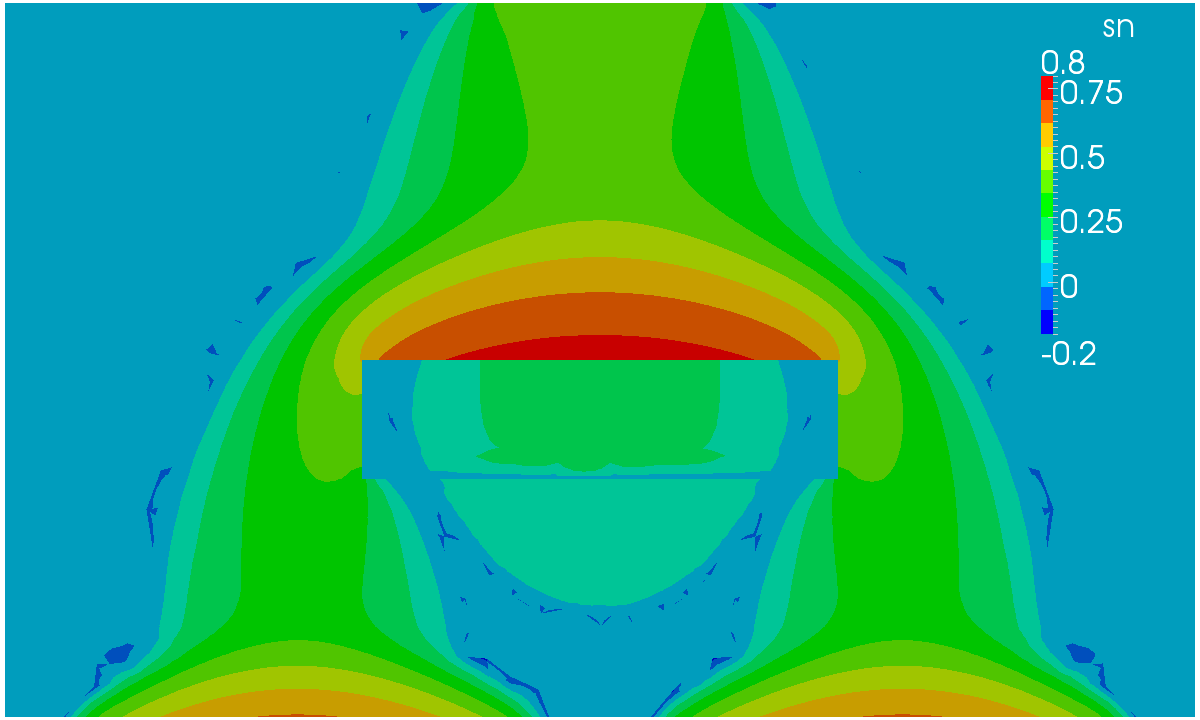}\hfill
\includegraphics[width=0.33\textwidth]{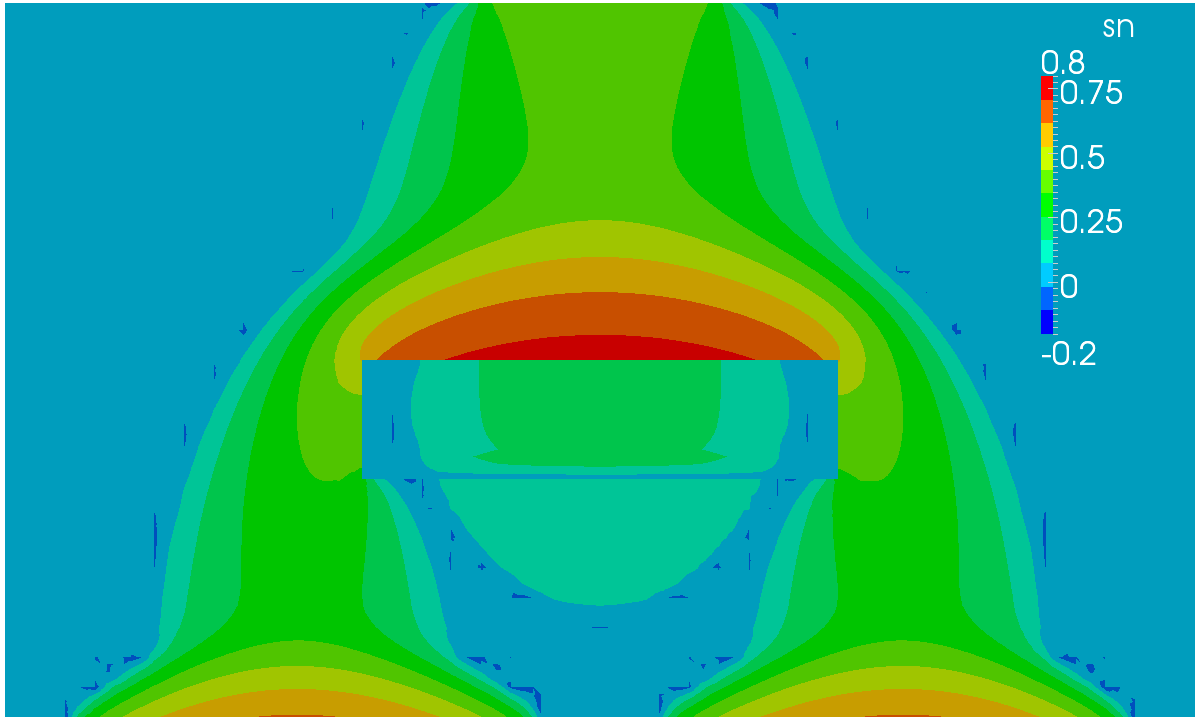}\hfill
\includegraphics[width=0.33\textwidth]{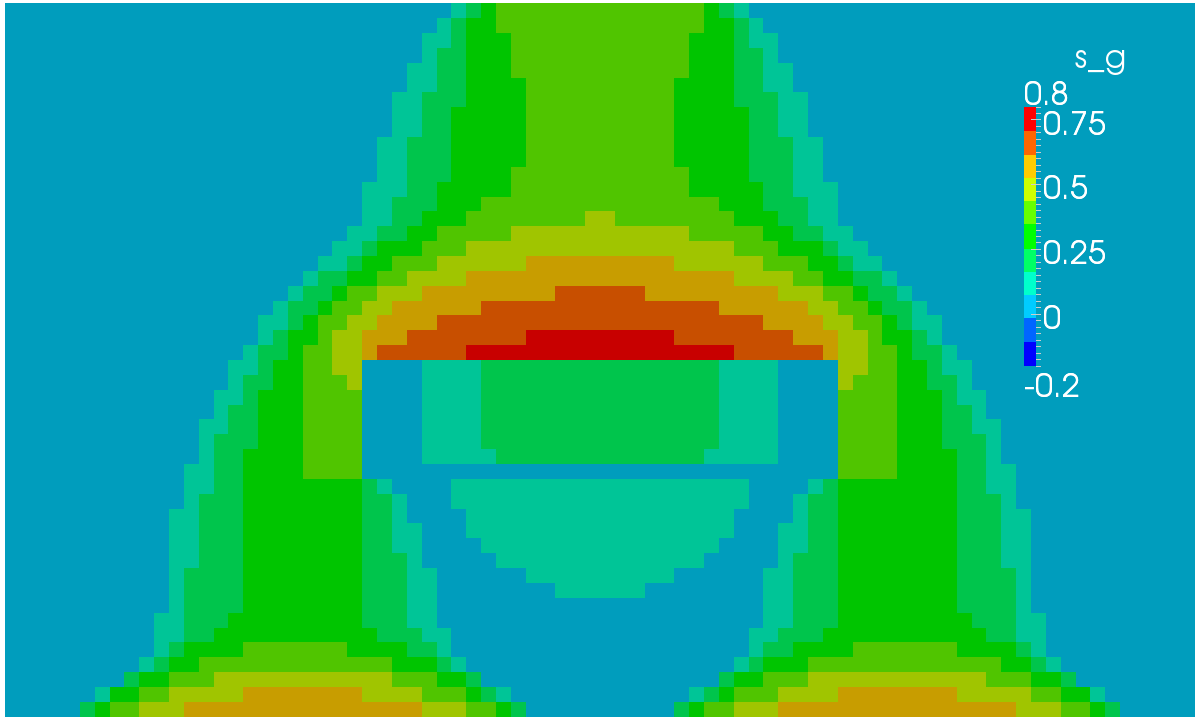}\\
\caption{Comparison of non-wetting phase saturation for different schemes on coarse grids. First set of three rows: 
Left column shows DG/$\mathbb{P}_1$ on unstructured triangular meshes with 120, 498 and 1474 elements, middle column 
shows DG/$\mathbb{Q}_1$ on structured, equidistant meshes with 60, 240 and 960 elements, 
right column shows upwind CCFV scheme on structured, equidistant meshes with 
240, 960 and 3840 elements. Second set of three rows: DG/$\mathbb{P}_2$ (left column),
DG/$\mathbb{Q}_2$ (middle column), second order CCFV (right column) on the same meshes as first set.}
\label{fig:dnapl2d_coarse_comparison}       
\end{figure*}

Discontinuous Galerkin schemes (of high order) are more expensive in terms 
of degrees of freedom and computation
time compared to simple low order schemes. Therefore it is of particular interest to see the performance 
on relatively coarse meshes. 
In Figure \ref{fig:dnapl2d_coarse_comparison} we compare 
the non-wetting phase saturation obtained with DG/$\mathbb{P}_1$, DG/$\mathbb{P}_2$,
DG/$\mathbb{Q}_1$ and DG/$\mathbb{Q}_2$ with a cell-centered finite volume scheme (CCFV) 
using either full-upwinding oder central evaluation of mobilities.
The CCFV scheme is based on a
wetting-phase pressure / capillary pressure formulation with harmonic permeability weighting
(where relative permeabilities are either evaluated with the upwind saturation
or the arithmetically averaged saturation, cf. \cite{co2_2011} for details).

In Figure \ref{fig:dnapl2d_coarse_comparison} consider the first row of images. From 
the left we have DG/$\mathbb{P}_1$ on an unstructured mesh consisting of 120 triangles generated
with Gmsh \cite{NME:NME2579} in the middle DG/$\mathbb{Q}_1$ on a structured, equidistant mesh 
consisting of 60 quadrilaterals and on the right the CCFV upwind scheme on the structured mesh
being uniformly refined to 240 elements. The meshes have been chosen such that they correspond
roughly to the same number of spatial degrees of freedom (in the case of  DG/$\mathbb{Q}_1$ 
and CCFV they are identical). In the second and third row the same schemes are shown on uniformly
refined meshes. In rows 4--6 we show results for DG/$\mathbb{P}_2$ (left column), 
DG/$\mathbb{Q}_2$ (middle column) and CCFV with central evaluation of capillary pressure (right column).
All results are shown for $T=3600s$ while a different number of timesteps have been
performed keeping $\Delta t/h$ fixed. In addition, implicit Euler has been employed for
upwind CCFV while the second order Alexander scheme \cite{alexander:77} was used for all
other spatial discretizations.
In Table \ref{tab:dnapl2d_coarse_comparison} the minimum
and maximum of the numerical solution for the different schemes and meshes are given. 

From these results we can conclude:
\begin{itemize}
\item On the coarsest meshes the lowest order DG schemes exhibit severe
undershoots. These negative saturations are always located in the vicinity of the free boundary
and they are reduced as the mesh is refined. The CCFV upwind scheme is monotone as expected
and the CCFV central scheme shows small undershoots that are quickly reduced as the mesh is refined.
\item On the coarsest meshes (first row) the DG schemes are quite accurate above, within and below the
low permeability lense in comparison to the CCFV upwind scheme. This is also supported
by the maximum of the numerical solutions (which is attained in the midpoint of the
upper boundary of the lense). Both CCFV schemes are less accurate
(given the same number of degrees of freedom) with respect to the maximum value.
\item Increasing the polynomial degree in the DG scheme gives a considerable
improvement of the solution, also with respect to the undershoots and the maximum. 
\end{itemize}

\begin{table*}
\caption{Minima and maxima of the non-wetting phase saturation corresponding to the
images shown in Figure \ref{fig:dnapl2d_coarse_comparison}. Note that for each scheme 
results for one additional level of refinement is given.}
\begin{center}
\begin{tabular}{rrrrrrrrr}
\hline\noalign{\smallskip}
\multicolumn{3}{c}{DG/$\mathbb{P}_1$} & \multicolumn{3}{c}{DG/$\mathbb{Q}_1$} & \multicolumn{3}{c}{CCFV upwind}\\
Elements & min & max &Elements & min & max & Elements & min & max\\
\noalign{\smallskip}\hline\noalign{\smallskip}
120 & -0.57 & 0.7416   & 60   & -0.51 & 0.7555 & 240     & 0 & 0.7022\\
498 & -0.32 & 0.7461   & 240 & -0.20 & 0.7489 & 960     & 0 & 0.7248\\
1474 & -0.20 & 0.7493 & 960 & -0.17 & 0.7501 & 3840   & 0 & 0.7367\\
7182 & -0.14 & 0.7503 & 3840 & -0.10&0.7505 & 15360 & 0 & 0.7434\\
\noalign{\smallskip}\hline\noalign{\smallskip}
\multicolumn{3}{c}{DG/$\mathbb{P}_2$} & \multicolumn{3}{c}{DG/$\mathbb{Q}_2$} & \multicolumn{3}{c}{CCFV central}\\
Elements & min & max &Elements & min & max & Elements & min & max\\
\noalign{\smallskip}\hline\noalign{\smallskip}
120 & -0.39 & 0.7488   & 60   & -0.19 & 0.7555 & 240     & -0.14 & 0.7243\\
498 & -0.23 & 0.7499   & 240 & -0.20 & 0.7489 & 960     & $-2.0\cdot 10^{-2}$ & 0.7347\\
1474 & -0.14 & 0.7505 & 960 & -0.10 & 0.7501 & 3840   & $-1.0\cdot 10^{-2}$ & 0.7417\\
7182 & -0.10 & 0.7507 & 3840 & $-8.5\cdot 10^{-2}$&0.7505 & 15360 & $-5.6\cdot 10^{-3}$ & 0.7459\\
\hline\noalign{\smallskip}
\end{tabular}
\end{center}
\label{tab:dnapl2d_coarse_comparison}       
\end{table*}

\subsubsection{Accuracy on Fine Meshes}\label{sec:fine_mesh}

The DNAPL infiltration problem is now solved on meshes with up to $2560\times 1536$ elements
for the CCFV scheme. Since no analytical solution is available we plot 1D profiles along the vertical
line $x=0.5$ at the final time $T=3600$ for various schemes, mesh resolutions and time step sizes.
The time step size is always chosen such that $\Delta t/h=const$. The number
of time steps is listed in Table \ref{tab:dnapl2d_scalability_comparison}.

\begin{figure*}
\begin{center}
\includegraphics[width=0.49\textwidth]{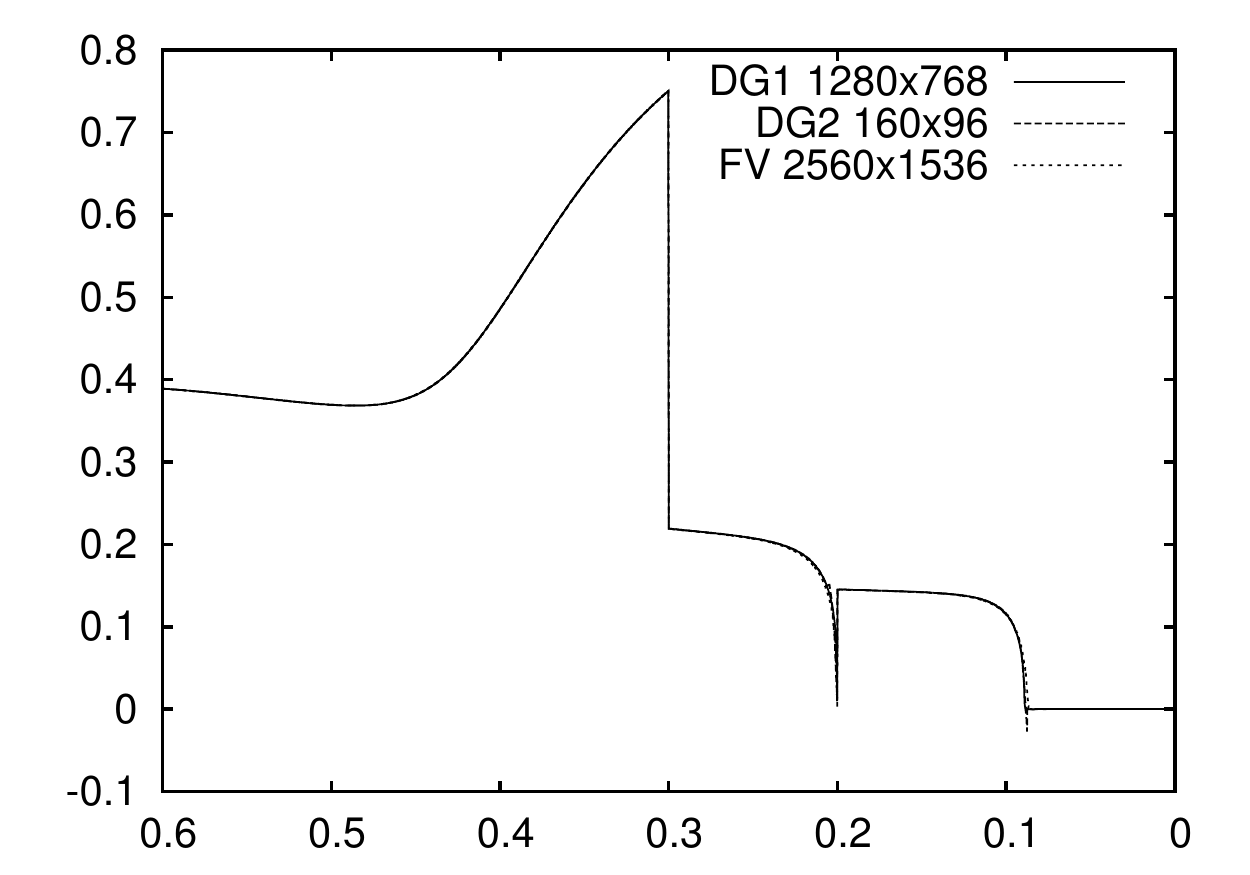} \hfill
\includegraphics[width=0.49\textwidth]{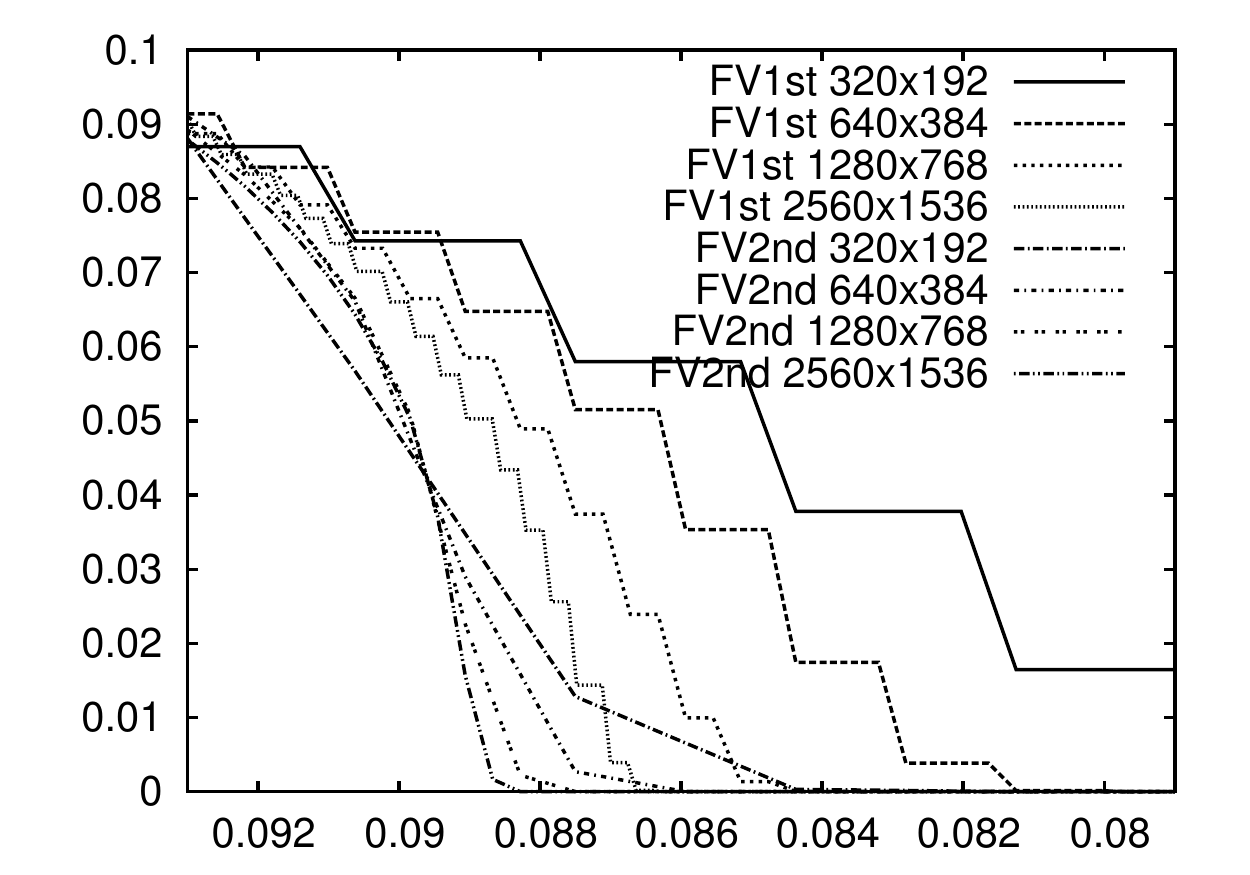}\\
\includegraphics[width=0.49\textwidth]{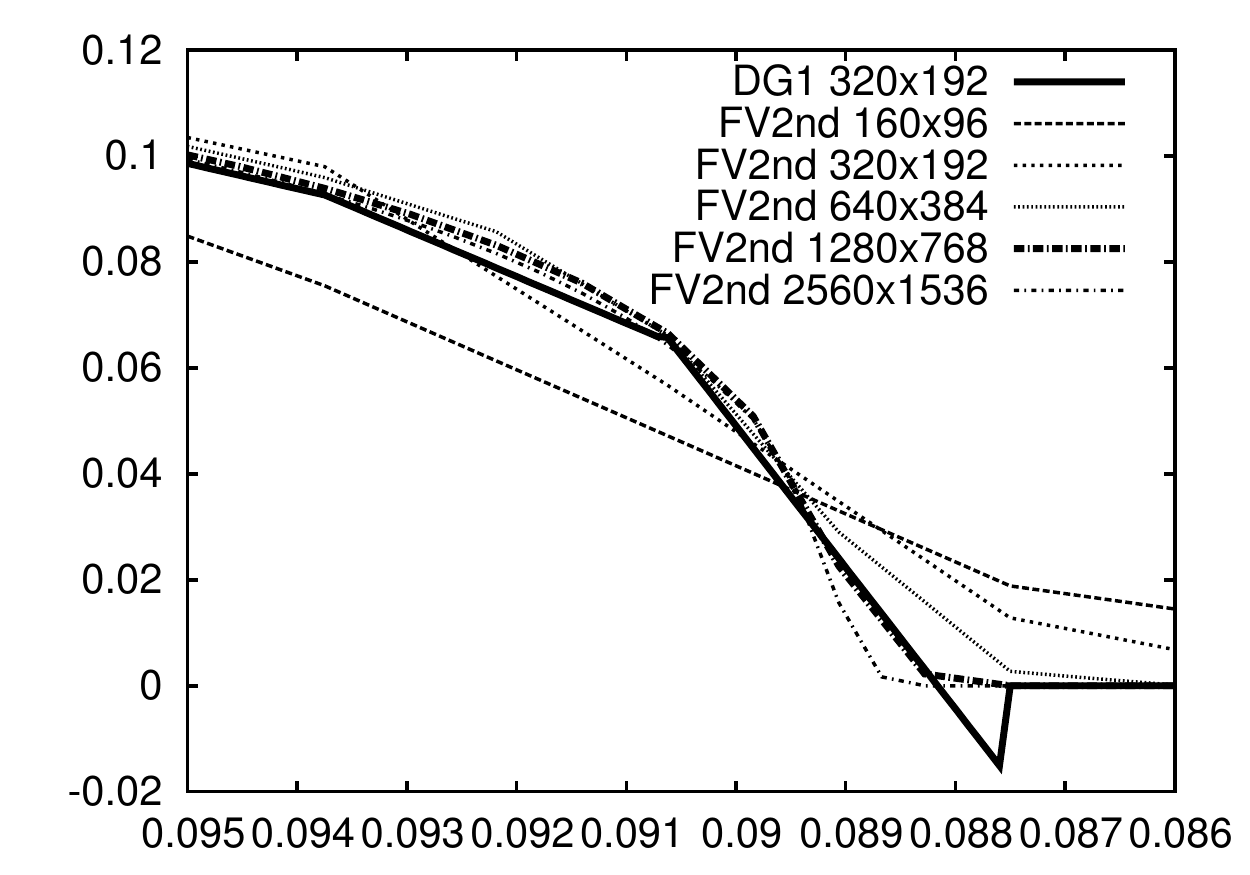} \hfill
\includegraphics[width=0.49\textwidth]{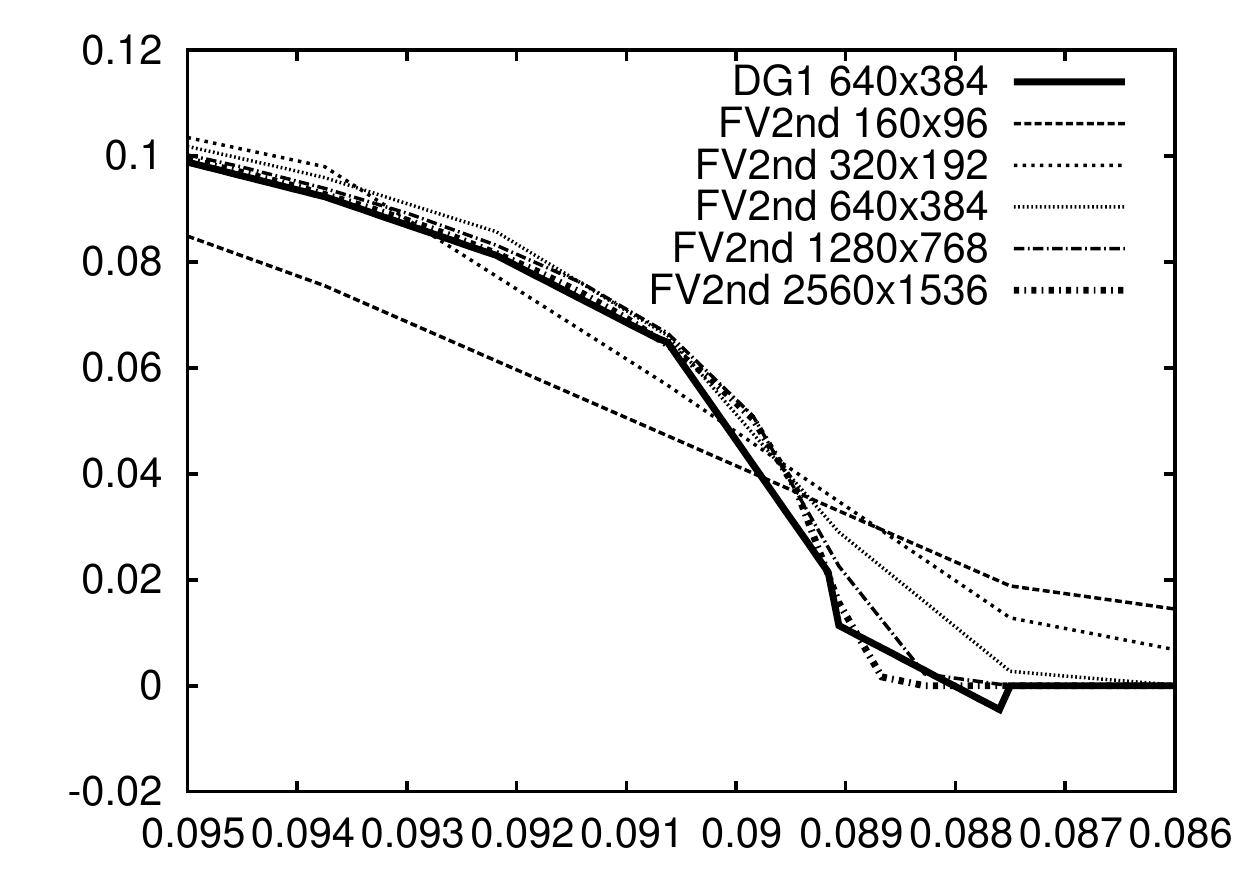}
\end{center}
\caption{Comparison of non-wetting phase saturation over $z$ position at $x=0.5$ in test case 3.
Top left: Complete profile along the line $x=0.5$ for DG and CCFV scheme
on fine meshes. 
Top right:  Zoom of region near the free boundary and comparison of first and second order
CCFV schemes.
Bottom left: Comparison of DG scheme on $320\times 192$ mesh with second order CCFV.
Bottom right: Comparison of DG scheme on $640\times 384$ mesh with second order CCFV.}
\label{fig:test_case_3_profiles}       
\end{figure*}

Figure \ref{fig:test_case_3_profiles} (top left) shows the overall profile with the DNAPL pooling
up in the interval $z\in (0.3,0.6)$, the upper interface at $z=0.3$ where DNAPL infiltrates the 
low permeability lense,
the region $z\in (0.2,0.3)$ within the low permeability lense, the lower
interface $z=0.2$ where DNAPL exfiltrates the lense and finally the free boundary near position 
$z\approx 0.09$.
At the lower interface the saturation $s_n$ is zero from inside the lense
and capillary pressure is discontinuous since
the critical saturation is not attained outside the lense. Clearly, at the
resolution shown, the solutions obtained with DG/$\mathbb{Q}_1$ and DG/$\mathbb{Q}_2$ 
as well as the second-order CCFV scheme
coincide well. In order to compare the schemes in detail we look at the solution close to the
free boundary.

The right plot in the upper row of Figure \ref{fig:test_case_3_profiles} provides a comparison of 
the first and second order CCFV schemes. The implicit Euler / full upwind scheme is very diffusive
and even the coarsest solution obtained with the second-order scheme that is shown $(320\times192)$
exhibits a better position of the free boundary than the first-order scheme on a three
times refined grid. Therefore we do not consider the first-order method any further in the sequel.

The plots in the bottom row of Figure \ref{fig:test_case_3_profiles} compare the second-order
DG/$\mathbb{Q}_1$ scheme with the second-order CCFV scheme. The results show that 
DG/$\mathbb{Q}_1$ on a given mesh is as accurate as the solution obtained with second-order
CCFV on a two-times refined mesh. Since CCFV on a two times refined mesh has four times
(eight in 3D) the number of degrees of freedom and needs two times more time steps (see
Table \ref{tab:dnapl2d_scalability_comparison}) the DG scheme is more efficient in terms
of number of degrees of freedom.

\subsubsection{Comparison of Solver Performance and Overall Computation Time}\label{sec:solver_results}

\begin{table*}
\caption{Comparison of two iterative solvers for the DNAPL infiltration problem.}
\label{tab:solver_comparison}       
\begin{center}
\begin{tabular}{rrrrrrr}
\hline\noalign{\smallskip}
       & \multicolumn{3}{c}{BiCGStab-ILU} & \multicolumn{3}{c}{BiCGStab-Hybrid AMG} \\
\noalign{\smallskip}\hline\noalign{\smallskip}
DOF & ALIN & MAXLIN & TT & ALIN & MAXLIN & TT \\
\noalign{\smallskip}\hline\noalign{\smallskip}
480       & 33.0   &   58 &      0.7 & 2.8 & 4 & 0.9 \\ 
1920     & 54.2   & 107 &        3  & 2.2 & 4 & 3.8 \\
7680     & 91.0   & 223 &       14 & 2.0 & 2 & 19.0 \\
30720   & 152.2   & 459 &     71 & 2.3 & 4 &  66.7\\
122880 & 231.0 & 773   &  408  & 2.6 & 8 &  272.8 \\
491520 & 353.4 & 1945 & 2083 & 2.8 & 8 &  1114.9\\
\hline\noalign{\smallskip}
\end{tabular}
\end{center}
\end{table*}

First we would like to illustrate the need for efficient linear solvers. For that
reason Table \ref{tab:solver_comparison} compares the performance 
of two different linear solvers, the BiCGStab method preconditioned by block ILU
(blocks corresponding to all degrees of freedom associated with a mesh element) 
and the hybrid AMG/DG preconditioner
presented in Section \ref{Sec:Solver}. 10 time steps of the DNAPL infiltration problem have
been computed with the DG/$\mathbb{Q}_1$ scheme and $\Delta t/h$ fixed. 
The table lists the number of spatial degrees of freedom, the average number of 
preconditioner evaluations per Newton step (ALIN), the maximum number of preconditioner
evaluations in one Newton step (MAXLIN) and the total computation time (TT)
in seconds (including Jacobian assembly)
on one Intel Core i7 processor running at 2.6 GHz. The single grid preconditioner
clearly shows the expected doubling of the number of iterations while the
AMG preconditioner needs a constant average number of iterations and the maximum
number slowly increasing. With respect to total computation time, the single grid method
is faster or comparable to the AMG preconditioner up to 30000 degrees of freedom. At 500000
degrees of freedom the AMG method is faster by a factor of two in total computation time.
Note that on the finest mesh about 700$s$ are used for Jacobian assembly meaning that
the AMG solver alone is about 3.5 times faster. 

Table \ref{tab:dnapl2d_scalability_comparison} now provides more details of the simulation
runs for test case 3 reported in Subsection \ref{sec:fine_mesh} including total computation times. 
All computations have been
performed with the cluster system Helics3a at Heidelberg university which consists of
32 nodes with four AMD Opteron 6212 (Interlagos) processors operating at 2.6 GHz and
connected by a Mellanox 40G QDR infiniband network. Each processor has eight cores resulting
in 1024 cores for the full machine.

In the previous Subsection we concluded that the DG solution on the $320\times 192$ mesh
is as accurate as the CCFV solution on the $1280\times 768$ mesh. The total computation
time for the DG scheme on 64 cores was $3251s$ compared to $6496s$ for the CCFV scheme
on the same number of cores. DG on the $640\times 392$ mesh compares to CCFV
on the $2560\times 1536$ mesh with corresponding total computation times of
$11233s$ and $12775s$ on 256 cores. This shows that the DG scheme can provide an
advantage compared to the CCFV scheme even in the most relevant measure which is total computation
time. 

The less favourable comparison of both schemes on the finer grid is due to the worse
scalability of the AMG/DG preconditioner compared the AMG preconditioner applied to the linear 
systems arising in the CCFV scheme. The column labeled ALIN reports the average number of
preconditioner steps per Newton step. Clearly, the AMG preconditioner is more robust
for the CCFV scheme. The time needed for one application of the preconditioner is shown
in the column labeled LTIT. It shows that the time per iteration scales slightly better for the
DG scheme than for CCFV which might be due to better data locality of the DG scheme. 

Finally, Table \ref{tab:dnapl2d_scalability_comparison} illustrates that 
the fully-coupled DG scheme with half the time step size needs roughly the same
number of Newton iterations as the fully-coupled CCFV scheme on the same mesh.
This confirms that the nonlinear systems arising from the DG discretization can be solved
as efficiently as in the finite volume case.

\begin{table*}
\caption{Weak scalability results for test case 3. $P$ is the number of processors used,
$h^{-1}$ is the mesh size, DOF is the number spatial degrees of freedom,
TS is the number of successful time steps need to reach the final time $T=3600s$,
ANL is the average number of Newton iterations per (successful) time step,
ALIN is the average number of preconditioner steps needed per linear solve, TT is
the total computation time in seconds and LTIT is the time for one application of the preconditioner.}
\begin{center}
\begin{tabular}{lrrrrrrrr}
\hline
Scheme & $P$ & $h^{-1}$ & DOF & TS & ANL & ALIN & TT & LTIT \\
\hline
DG/$\mathbb{Q}_1$ & 1 & 40 & $7.7\cdot 10^3$ & 30 & 11.4 & 1.8 & 156.8 & 0.144\\
      & 4  & 80 & $3.1\cdot 10^4$ & 71  & 11.0 & 3.4 & 466.1 & 0.052\\
      &16  & 160& $1.2\cdot 10^5$ & 125  & 12.0 & 5.7 & 1187.3 & 0.062\\
      &64 & 320 &$4.9\cdot 10^5$ & 263  & 11.4 & 9.1 &  3250.5 & 0.067\\
      &256 & 640 & $2.0\cdot 10^6$ & 563  & 10.7 & 15.3 & 11233.0 & 0.082\\
      &1024 & 1280 & $7.9\cdot 10^6$ & 1369  & 9.2 & 17.0 & 41917.4 & 0.112\\
\hline
2nd order & 1  & 160 & $3.1\cdot 10^4$ & 60 & 7.7 & 1.8 & 191.2 & 0.043 \\
CCFV    &  4  & 320& $1.2\cdot 10^5$ & 120 & 8.2 & 2.2 & 592.1 & 0.069\\
    & 16  & 640& $4.9\cdot 10^5$ & 246 & 8.4 & 2.6 & 1453.7 & 0.078\\
    & 64 & 1280 & $2.0\cdot 10^6$ & 491 & 8.9 & 3.3 & 6496.1 & 0.151\\
    &256 & 2560&$7.9\cdot 10^6$ & 1021 & 9.7 & 4.5 & 12774.9 & 0.156\\
\hline
\end{tabular}
\end{center}
\label{tab:dnapl2d_scalability_comparison}       
\end{table*}

\subsection{Test Case 4: DNAPL Infiltration in Random Porous Medium}

In order to illustrate that the DG scheme is able to handle more difficult problems 
and also performs well in three space dimensions we applied it to DNAPL infiltration
into a random porous medium with log-normally distributed absolute permeability
with correlation length $6h$ in $x$-direction and $3h$ in $z$-direction ($h$ is the mesh size).
The variance has been choosen such that permeability varied about by about 1.5 
orders of magnitude in the two-dimensional example and one order of magnitude in the
three-dimensional example. The capillary pressure-saturation function is of Brooks-Corey
type with $\lambda=2.5$ and entry pressure is scaled with permeability
$$\pi(s_w,x) = \sqrt{\bar{K}/K(x)} s_w^{-1/\lambda}$$
where $\bar{K}$ is the mean of the permeability field. This results in a different entry pressure
for each mesh element.

\begin{figure*}
\includegraphics[width=0.333\textwidth]{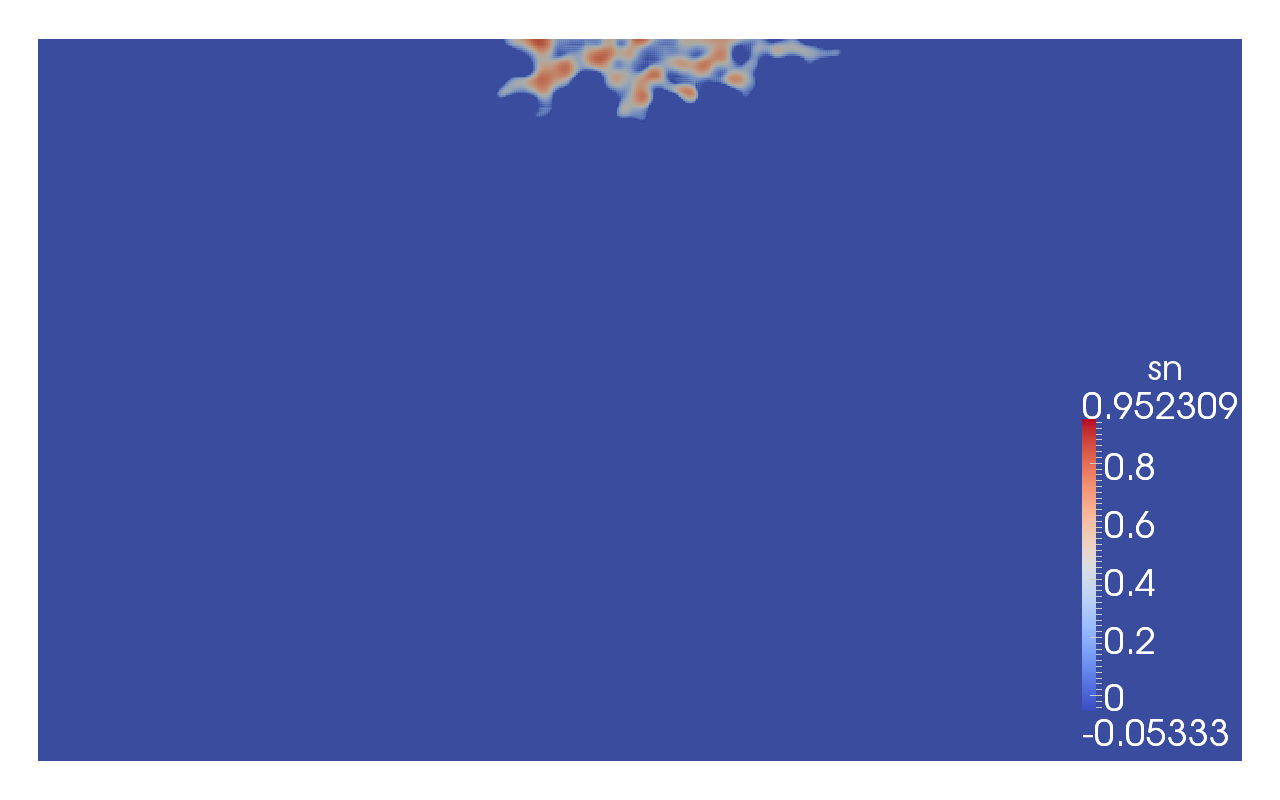}\hfill
\includegraphics[width=0.333\textwidth]{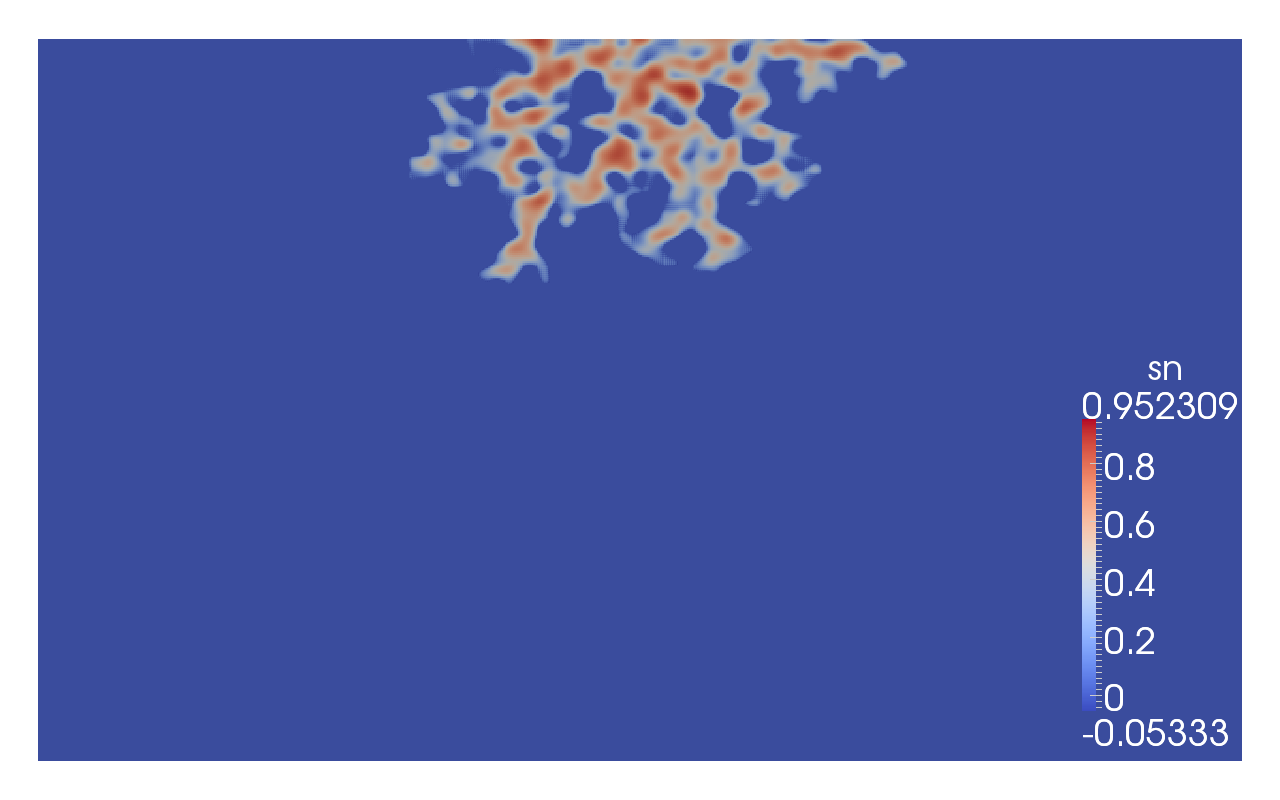}\hfill
\includegraphics[width=0.333\textwidth]{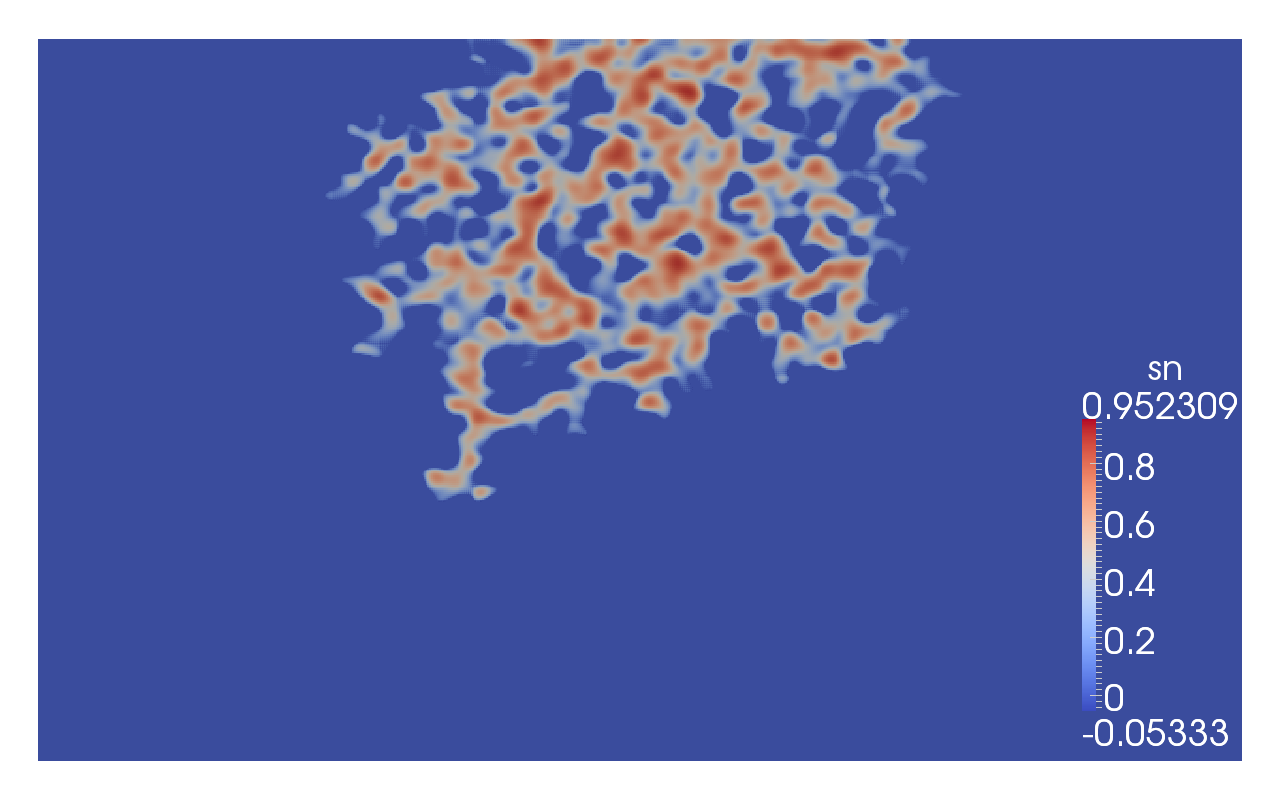}\\
\includegraphics[width=0.333\textwidth]{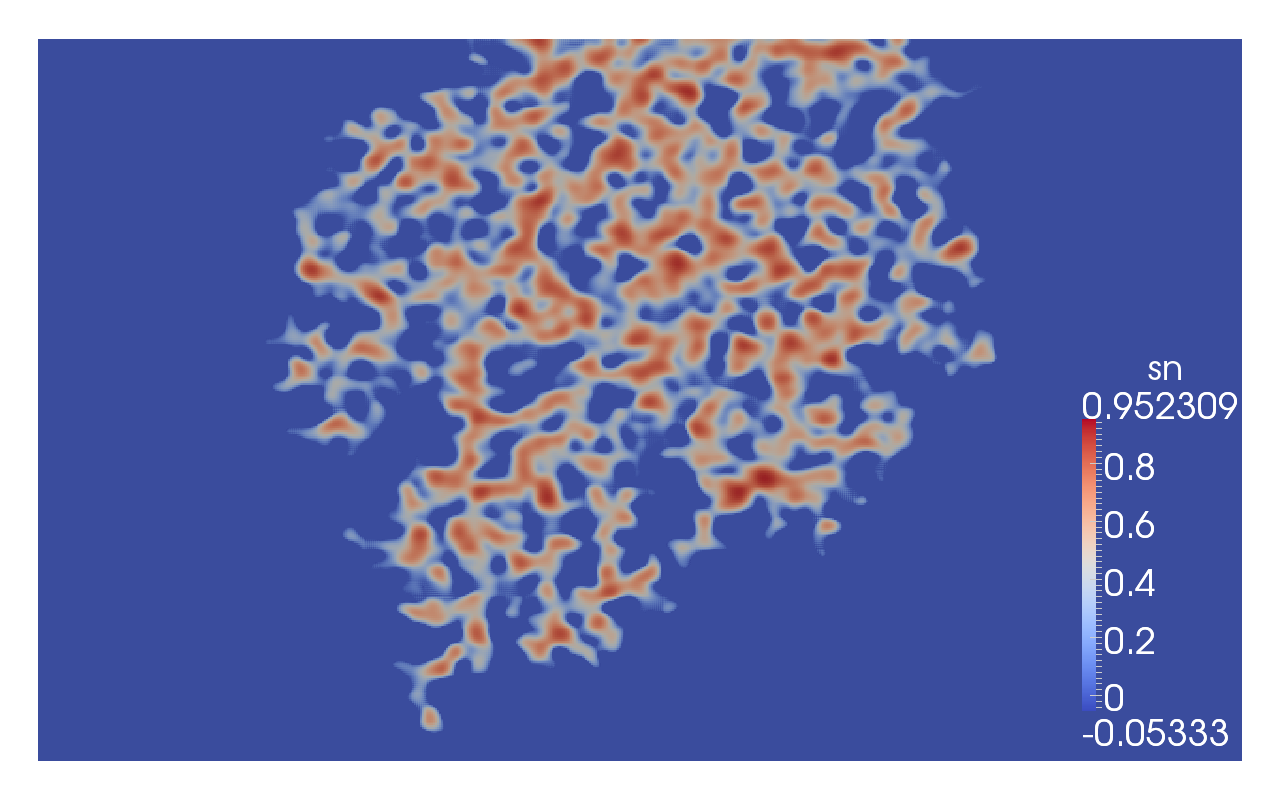}\hfill
\includegraphics[width=0.333\textwidth]{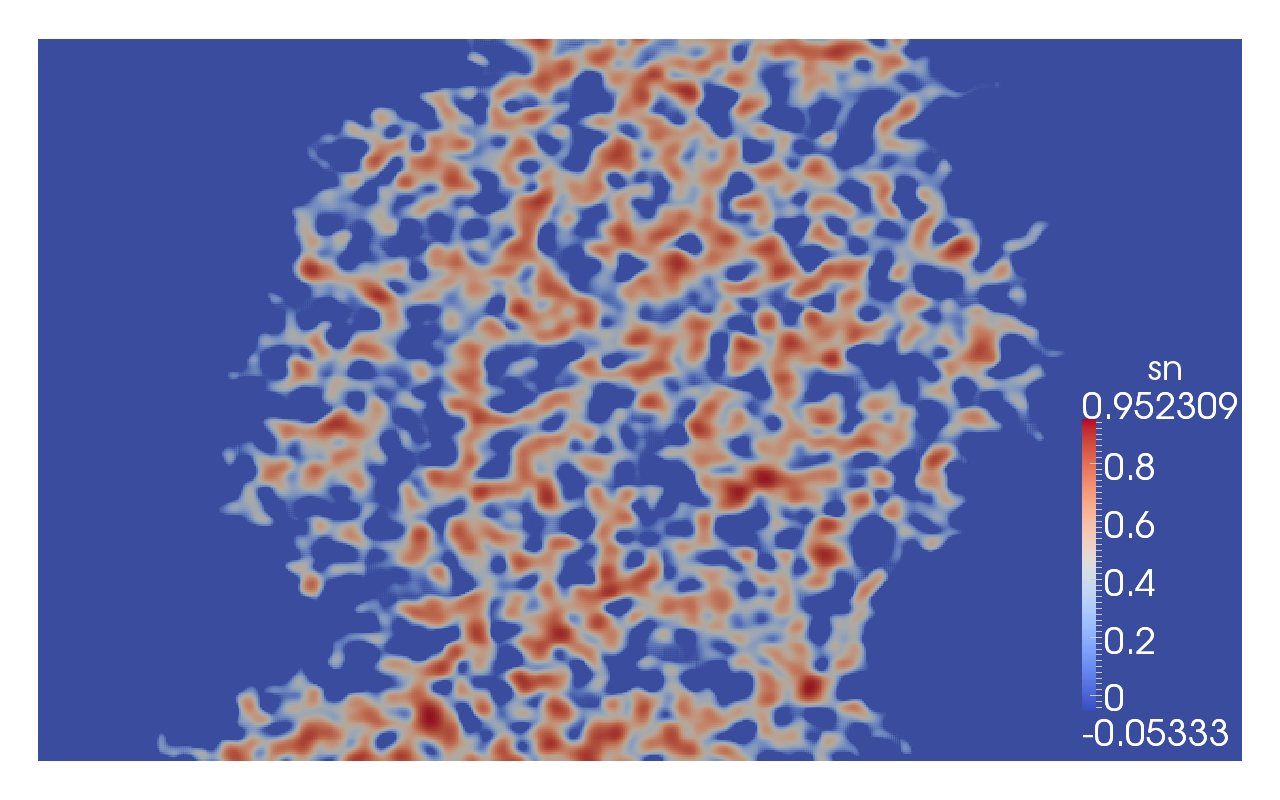}\hfill
\includegraphics[width=0.333\textwidth]{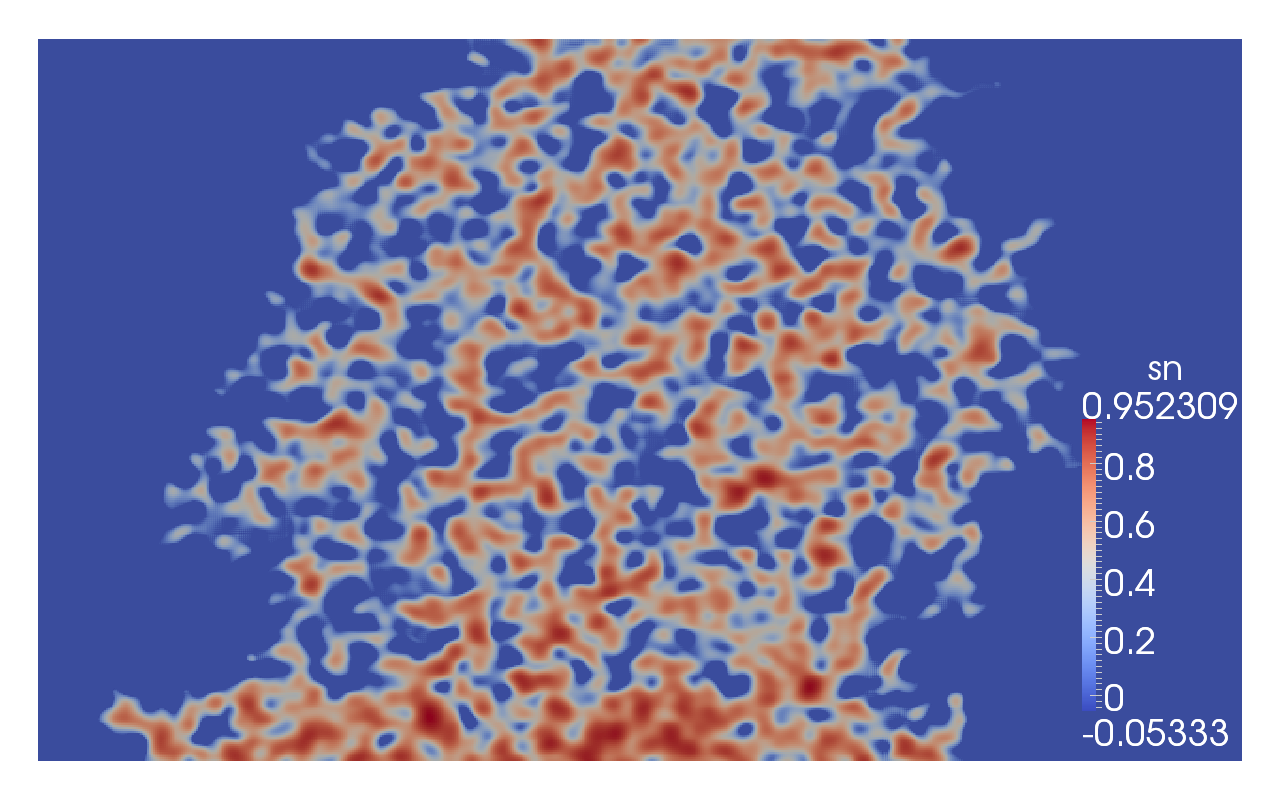}
\caption{2D DNAPL infiltration into a porous medium with random permeability and entry pressure. Images
show saturation of non-wetting phase.}
\label{fig:random_2d}       
\end{figure*}

\begin{figure*}
\includegraphics[width=0.24\textwidth]{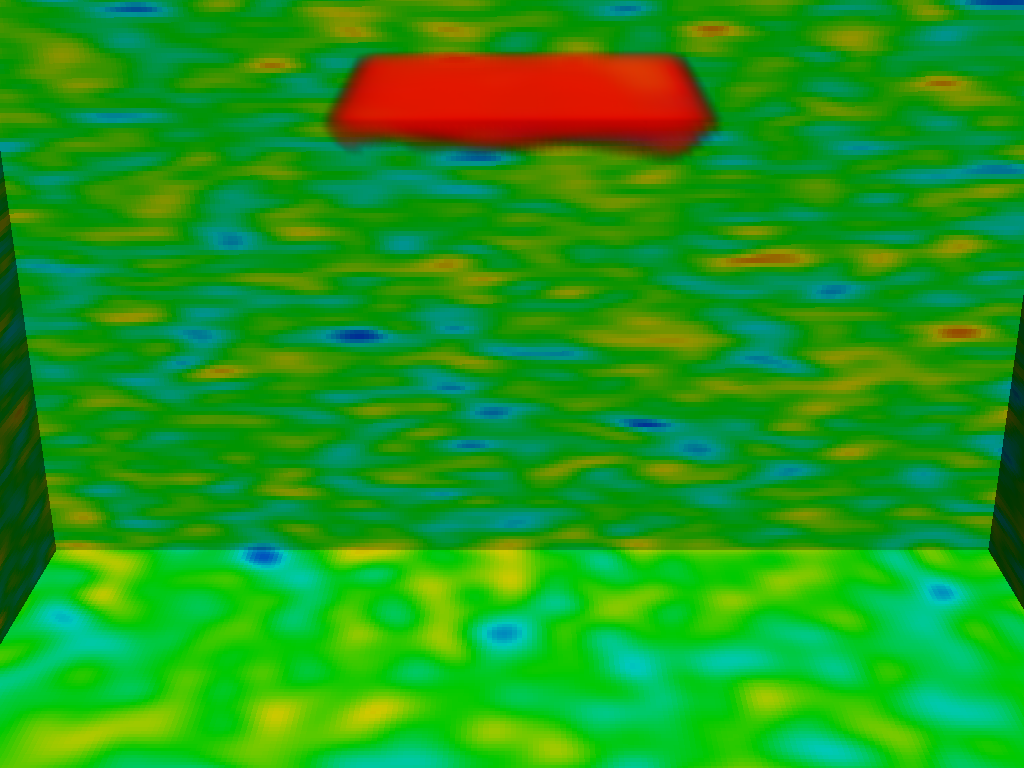}\hfill
\includegraphics[width=0.24\textwidth]{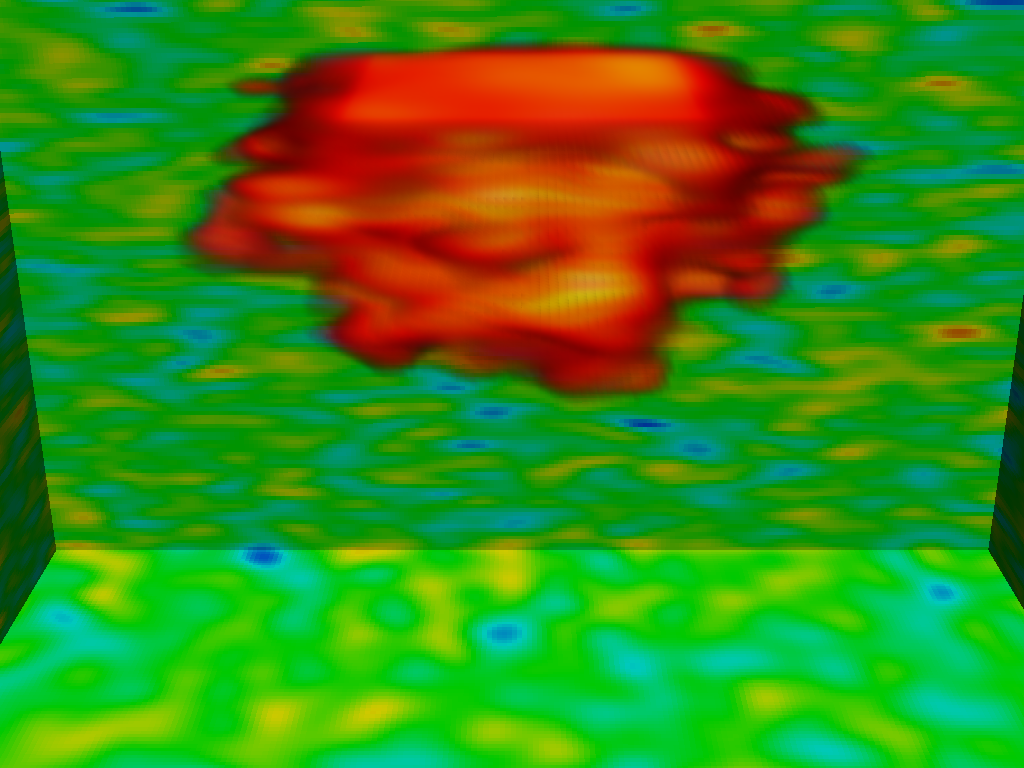}\hfill
\includegraphics[width=0.24\textwidth]{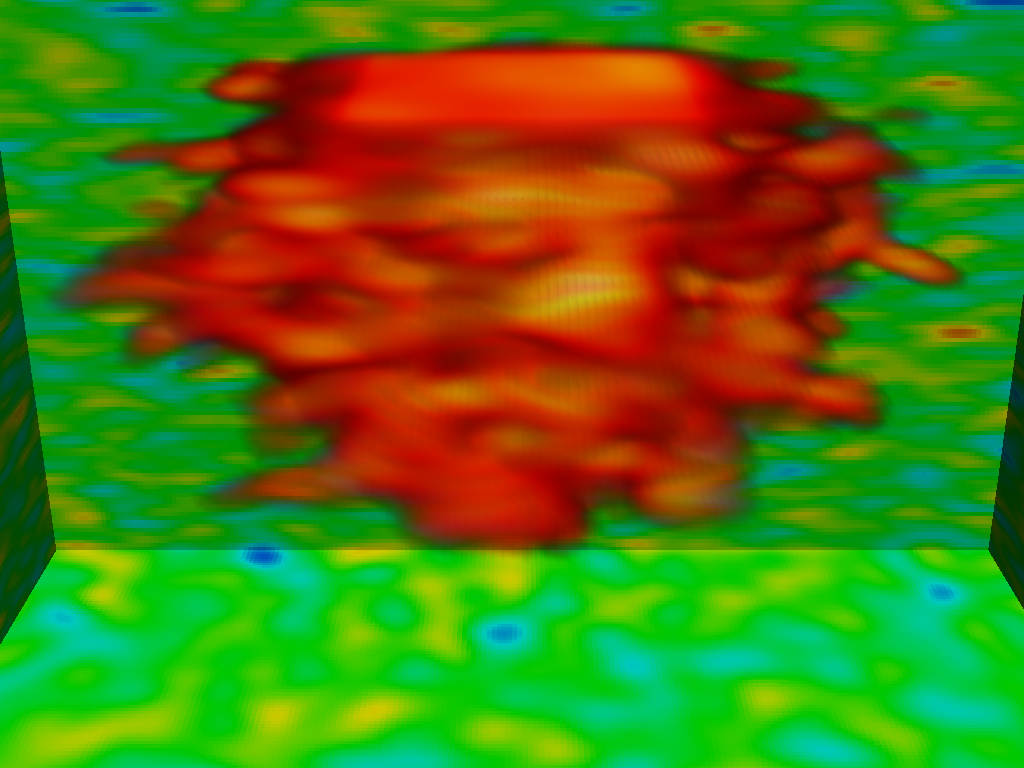}\hfill
\includegraphics[width=0.24\textwidth]{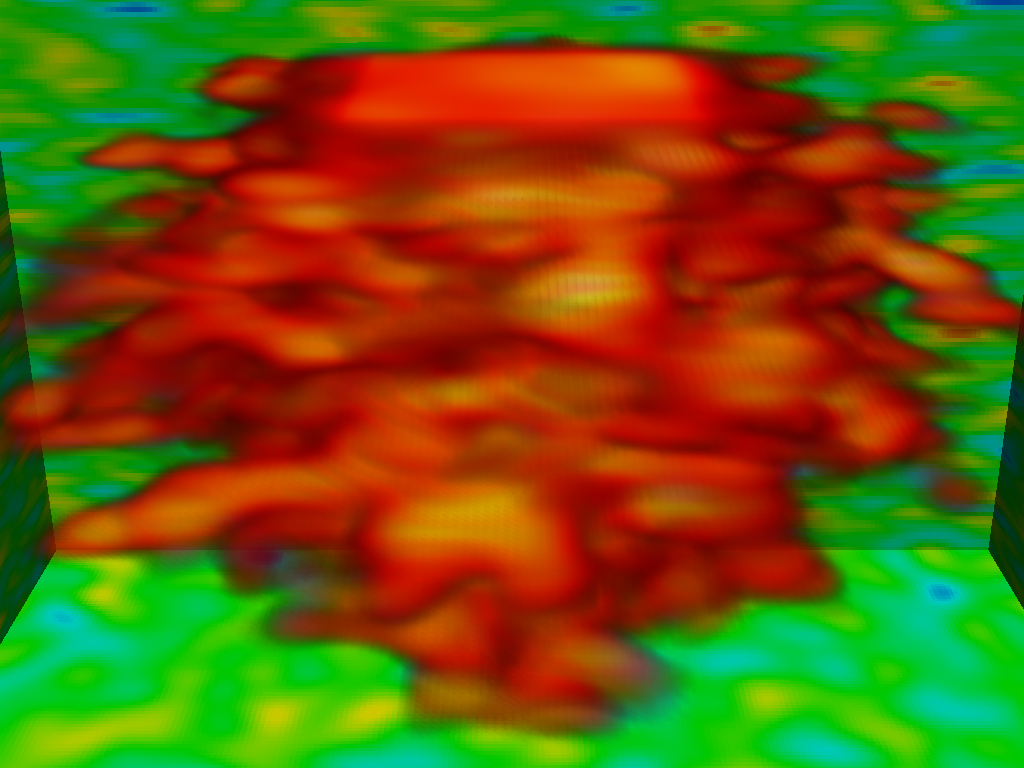}
\caption{3D DNAPL infiltration into a porous medium with random permeability and entry pressure.
Images show volume rendering of non-wetting phase saturation in red and the absolute
permability on the sides of the cube.}
\label{fig:random_3d}       
\end{figure*}

Figures \ref{fig:random_2d} and \ref{fig:random_3d} show the results obtained with 
the DG/$\mathbb{Q}_1$ scheme at various times. Clearly, capillary heterogeneity has
a strong influence on the saturation distribution even for relatively mild variations
in absolute permeability. Additional information on these simulations is shown 
in Table \ref{tab:test_case_4_simulation_data}. The 2D simulation used two million
spatial degrees of freedom and computed about 12000 time steps on 512 processors
in about $8h$ total computation time. The 3D simulation used nearly 100 million
spatial degrees of freedom and performed about 1000 time steps which took
about $100h$ total computation time on 1024 processors. It is interesting to note that
in the 3D simulation the ratio of time spent for assembling the Jacobians to time
spent for solving the linear systems (last column in Table \ref{tab:test_case_4_simulation_data})
was $1.4$ while for the 2D simulation it was $0.5$.

\begin{table*}
\caption{Simulation data for the random DNAPL infiltration problem (test case 4).}
\label{tab:test_case_4_simulation_data}       
\begin{center}
\begin{tabular}{lrrrrrrrr}
\hline
Problem  & $P$ & DOF & TS & ANL & ALIN & TT [h] & ASS/SLV\\
\hline
2D & 512   & $2\cdot 10^6$    & 12019 & 4.1 & 2.5 & 8.3 & 0.5\\
3D & 1024 & $8.8\cdot 10^7$ &   1026 & 6.0 & 4.3 & 107.0 & 1.4\\
\hline
\end{tabular}
\end{center}
\end{table*}

\section{Conclusion}

In this work a new fully-coupled discontinuous Galerkin scheme for the two-phase flow problem based on 
a formulation with wetting-phase potential and capillary potential as primary variables is presented.
By way of numerical experiment it is shown that (i) the scheme is as
accurate as a cell-centered finite volume scheme on a two times refined grid, (ii) no
$H(\text{div})$ reconstruction of the velocity is necessary in the fully-coupled scheme in contrast
to some decoupled schemes and (iii) an efficient parallel algebraic multigrid preconditioner
for the fully-coupled two-phase DG system is available. 
Even when compared to the very cheap cell-centered scheme
significant speedups w.r.t. total computation time can be achieved. In this comparison it should be
taken into account that DG is a much more flexible approach that is able to handle
unstructured, non-conforming grids, $hp$-adaptivity and full tensor permeabilities.
Problems with up to about 100 million degrees of freedom in three space
dimension are solved on 1000 cores. Future work should include a numerical
comparison with pressure-saturation based formulations as well as decoupled formulations.

\paragraph{Acknowledgements}

This work would not have been possible without the wonderful effort of
the DUNE developer community during the last decade. There devotion to producing high quality
open-source software is greatly appreciated. I also thank Olaf Ippisch for many
fruitful discussions about pressure-pressure based methods for two-phase flow.



\small
\bibliographystyle{plain}
\bibliography{dg_two_phase}   

\begin{thebibliography}{10}

\bibitem{AizingerDawsonCockburnCastillo01}
V.~Aizinger, C.~Dawson, B.~Cockburn, and P.~Castillo.
\newblock The local discontinuous {G}alerkin method for contaminant transport.
\newblock {\em Adv. Wat. Res.}, 24:73--87, 2001.

\bibitem{alexander:77}
R.~Alexander.
\newblock Diagonally implicit {R}unge-{K}utta methods for stiff {O. D. E.'s}.
\newblock {\em SIAM Journal on Numerical Analysis}, 14(6):1006--1021, 1977.

\bibitem{DGUnified02}
{D. N.} Arnold, F.~Brezzi, B.~Cockburn, and {L. D.} Marini.
\newblock Unified analysis of discontinuous {G}alerkin methods for elliptic
  problems.
\newblock {\em SIAM J. Numer. Anal.}, 39(5):1749--1779, 2002.

\bibitem{ayuso11:multilevel}
B.~Ayuso De~Dios, M.~Holst, Yunrong Zhu, and L.T. Zikatanov.
\newblock Multilevel preconditioners for discontinuous {G}alerkin
  approximations of elliptic problems with jump coefficients.
\newblock Technical report, arXiv:1012.1287, 2011.

\bibitem{AzizSettari}
K.~Aziz and A.~Settari.
\newblock {\em Petroleum Reservoir Simulation}.
\newblock Elsevier, 1979.

\bibitem{Bastian99}
P.~Bastian.
\newblock Numerical computation of multiphase flow in porous media.
\newblock Habilitationsschrift, 1999.

\bibitem{Cisc2002}
P.~Bastian.
\newblock Higher order discontinuous {G}alerkin methods for flow and transport
  in porous media.
\newblock In E.~B{\"{a}}nsch, editor, {\em Challenges in Scientific Computing
  -- CISC 2002}, volume~35 of {\em Lecture Notes in Computational Science and
  Engineering}, pages 1--22. Springer, 2003.

\bibitem{Dune2008b}
P.~Bastian, M.~Blatt, A.~Dedner, C.~Engwer, R.~Kl{\"{o}}fkorn, R.~Kornhuber,
  M.~Ohlberger, and O.~Sander.
\newblock A generic grid interface for parallel and adaptive scientific
  computing. part {II}: implementation and tests in {DUNE}.
\newblock {\em Computing}, 82(2-3):121--138, 2008.

\bibitem{Dune2008a}
P.~Bastian, M.~Blatt, A.~Dedner, C.~Engwer, R.~Kl{\"{o}}fkorn, M.~Ohlberger,
  and O.~Sander.
\newblock A generic grid interface for parallel and adaptive scientific
  computing. part {I}: abstract framework.
\newblock {\em Computing}, 82(2-3):103--119, 2008.

\bibitem{pdelabalgoritmy}
P.~Bastian, F.~Heimann, and S.~Marnach.
\newblock Generic implementation of finite element methods in the distributed
  and unified numerics environment ({DUNE}).
\newblock {\em Kybernetika}, 46(2):294--315, 2010.

\bibitem{BH99}
P.~Bastian and R.~Helmig.
\newblock Efficient fully-coupled solution techniques for two-phase flow in
  porous media: {P}arallel multigrid solution and large scale computations.
\newblock {\em Adv. Water Res.}, 23:199--216, 1999.

\bibitem{BastianLangCouplex}
P.~Bastian and S.~Lang.
\newblock Couplex benchmark computations with {UG}.
\newblock {\em Computational Geosciences}, 8(2):125--147, 2004.

\bibitem{BastianRiviere}
P.~Bastian and B.~Rivi{\`{e}}re.
\newblock Superconvergence and {H}(div)-projection for discontinuous {G}alerkin
  methods.
\newblock {\em Int. J. Numer. Meth. Fluids.}, 42(10):1043--1057, 2003.

\bibitem{BastianRiviereTwoPhase}
P.~Bastian and B.~Rivi{\`{e}}re.
\newblock Discontinuous galerkin methods for two-phase flow in porous media.
\newblock Technical Report 2004-28, IWR, University of Heidelberg, 2004.

\bibitem{amg4dg}
Peter Bastian, Markus Blatt, and Robert Scheichl.
\newblock Algebraic multigrid for discontinuous galerkin discretizations of
  heterogeneous elliptic problems.
\newblock {\em Numerical Linear Algebra with Applications}, 19(2):367--388,
  2012.

\bibitem{Bear72}
J.~Bear.
\newblock {\em Dynamics of Fluids in Porous Media}.
\newblock Dover Publications, 1972.

\bibitem{Bertsch2003}
M.~Bertsch, R.~Passo, and C.~Van~Duijn.
\newblock Analysis of oil trapping in porous media flow.
\newblock {\em SIAM Journal on Mathematical Analysis}, 35(1):245--267, 2003.

\bibitem{blattamg}
Markus Blatt.
\newblock {\em A Parallel Algebraic Multigrid Method for Elliptic Problems with
  Highly Discontinuous Coefficients}.
\newblock PhD thesis, Ruprecht--Karls--Universit\"at Heidelberg, 2010.

\bibitem{BourgeatJurakSmai2009}
Alain Bourgeat, Mladen Jurak, and Farid Sma\"{i}.
\newblock Two-phase, partially miscible flow and transport modeling in porous
  media; application to gas migration in a nuclear waste repository.
\newblock {\em Computational Geosciences}, 13(1):29--42, 2009.

\bibitem{braess95:amg}
D.~Braess.
\newblock Towards algebraic multigrid for elliptic problems of second order.
\newblock {\em Computing}, 55:379--393, 1995.

\bibitem{Brenner2013}
Konstantin Brenner, Cl\'{e}ment Canc\`{e}s, and Danielle Hilhorst.
\newblock Finite volume approximation for an immiscible two-phase flow in
  porous media with discontinuous capillary pressure.
\newblock {\em Computational Geosciences}, 17(3):573--597, 2013.

\bibitem{brenner05:_conver_multig_algor_inter_penal_method}
Susanne~C. Brenner and Jie Zhao.
\newblock Convergence of multigrid algorithms for interior penalty methods.
\newblock {\em Applied Numerical Analysis and Computational Mathematics},
  2(1):3--18, 2005.

\bibitem{BC}
{R. H.} Brooks and {A. T.} Corey.
\newblock {\em Hydraulic Properties of Porous Media}, volume~3 of {\em Colorado
  State University Hydrology Paper}.
\newblock Colorado State University, 1964.

\bibitem{Buzzi2009}
Fulvia Buzzi, Michael Lenzinger, and Ben Schweizer.
\newblock Interface conditions for degenerate two-phase flow equations in one
  space dimension.
\newblock {\em Analysis}, 29(3):299--316, 2009.

\bibitem{Cances2009}
C.~Canc\`{e}s, T.~Gallou\"{e}t, and A.~Porretta.
\newblock Two-phase flows involving capillary barriers in heterogeneous porous
  media.
\newblock {\em Interfaces and Free Boundaries}, 11:239--258, 2009.

\bibitem{doi:10.1137/11082943X}
C.~Canc\`{e}s and M.~Pierre.
\newblock An existence result for multidimensional immiscible two-phase flows
  with discontinuous capillary pressure field.
\newblock {\em SIAM Journal on Mathematical Analysis}, 44(2):966--992, 2012.

\bibitem{MZA:8194530}
Cl\'{e}ment Canc\`{e}s.
\newblock Finite volume scheme for two-phase flows in heterogeneous porous
  media involving capillary pressure discontinuities.
\newblock {\em ESAIM: Mathematical Modelling and Numerical Analysis},
  43:973--1001, 9 2009.

\bibitem{chavent}
G.~Chavent and J.~Jaffr{\'{e}}.
\newblock {\em Mathematical Models and Finite Elements for Reservoir
  Simulation}.
\newblock North--Holland, 1978.

\bibitem{CO2Benchmark2009}
Holger Class, Anozie Ebigbo, Rainer Helmig, Helge~K. Dahle, Jan~M. Nordbotten,
  Michael~A. Celia, Pascal Audigane, Melanie Darcis, Jonathan Ennis-King,
  Yaqing Fan, Bernd Flemisch, Sarah~E. Gasda, Min Jin, Stefanie Krug, Diane
  Labregere, Ali Naderi~Beni, Rajesh~J. Pawar, Adil Sbai, Sunil~G. Thomas,
  Laurent Trenty, and Lingli Wei.
\newblock A benchmark study on problems related to {CO$_2$} storage in geologic
  formations.
\newblock {\em Computational Geosciences}, 13(4):409--434, 2009.

\bibitem{DGProceedings00}
B.~Cockburn, {S. Y.} Lin, and {C.-W.} Shu, editors.
\newblock {\em Discontinuous {G}alerkin methods. {T}heory, computation and
  applications}, volume~11 of {\em Lecture Notes in Computational Science and
  Engineering}. Springer-Verlag, 2000.

\bibitem{LDG98}
B.~Cockburn and {C.-W.} Shu.
\newblock The local discontinuous {G}alerkin finite element method for
  convection-diffusion systems.
\newblock {\em SIAM J. Numer. Anal.}, 35:2440--2463, 1998.

\bibitem{DGV}
B.~Cockburn and {C.-W.} Shu.
\newblock The {R}unge-{K}utta discontinuous {G}alerkin method for conservation
  laws {V}: {M}ultidimensional systems.
\newblock {\em J. Comput. Phys.}, 141:199--224, 1998.

\bibitem{Molenaar97}
{M. J.} de~Neef and J.~Molenaar.
\newblock Analysis of {DNAPL} infiltration in a medium with a low permeable
  lense.
\newblock {\em Computational Geosciences}, 1:191--214, 1997.

\bibitem{DiPietroErn2012}
{Daniele Antonio} {Di Pietro} and Alexandre Ern.
\newblock {\em Mathematical Aspects of Discontinuous {G}alerkin Methods}.
\newblock Springer, 2012.

\bibitem{Ern2008}
{Daniele Antonio} {Di Pietro}, Alexandre Ern, and Jean-Luc Guermond.
\newblock Discontinuous galerkin methods for anisotropic semidefinite diffusion
  with advection.
\newblock {\em SIAM J. Numer. Anal.}, 2008.

\bibitem{VanDuijn1995}
C.J. Duijn, J.~Molenaar, and M.J. {de Neef}.
\newblock The effect of capillary forces on immiscible two-phase flow in
  heterogeneous porous media.
\newblock {\em Transport Porous Media}, 21:71--93, 1995.

\bibitem{VanDuijn1998}
C.J. Duijn, J.~Molenaar, and M.J. {de Neef}.
\newblock Similarity solution for capillary redistribution of two phases in a
  porous medium with a single discontinuity.
\newblock {\em Adv. Water Res.}, 21:451--461, 1998.

\bibitem{Epshteyn2007383}
Y.~Epshteyn and B.~Rivi\`{e}re.
\newblock Fully implicit discontinuous finite element methods for two-phase
  flow.
\newblock {\em Applied Numerical Mathematics}, 57(4):383 -- 401, 2007.

\bibitem{ErnMozolevskiSchuh2009}
A.~Ern, I.~Mozolevski, and L.~Schuh.
\newblock Accurate velocity reconstruction for discontinuous galerkin
  approximations of two-phase porous media flows.
\newblock {\em C.R. Acad. Sci. Paris}, Ser. I(9--10):551--554, 2009.

\bibitem{Ern20101491}
A.~Ern, I.~Mozolevski, and L.~Schuh.
\newblock Discontinuous galerkin approximation of two-phase flows in
  heterogeneous porous media with discontinuous capillary pressures.
\newblock {\em Computer Methods in Applied Mechanics and Engineering},
  199(23--24):1491 -- 1501, 2010.

\bibitem{Ern2012348}
A.~Ern, I.~Mozolevski, and L.~Schuh.
\newblock Corrigendum to ``{D}iscontinuous {G}alerkin approximation of
  two-phase flows in heterogeneous porous media with discontinuous capillary
  pressures'' [comput. methods appl. mech. engrg. 199 (2010) 1491-1501].
\newblock {\em Computer Methods in Applied Mechanics and Engineering},
  245-246(0):348 -- 349, 2012.

\bibitem{ErnHdiv2007}
A.~Ern, S.~Nicaise, and M.~Vohral\'{i}k.
\newblock An accurate ${H}(\text{div})$ flux reconstruction for discontinuous
  galerkin approximations of elliptic problems.
\newblock {\em C. R. Math. Acad. Sci. Paris}, 2007.

\bibitem{ErnMozolevski2012}
Alexandre Ern and Igor Mozolevski.
\newblock Discontinuous {G}alerkin method for two-component liquid-gas porous
  media flows.
\newblock {\em Computational Geosciences}, 16(3):677--690, 2012.

\bibitem{Ern01042009}
Alexandre Ern, Annette~F. Stephansen, and Paolo Zunino.
\newblock A discontinuous {G}alerkin method with weighted averages for
  advection-diffusion equations with locally small and anisotropic diffusivity.
\newblock {\em IMA Journal of Numerical Analysis}, 29(2):235--256, 2009.

\bibitem{Eslinger2005}
{O.J.} Eslinger.
\newblock {\em Discontinuous Galerkin finite element methods applied to
  two-phase, air-water flow problems}.
\newblock PhD thesis, University of Texas at Austin,, 2005.

\bibitem{FVCA6}
J.~Fo\v{r}t, J.~F\"{u}rst, J.~Halama, R.~Herbin, and F.~Hubert, editors.
\newblock {\em Finite volumes for complex applications {VI}: {P}roblems \&
  perspectives}, volume~4 of {\em Proceedings in mathematics}. Springer, 2011.

\bibitem{NME:NME2579}
Christophe Geuzaine and Jean-Fran\c{c}ois Remacle.
\newblock Gmsh: A 3-d finite element mesh generator with built-in pre- and
  post-processing facilities.
\newblock {\em International Journal for Numerical Methods in Engineering},
  79(11):1309--1331, 2009.

\bibitem{kanschat03:dg_multilevel}
J.~Gopalakrishnan and G.~Kanschat.
\newblock A multilevel discontinuous {G}alerkin method.
\newblock {\em Numerische Mathematik}, 95:527--550, 2003.

\bibitem{Bibel}
W.~Hackbusch.
\newblock {\em Multi--Grid Methods and Applications}.
\newblock Springer--Verlag, 1985.

\bibitem{Helmig97}
R.~Helmig.
\newblock {\em Multiphase Flow and Transport Processes in the Subsurface -- A
  Contribution to the Modeling of Hydrosystems}.
\newblock Springer--Verlag, 1997.

\bibitem{Helmig1998697}
Rainer Helmig and Ralf Huber.
\newblock Comparison of galerkin-type discretization techniques for two-phase
  flow in heterogeneous porous media.
\newblock {\em Advances in Water Resources}, 21(8):697 -- 711, 1998.

\bibitem{Hoteit200856}
Hussein Hoteit and Abbas Firoozabadi.
\newblock Numerical modeling of two-phase flow in heterogeneous permeable media
  with different capillarity pressures.
\newblock {\em Advances in Water Resources}, 31(1):56 -- 73, 2008.

\bibitem{HoustonHartmann2008}
P.~Houston and R.~Hartmann.
\newblock An optimal order interior penalty discontinuous {G}alerkin
  discretization of the compressible {N}avier-{S}tokes equations.
\newblock {\em J. Comp. Phys.}, 227:9670--9685, 2008.

\bibitem{johannsen05:nipg_multigrid}
Klaus Johannsen.
\newblock Multigrid methods for nonsymmetric interior penalty discontinuous
  {G}alerkin methods.
\newblock ICES Report 05-23, University of Texas at Austin, 2005.

\bibitem{KlieberRiviere2006}
W.~Klieber and B.~Rivi{\`{e}}re.
\newblock Adaptive simulations of two-phase flow by discontinuous galerkin
  methods.
\newblock {\em Comput. Meth. Appl. Mech. Engrg.}, 2006.

\bibitem{Kueper1}
{B. H.} Kueper and {E. O.} Frind.
\newblock Two--phase flow in heterogeneous porous media 1. model development.
\newblock {\em Water Resources Research}, 27(6):1049--1057, 1991.

\bibitem{Kueper2}
{B. H.} Kueper and {E. O.} Frind.
\newblock Two--phase flow in heterogeneous porous media 2. model application.
\newblock {\em Water Resources Research}, 27(6):1059--1070, 1991.

\bibitem{Eike2013}
{E.H.} Mueller and R.~Scheichl.
\newblock Massively parallel solvers for elliptic {PDE}s in numerical weather-
  and climate prediction.
\newblock arXiv:1307.2036, 2013.

\bibitem{Nayagum2004}
D.~Nayagum, G.~Sch\"{a}fer, and R.~Mos\'{e}.
\newblock Modelling two-phase incompressible flow in porous media using mixed
  hybrid and discontinuous finite elements.
\newblock {\em Computational Geosciences}, 2004.

\bibitem{co2_2011}
Rebecca Neumann, Peter Bastian, and Olaf Ippisch.
\newblock Modeling and simulation of two-phase two-component flow with
  disappearing nonwetting phase.
\newblock {\em Computational Geosciences}, pages 1--11, 2012.

\bibitem{olson_schroder_2011_jcp}
Luke~N. Olson and Jacob~B. Schroder.
\newblock Smoothed aggregation multigrid solvers for high-order discontinuous
  galerkin methods for elliptic problems.
\newblock {\em Journal of Computational Physics}, 230(18):6959 -- 6976, 2011.

\bibitem{Peaceman}
D.~W. Peaceman.
\newblock {\em Fundamentals of Numerical Reservoir Simulation}.
\newblock Elsevier, 1977.

\bibitem{PLH09a}
Florian Prill, Maria Luk\'{a}\v{c}ov\'{a}-Medvid'ov\'{a}, and Ralf Hartmann.
\newblock Smoothed aggregation multigrid for the discontinuous {G}alerkin
  method.
\newblock {\em SIAM Journal on Scientific Computing}, 31:3503--3528, 2009.

\bibitem{Raw}
M.~Raw.
\newblock Robustness of coupled algebraic multigrid for the {N}avier--{S}tokes
  equations.
\newblock Technical Report 96--0297, AIAA, 1996.

\bibitem{Kluft2004}
V.~Reichenberger, H.~Jakobs, P.~Bastian, and R.~Helmig.
\newblock A mixed-dimensional finite volume method for multiphase flow in
  fractured porous media.
\newblock {\em Adv. Wat. Res.}, 29(7):1020--1036, 2006.

\bibitem{RiviereEccomas2004}
B.~Rivi{\`{e}}re.
\newblock Numerical study of a discontinuous galerkin method for incompressible
  two-phase flow.
\newblock In {\em Proceedings of ECCOMAS 2004}, volume~2, 2004.

\bibitem{RiviereWheelerGirault99}
B.~Rivi{\`{e}}re, {M. F.} Wheeler, and V.~Girault.
\newblock Improved energy estimates for interior penalty, constrained and
  discontinuous {G}alerkin methods for elliptic problems {I}.
\newblock {\em Comput. Geosci.}, 3:337--360, 1999.

\bibitem{Riviere2008}
B\'{e}atrice Rivi\`{e}re.
\newblock {\em Discontinuous {G}alerkin methods for solving elliptic and
  parabolic equations}.
\newblock SIAM, 2008.

\bibitem{ruge87:multig_method_amg}
J.W. Ruge and K.~St\"uben.
\newblock Algebraic multigrid.
\newblock In S.~F. McCormick, editor, {\em Multigrid Methods}, chapter~4, pages
  73--130. SIAM Philadelphia, 1987.

\bibitem{TrottenbergBook2001}
U.Trottenberg, {C. W.} Oosterlee, and A.~Schu\"{u}ller.
\newblock {\em Multigrid}.
\newblock Academic Press, 2001.

\bibitem{PVanek_JMandel_MBrezina_1996a}
P.~Van{\v e}k, J.~Mandel, and M.~Brezina.
\newblock Algebraic multigrid based on smoothed aggregation for second and
  fourth order problems.
\newblock {\em Computing}, 56:179--196, 1996.

\bibitem{Wolff2013}
Markus Wolff, Yufei Cao, Bernd Flemisch, Rainer Helmig, and Barbara Wohlmuth.
\newblock Multipoint flux approximation {L}-method in 3d: numerical convergence
  and application to two-phase flow in porous media.
\newblock In P.~Bastian, J.~Kraus, R.~Scheichl, and M.~Wheeler, editors, {\em
  Simulation of Flow in Porous Media}, volume~12 of {\em Radon Series on
  Computational and Applied Mathematics}, pages 39--80. de Gruyter, 2013.

\bibitem{xu92:iterative_methods_by_space_decomp}
Jinchao Xu.
\newblock {Iterative Methods by Space Decomposition and Subspace Correction}.
\newblock {\em SIAM Review}, 34(4):581--613, 1992.

\end{thebibliography}


\end{document}